\documentclass[%
 reprint, 
 amsmath,amssymb,
 aps, physrev,
 nofootinbib,
]{revtex4-2}

\usepackage{graphicx}
\usepackage{dcolumn}
\usepackage{bm}
\usepackage{hyperref}

\usepackage{mathrsfs}

\usepackage{xcolor}

\begin{document}


\title{\textbf{Alternative Boundary Conditions in Asymptotically Anti-de Sitter Spacetimes} 
}%

\author{Robert M. Wald}
 \email{rmwa@uchicago.edu}
\author{Xinkang Wang}%
 \email{xinkangwang@uchicago.edu}
\affiliation{%
  Leinweber Institute for Theoretical Physics, University of Chicago, Chicago, IL 60637\\
  Department of Physics, University of Chicago, Chicago, IL 60637
}%


\begin{abstract}

Anti-de Sitter (AdS) spacetimes are not globally hyperbolic, so, in general, boundary conditions must be imposed on fields so as to have well-defined dynamics. Ishibashi and Wald (IW) determined the boundary conditions for scalar, electromagnetic, and linearized gravitational fields in AdS$_{(1,d)}$ that give rise to well-posed dynamics with a conserved, positive energy. However, the IW prescription involved the introduction of nonlocal potentials and the prescription was given only for individual spherical harmonic modes. In this paper, we determine which of the IW boundary conditions are local in the primary fields and which of these local conditions are AdS-invariant. In 4 spacetime dimensions, the standard AdS-invariant boundary conditions in the electromagnetic and gravitational cases involve setting the rescaled magnetic field $\tilde{B}_I$ and the magnetic part of the Weyl tensor $\tilde{B}_{IJ}$ to zero on the conformal boundary (where ``magnetic'' is defined relative to the boundary normal). We show that these boundary conditions can be generalized by replacing ``magnetic" with arbitrary linear combinations of ``magnetic'' and ``electric,'' and we conjecture that these are the most general AdS-invariant, conservative boundary conditions in AdS$_{(1,3)}$. The electromagnetic boundary conditions can be generalized to Yang-Mills theory, and the gravitational boundary conditions can be generalized to nonlinear gravity in a general Fefferman-Graham setting, where prior work of Friedrich and Sou\^{e}tre shows that they give rise to well-posed evolution. We construct a phase space for electromagnetism, Yang-Mills theory, and nonlinear gravity with our general boundary conditions. In the Maxwell and Yang-Mills cases, the phase space for the nonstandard boundary conditions is naturally associated with a Lagrangian that has a Pontryagin term. In the gravitational case, the associated Lagrangian has both Gauss-Bonnet and gravitational Pontryagin terms. In gravity, with the exception of the standard boundary conditions, we find that there are no asymptotic symmetries, so all ``conserved charges'' (such as ADM mass) vanish identically, similarly to the case of a closed universe. For AdS itself, this results in a severe linearization instability in the sense of Fischer and Marsden and Moncrief: For all but the standard boundary conditions, all nontrivial linearized perturbations are spurious, i.e., they do not correspond to exact solutions.  In addition, the modified symplectic structure modifies the formula for the entropy of a black hole.

\end{abstract}

\maketitle
\hypersetup {linktocpage}

\tableofcontents

\newcommand{\CT}{\boldsymbol{C}}

\section{Introduction}
\label{sec:intro}

As is well known, anti-de Sitter (AdS) spacetime fails to be globally hyperbolic. Consequently, it would be expected that in order for a field to have well-posed, deterministic dynamics in AdS spacetime, one would need to supplement the equations of motion of the field with boundary conditions at infinity. There has been considerable prior work on this issue.

Early work on boundary conditions by Avis, Isham and Storey \cite {AIS1978} was motivated by considering the stress-energy flux through infinity and imposing a time-averaged conservation of energy. For the conformally-coupled massless scalar field in AdS$_{1, 3}$, they found two ``reflective" boundary conditions, which are the Dirichlet boundary condition and the Neumann boundary condition. A more comprehensive analysis on scalar fields as well as spin-$\frac{1}{2}$ and spin-1 fields was then given by Breitenlohner and Freedman \cite {BF1982} in the context of supergravity in AdS$_{1, 3}$. For generic scalar fields, they found that, within a certain mass window, there are two boundary conditions (Dirichlet and Neumann) that are invariant under supersymmetry transformations and, in particular, invariant under AdS symmetries. Also in the context of supergravity in AdS$_{1, 3}$, Hawking \cite {Hawking1983} showed that only one of the two Breitenlohner and Freedman boundary conditions is such that the conformal symmetry group on the boundary is the full AdS symmetry group. He showed that this boundary condition implies that magnetic parts of the spin-one field and the Weyl tensor fall off faster than the electric parts (where ``magnetic'' and ``electric'' parts are defined with respect to the spacelike normal to infinity---see \eqref{EIBI} and \eqref{EIJBIJ} below). Ashtekar and Magnon \cite {AM1984} then showed that in the pure Einstein gravity case in 4 spacetime dimensions, the requirement that the conformal symmetry group of infinity be the full AdS$_{1, 3}$ symmetry group implies that the magnetic part of the rescaled Weyl tensor must vanish at infinity\footnote{Ashtekar and Das \cite{AD1999} further extended the boundary condition result for Einstein gravity to arbitrary spacetime dimension $D \geq 4$.}. Shortly thereafter Henneaux and Teitelboim \cite {HT1985} proposed metric falloff conditions based on the behavior of Kerr-AdS solutions and AdS$_{1, 3}$ symmetry. As we shall see in section \ref{sec:NONLINGRAVBC}, the Henneaux and Teitelboim metric falloff conditions are equivalent to the Hawking and Ashtekar-Magnon conditions on the magnetic part of the Weyl tensor. The Hawking-Ashtekar-Magnon-Henneaux-Teitelboim conditions have been explicitly or implicitly imposed in most works on asymptotically AdS spacetimes. We will refer to them in this paper as the ``standard'' boundary conditions. However, alternative boundary conditions have been proposed by Comp\`{e}re and Marolf \cite {CM2008}, Comp\`ere, Fiorucci and Ruzziconi \cite{CFR2019}, and others.

The above works did not consider whether the proposed boundary conditions would lead to well-posed dynamical evolution for Maxwell theory or Einstein gravity. The well-posedness of the initial value formulation for nonlinear gravity was considered by Friedrich \cite {Friedrich1995} in the context of the vacuum Einstein equation with negative cosmological constant in 4 spacetime dimensions with a regular timelike conformal boundary. He proposed a class of ``maximally dissipative boundary conditions'' at infinity---in which there is an outward flux of energy through the boundary---and proved that Einstein's equations are well-posed with these boundary conditions. As will be discussed further in appendix \ref{sec:appInh}, although Friedrich's focus was on dissipative boundary conditions, certain limiting cases of his boundary conditions are conservative and correspond closely to the general boundary conditions we will be discussing in this paper (and the boundary condition proposed in \cite{CFR2019}). More recently, Holzegel, Luk, Smulevici, and Warnick \cite {Four2015} proposed a different set of ``optimally dissipative boundary conditions'' for scalar fields, Maxwell fields and linearized gravity. Finally, very recently Sou\^{e}tre \cite{Souetre2025} considered conservative boundary conditions of the type we shall be discussing further in this paper and proved that they give rise to a well-posed initial value formulation. 

A more systematic approach to finding all conservative boundary conditions with well-posed evolution was taken by Ishibashi and Wald \cite {IW2004}. Their approach was based on a proposal of Wald \cite {Wald1979} for defining dynamics for a wave equation for a scalar field $\phi$ in a general static, non-globally hyperbolic spacetime. As we shall review in section \ref{sec:oview}, this approach involves introducing a suitable $L^2$ Hilbert space $\mathcal H$ and viewing the wave equation as an evolution equation in this Hilbert space involving a symmetric operator $A:\mathcal{H} \to \mathcal{H}$. The possible self-adjoint extensions of $A$ then correspond to possible choices of boundary conditions for $\phi$, and the dynamics with any choice of extension is well posed. In \cite{IW2003}, Ishibashi and Wald showed that---subject to some assumptions stated in that reference---any well-posed prescription for dynamics that has a suitable associated conserved energy must correspond to a choice of self-adjoint extension of $A$. Therefore, the possible choices of self-adjoint extensions of $A$ should exhaust the possible choices of boundary conditions that yield well-posed evolution with a suitable conserved energy. However, this approach is strictly limited to linear equations (since the operator $A$ must be linear) and, in its stated form, the prescription of \cite {Wald1979} applies only to scalar wave equations. 

In \cite{IW2004}, Ishibashi and Wald (IW) considered scalar, electromagnetic, and linearized gravitational perturbations on a global, Lorentzian AdS background in $D \geq 4$ spacetime dimensions. The approach of \cite {Wald1979} can be applied directly in the scalar case, but the problem of finding self-adjoint extensions simplifies considerably if one first expands in spherical harmonics, so that the wave equation reduces to a family of wave equations in $1+1$ dimensions and, thereby, the corresponding operator $A$ is one-dimensional. In order to treat the electromagnetic and gravitational perturbations in this framework, IW decomposed the electromagnetic and gravitational perturbation into scalar, vector, and tensor parts (relative to $(D-2)$-spheres) and expanded each part in spherical harmonics of the appropriate type. (The tensor part is nontrivial only in the gravitational case for $D > 4$.) The coefficients of the tensor part in linearized gravity, $\Phi_T^{\rm G}$, are automatically gauge-invariant, as are the coefficients of the vector part in the electromagnetic case $\Phi_V^{\rm EM}$. For the vector part in the gravitational case and the scalar part in both cases, IW introduced gauge-invariant quantities that characterize the perturbations. Thus, the electromagnetic and gravitational perturbations are each described by gauge-invariant quantities $\Phi_T$, $\Phi_V$, and $\Phi_S$, each of which satisfies a $1+1$ dimensional wave equation of the same general form as the scalar field. 

The possible self-adjoint extensions of the operator $A$ appearing in this wave equation were then analyzed by IW \cite{IW2004} using von Neumann deficiency index theory. Since $A$ is a real second-order differential operator in one dimension, it can be seen that the deficiency indices are always equal (so self-adjoint extensions always exist) and---depending on a parameter appearing in the $1+1$ dimensional wave equation---the dimension of the deficiency index subspaces is either 0 (so $A$ has a unique self-adjoint extension and there is no freedom in the choice of boundary conditions) or 1 (so there is a one-parameter family of self-adjoint extensions). Further restrictions on the self-adjoint extension occur if one requires the extended operator to be positive, so that the dynamics is stable. The final results obtained by IW are as follows: 

\begin{itemize}
\item For the scalar field in any dimension, for effective mass squared $\mu^2$ below the Breitenlohner-Freedman bound \cite {BF1982}, $A$ is unbounded below and the dynamics is (highly) unstable. Within the Breitenlohner-Freedman window for $\mu^2$, there is a one-parameter family (comprising a circle) of  Robin boundary conditions for each spherical harmonic mode, with the Dirichlet and Neumann conditions obtained by Breitenlohner and Freedman lying at antipodes of the circle, and with a restriction on this parameter on one side of the circle if one imposes positivity. For $\mu^2$ above the Breitenlohner-Freedman window, $A$ has a unique self-adjoint extension. 

\item For the electromagnetic field, for $D=4$, there is a similar one-parameter family of Robin boundary conditions for both $\Phi_V^{\rm EM}$ and $\Phi_S^{\rm EM}$---which include Dirichlet and Neumann as special cases---with a restriction of this parameter if one imposes positivity. For $D=5$ or $6$, the vector part of the operator $A_V$ has a unique self-adjoint extension, but there is a one-parameter family of Robin boundary conditions for $\Phi_S^{\rm EM}$. For $D > 6$, both $A_V$ and $A_S$ have unique self-adjoint extensions.

\item For the linearized gravitational field, the operator $A_T$ for the tensor part, $\Phi_T^{\rm G}$, (which is present only for $D>4$) always has a unique self-adjoint extension. For the vector and scalar parts, the situation is the same as in the electromagnetic case: For $D=4$, one has a one-parameter family of Robin boundary conditions for both $\Phi_V^{\rm G}$ and $\Phi_S^{\rm G}$; for $D=5$ or $6$, one has a one-parameter family of Robin boundary conditions only for $\Phi_S^{G}$; for $D>6$, no boundary conditions are needed.

\end{itemize}

In principle, the IW analysis gives a comprehensive answer to the question of what boundary conditions for the scalar, electromagnetic, and linearized gravitational fields in AdS$_{1, d}$ give rise to well-posed, energy-conservative dynamics. However, one shortcoming is that it treats each spherical harmonic mode separately and independently. Consequently, in cases where different self-adjoint extensions are possible, one could choose different Robin parameters for different spherical harmonic modes. This would lead to boundary conditions that are nonlocal in the fields, which presumably would not be of physical interest. Thus, it would be of interest to restrict consideration to the IW boundary conditions that can be expressed as local conditions on the fields. It is also of interest to determine which of these local conditions are invariant under AdS isometries, so that the space of solutions satisfying these boundary conditions is mapped into itself under AdS isometries. In the case of a scalar field, $\phi$, it is easily seen that the requirement that every spherical harmonic mode has the same Robin parameter will give rise to boundary conditions that are local in $\phi$. It also is not difficult to see that the Dirichlet and Neumann conditions are invariant under AdS isometries, but the general Robin conditions are not. However, in the electromagnetic and gravitational cases, the potentials $\Phi_V$ and $\Phi_S$ are defined in a nonlocal manner in terms of the fields, so it is less obvious which of the IW boundary conditions can be expressed as local conditions on the fields. It is considerably less obvious which of the conditions are invariant under AdS isometries. 

One of the main purposes of this paper is to determine which of the IW boundary conditions in the electromagnetic and linearized gravitational cases are local in the fields, and which of these local conditions are AdS-invariant. In the body of the text, we will focus mainly on the case $D=4$, but in Appendix \ref{sec:apphigher} we will treat the other cases ($D=5,6$) where boundary conditions are needed. 

Before summarizing our results, we need to address a further point special to $D=4$. As described above, IW treated $\Phi_V$ and $\Phi_S$ as independent fields and thereby imposed boundary conditions for them separately. However, for $D=4$ in both the electromagnetic and gravitational cases, the potentials $\Phi_V$ and $\Phi_S$ satisfy exactly the same equations of motion. This is not an ``accident'' in electromagnetism because $\Phi_V^{\rm EM}$ and $\Phi_S^{\rm EM}$ transform into each other under duality rotations, so the fact that the potentials satisfy the same equation is a consequence of duality invariance of Maxwell's equations. In linearized gravity off of AdS, it is not obvious, a priori, that there is such a thing as a duality rotation, but one can understand the equality of the equations for $\Phi_S$ and $\Phi_V$ as implying the existence of a similar duality rotation transformation of the metric perturbations taking the linearized Weyl tensor into its dual. A consequence of $\Phi_V$ and $\Phi_S$ satisfying the same equation is that one can apply the IW boundary conditions to linear combinations of the potentials instead of the potentials themselves. Consequently, taking this duality invariance into account, we will obtain a larger set of AdS-invariant boundary conditions than would be obtained by applying the IW boundary conditions to $\Phi_V$ and $\Phi_S$ separately.

Our results on the boundary conditions for the electromagnetic and linearized gravitational cases in $D=4$ that are both local and AdS-invariant are as follows. For the electromagnetic field, we define the (rescaled) electric and magnetic fields at infinity by
\begin {align}
\tilde{E}_I \equiv F_{I \mu} n^\mu, \quad \tilde{B}_I \equiv {}^*F_{I \mu} n^\mu 
\label{EIBI}
\end {align}
Here $F_{\mu \nu}$ is the Maxwell field tensor, ${}^*F_{\mu \nu}$ is its dual, $n^\mu$ is the unit outward pointing normal in the conformally completed spacetime, and capital latin indices such as  ``$I$'' denote tensor indices in the boundary. This corresponds to the usual definition of electric and magnetic fields (up to a radial rescaling) except that they are usually defined relative to a unit timelike vector rather than the spacelike vector $n^\mu$. In the electromagnetic case, the general IW boundary conditions (applied separately to $\Phi_V^{\rm EM}$ and $\Phi_S^{\rm EM}$) that are local in the fields are
\begin {align}
& \big[  n^\mu \partial_\mu \big(\tilde{E}_t \big) + q^S \big( \tilde{E}_t \big) \big]_{\mathcal{I}} = 0 \nonumber \\ 
& \big[ n^\mu \partial_\mu \big( \tilde{B}_t \big)  + q^V \big( \tilde{B}_t \big)  \big]_{\mathcal{I}} = 0 
\label{eq:EMBC1}
\end {align}
where $q^S$ and $q^V$ are allowed to take arbitrary real values (including $\infty$) but are restricted from becoming too negative for stable evolution. In accord with the previous paragraph, duality rotations of these boundary conditions are also allowed. 

The general boundary conditions \eqref{eq:EMBC1} are not AdS-invariant, nor are they conformally invariant (since $n^\mu$ will rescale under a change in the choice of conformal factor $\Omega$). As we shall show, the only ones that are AdS-invariant (and conformally invariant) are $q^V = \infty,\, q^S = 0$ and $q^S = \infty,\, q^V = 0$. These can be re-expressed more simply as $\tilde{B}_I = 0$ (or, equivalently, $F_{IJ} = 0$) and $\tilde{E}_I = 0$, respectively. Taking into account the ability to perform duality rotations, the general AdS-invariant local boundary conditions for the electromagnetic field in $D=4$ can be compactly expressed as \footnote{For $D=5$ and $6$, the only AdS-invariant local boundary condition is $F_{IJ} = 0$.}
\begin {align}
\big[\tilde{E}_{I} + \alpha\tilde{B}_{I}\big]_{\mathcal{I}} = 0
\label{emadsbc}
\end {align}
where $\alpha$ can take any real value (including $\alpha = \infty$). (There are no restrictions on $\alpha$ arising from stability.) In view of the discussion above, \eqref{emadsbc} should be the most general set of local boundary conditions for the electromagnetic field in $D=4$ that admits a well-posed initial value formulation, has a conserved energy, and is AdS-invariant.

Although the IW derivation of well-posedness applies only in the case of linear theories, the boundary conditions \eqref{emadsbc} are well defined in the case of general Yang-Mills theories in $D=4$. We believe that these boundary conditions will give rise to well-posed evolution in the Yang-Mills case, and that \eqref{emadsbc} is the most general local boundary condition for a Yang-Mills field in $D=4$ that admits a well-posed initial value formulation, has a conserved energy, and is AdS-invariant.

The linearized gravitational case closely parallels the electromagnetic case. We define the (rescaled) electric and magnetic parts of the physical linearized Weyl tensor $C^{(1)}_{\lambda \rho J \nu}$ at infinity by 
\begin {align}
\tilde{E}^{(1)}_{IJ} \equiv \Omega C^{(1)}_{I \mu J \nu} n^\mu n^\nu, \quad
\tilde{B}^{(1)}_{IJ}  \equiv \Omega \, {}^*C^{(1)}_{I \mu J \nu} n^\mu n^\nu 
\label{EIJBIJ}
\end {align}
where $\Omega$ denotes the conformal factor and the limit to conformal infinity is taken. The superscript ``$(1)$'' denotes that these are linearized quantities.
The general IW boundary conditions (applied separately to $\Phi_V^G$ and $\Phi_S^G$) that are local in the fields are
\begin {align}
& \big[  n^\mu \partial_\mu \big(\tilde{E}^{(1)}_{tt} \big) + q^S \big( \tilde{E}^{(1)}_{tt} \big) \big]_{\mathcal{I}} = 0 \nonumber \\ 
& \big[ n^\mu \partial_\mu \big( \tilde{B}^{(1)}_{tt} \big)  + q^V \big( \tilde{B}^{(1)}_{tt} \big)  \big]_{\mathcal{I}} = 0 
\label{eq:gravBC1}
\end {align}
where $q^S$ and $q^V$ are allowed to take arbitrary real values (including $\pm \infty$) but are restricted from becoming too negative for stable evolution. As previously explained, duality rotations of these boundary conditions are also allowed. Of the local conditions \eqref{eq:gravBC1}, the only ones that are AdS-invariant are $q^V = \infty,\, q^S = 0$ and $q^S = \infty,\, q^V = 0$. These can be re-expressed more simply as $\tilde{B}^{(1)}_{IJ} = 0$ and $\tilde{E}^{(1)}_{IJ} = 0$, respectively. Taking into account the ability to perform duality rotations, the general AdS-invariant local boundary conditions for linearized gravity for $D=4$ are \footnote{For $D=5$ and $6$, the general AdS-invariant local boundary condition is $\Omega n^\mu C_{\mu IJK} = 0$.}
\begin {align}
\big[ \tilde{E}^{(1)}_{IJ} + \alpha\tilde{B}^{(1)}_{IJ} \big]_{\mathcal{I}} = 0
\label{linadsbc}
\end {align}
where $\alpha$ can take any real value (including $\alpha = \infty$). We believe that \eqref{linadsbc} provides the most general set of local boundary conditions for linearized gravity about AdS$_{1,3}$ that admits a well-posed initial value formulation, has a conserved energy, and is AdS-invariant.

Although the derivation by IW of well-posedness with the boundary conditions \eqref{linadsbc} applies only to linearized gravity off of an AdS background spacetime, the boundary conditions
\begin {align}
\big[\tilde{E}_{IJ} + \alpha\tilde{B}_{IJ} \big]_{\mathcal{I}} = 0 
\label{gengravbc}
\end {align}
are well defined in the context of fully nonlinear gravity\footnote{Similarly, the boundary conditions \eqref{emadsbc} are well defined in the context of Yang-Mills theory, as we shall discuss further in section \ref{sec:YMBC}.} in the setting of general Fefferman-Graham metrics. Indeed, as we shall discuss further in section \ref{sec:nonGRAphase} and appendix \ref{sec:appInh}, these boundary conditions can be obtained as special limiting cases of Friedrich's \cite{Friedrich1995} general dissipative boundary conditions and they are also precisely equivalent to the conservative boundary conditions considered by Sou\^{e}tre \cite{Souetre2025}. Both Friedrich and Sou\^{e}tre proved well-posedness of the initial value formulation with the boundary conditions \eqref{gengravbc}. Based upon the considerations given above, we believe that these are the most general conservative boundary conditions for Einstein's equation that admit a well-posed initial value formulation and are ``covariant'' in the sense that they do not require the introduction of a special frame to define. Note that the boundary condition $\tilde{B}_{IJ} = 0$ (i.e., $\alpha = \infty$) corresponds to the standard boundary conditions mentioned at the beginning of this section. 

Although the boundary conditions \eqref{emadsbc} in the electromagnetic case and \eqref{gengravbc} in the (nonlinear) gravitational case give rise to well-posed dynamics, that is not the end of the story: We also would like to define a phase space structure on the space of solutions satisfying these boundary conditions. In order for this to be possible, we need to have a conserved symplectic current whose integral over spacelike slices going to infinity is finite (so we have a well-defined symplectic product) and whose flux through infinity vanishes (so that this symplectic product is independent of choice of slice). In the electromagnetic case with the usual boundary condition $\tilde{B}_I=0$ or with the boundary condition $\tilde{E}_I=0$ (i.e., $\alpha = \infty$ or $0$), the usual Maxwell symplectic current has the desired properties, but this is not the case for other $\alpha$. Exactly analogous results hold in the Yang-Mills case. In the nonlinear gravitational case with the usual boundary condition $\tilde{B}_{IJ}=0$ (i.e., $\alpha = \infty$), the usual Einstein symplectic current has the desired properties, but this is not the case for other values of $\alpha$. Nevertheless, we will show in section \ref{sec:phase} that we can obtain a symplectic current with the desired properties by adding certain ``topological terms'' to the Maxwell, Yang-Mills, and Einstein-Hilbert Lagrangians. These terms do not affect the equations of motion but they do affect the symplectic potential and the symplectic current. In the electromagnetic case, we add a Pontryagin term to the Maxwell Lagrangian. An analogous procedure also works in the Yang-Mills case. In the gravitational case, we add both a Gauss-Bonnet term and a Pontryagin term to the Einstein-Hilbert Lagrangian. The Gauss-Bonnet term in the local Lagrangian effectively replaces both the usual Gibbons-Hawking-York boundary term and the ``counterterm action'' boundary term given by Balasubramanian and Kraus \cite{BK1999} (see also \cite{HeSk1998}), so the action has no explicit boundary contributions. Therefore, we can directly apply the usual covariant phase space methods in our constructions. The local and covariant symplectic current thereby obtained has vanishing flux at infinity and finite integral over spacelike slices going to infinity. Thus, it allows us to define the desired phase space in the electromagnetic, Yang-Mills, and gravitational cases.

As we shall discuss in section \ref{ASCC}, the electromagnetic and nonlinear gravitational phase spaces with the non-standard boundary conditions $\alpha \neq \infty$ have a trivial asymptotic gauge symmetry group. In particular, for the nonlinear gravitational phase space with $\alpha \neq \infty$, all boundary-condition-preserving diffeomorphisms are gauge, including the diffeomorphisms that are nonvanishing at infinity. All Hamiltonian charges---such as the ADM mass with respect to a timelike vector field at infinity---vanish identically. A consequence of this fact is that for $\alpha \neq \infty$, any spacetime with a Killing field will suffer from a ``linearization instability'' in the sense of Fischer and Marsden \cite{FM1979} and Moncrief \cite{Mon1975} similar to that occurring in a closed universe: For each Killing field, linearized perturbations will have to satisfy a quadratic constraint in order to correspond to a one-parameter family of exact solutions. This linearization instability is particularly severe for AdS itself because the quadratic constraint arising from the globally timelike Killing field is positive definite for non-gauge linearized perturbations. This implies that for all but the standard boundary conditions, all nontrivial linearized perturbations are spurious in the sense that they do not correspond to the linearization of a one-parameter family of exact solutions to Einstein's equation. Thus, for non-standard boundary conditions, AdS is an isolated solution in this sense.

Finally, we note at the end of section \ref{ASCC} that in the nonlinear gravitational case, the modified symplectic structure for $\alpha \neq \infty$ needed to define a phase space alters the formula for Noether charge. This results in a modified formula for the entropy of a black hole with a bifurcate Killing horizon in an asymptotically AdS spacetime with non-standard boundary conditions.

In section \ref{sec:oview}, we will give a review of the approach and results of the IW analysis \cite {IW2004}. In section \ref{sec:BC}, we will determine which of the IW boundary conditions give rise to boundary conditions that are local in the fields and we will determine which are AdS-invariant. This will motivate the boundary conditions \eqref{emadsbc} and \eqref{gengravbc} that will be considered in the remainder of the paper. The construction of phase space is given in section \ref{sec:phase}. Asymptotic symmetries are discussed in section \ref{ASCC} as well as consequences for linearization instability and black hole entropy. Appendix \ref{sec:appfall} relates the IW potentials for linearized gravity to the linearized Weyl tensor. The local and AdS invariant IW boundary conditions in the $D=5,6$ dimensional cases are treated in appendix \ref{sec:apphigher}. (The IW boundary conditions for $D > 6$ are unique.) A comparison of our boundary conditions with the conservative limit of Friedrich's conditions \cite{Friedrich1995}---as well as those of Comp\`{e}re et al.~\cite{CFR2019}---is given in appendix \ref{sec:appInh}.

\vspace{5 pt}
\textit{Conventions and Notation}. In this paper, we work in Lorentzian global AdS spacetime, with metric
\begin {align}
ds^2 = - \big(L^2 + r^2 \big) dt^2 + \frac{1} {1 + (\frac{r}{L})^2} dr^2 + r^2 d\Omega_{d-1}^2
\label{eq:AdSmetric1}
\end {align}
where $L$ is related to the cosmological constant $\Lambda$ by $\Lambda = - d (d-1)/2L^2$. We choose the orientation of spacetime so that $\epsilon_{t r z^1 \cdots z^{d-1}} > 0$ in the coordinates
of \eqref{eq:AdSmetric1}, with
$\epsilon_{tr} > 0$ and $\epsilon_{z^1 \cdots z^{d-1}} > 0$ for the volume elements of
the $(t,r)$ subspace and of the spheres; volume elements induced on hypersurfaces are
obtained by contracting with the outward unit normal. It is also convenient for certain purposes to introduce the coordinate $x$ defined by $r = L \cot (x)$, so that the timelike boundary at infinity is at $x=0$ rather than at $r \rightarrow \infty$, in which case the metric takes the form
\begin{equation}
ds^2 = \frac{L^2}{\sin^2 x} \left[-dt^2 + dx^2 + \cos^2x \, d \Omega_{d-1}^2 \right]
\label{eq:AdSmetric2}
\end{equation}
We will use Greek letters to denote spacetime indices, so, e.g., $g_{\mu \nu}$ denotes the spacetime metric and $\nabla_{\mu}$ denotes the covariant derivative compatible with $g_{\mu \nu}$. Following IW, when working in AdS spacetime, we use latin indices, $a,b,c, \dots$, from the early alphabet to denote tensors in the two-dimensional subspace orthogonal to the spheres, so $g_{ab}$ denotes the metric in that subspace and $\hat{\nabla}_a$ is its derivative operator. We use latin indices, $i,j,k, \dots$, from the middle alphabet to denote tensors on the $(d-1)$-spheres. In particular, $\gamma_{ij}$ denotes the round metric on the unit sphere $S^{d-1}$ and
$\hat{D}_i$ its compatible derivative operator.  Thus, for example, $F_{ij}$ denotes pullback of the electromagnetic field tensor $F_{\mu \nu}$ to $(d-1)$-spheres. Finally, we will use capital latin indices, $I,J,K, \dots$, from the middle alphabet to denote tensors on the boundary at infinity, $\mathcal I$, as we have already done in equations \eqref{emadsbc} and \eqref{gengravbc} above. Thus, the index $I$ runs over time, $t$, as well as the angular indices $i$.

\section{Review of the Ishibashi-Wald (IW) Approach and Results}
\label{sec:oview}

\subsection{General Approach to Obtaining Boundary Conditions}

As discussed in the Introduction, our work is based on the analysis of IW \cite {IW2004}, which, in turn, is based on a proposal of \cite {Wald1979} for defining well-posed dynamics for a field satisfying a linear wave equation in a static spacetime, so that there is a hypersurface orthogonal timelike Killing field $t^\mu$ with complete orbits. We also assume that the spacetime possesses a hypersurface, $\Sigma$, orthogonal to $t^\mu$ such that each orbit of $t^\mu$ intersects $\Sigma$ once and only once, but we do not assume that the spacetime is globally hyperbolic, i.e., it may have (singular or nonsingular) causal boundaries. Consider a field $\phi$ on this spacetime satisfying an equation of the form
\begin {align}
\nabla^\mu \nabla_\mu \phi - U \phi = 0
\label{sceq}
\end {align}
where $U$ is a real function on spacetime that does not depend on the static time coordinate $t$. This equation can be rewritten as
\begin {align}
\frac{\partial^2 \phi}{\partial t^2} = - A \phi
\label{hilbev}
\end {align}
with 
\begin {align}
 A = - N D^i(N D_i)+ N^2 U
 \label{eq:operatorA}
\end {align}
where $N = (-t^\mu t_\mu)^{1/2}$ is the norm of the static Killing field and $D_i$ is the derivative operator on $\Sigma$. 

The key idea is to view $A$ as an operator on the Hilbert space, $L^2(\Sigma, N^{-1} d \Sigma)$, of square integrable functions on $\Sigma$ with volume element $N^{-1}$ times the metric compatible volume element on $\Sigma$. The initial domain of $A$ is taken to be $C^\infty_0 (\Sigma)$, i.e., compactly supported smooth functions on $\Sigma$, and it is easily seen that $A$ is symmetric on this domain. It is possible that $A$ is essentially self-adjoint on this domain, i.e., it has a unique self-adjoint extension. However, even if it is not essentially self-adjoint, the operator $A$ is real---i.e., it commutes with complex conjugation---so its deficiency subspaces are of equal dimension, implying that self-adjoint extensions always exist. Now, suppose further that $A$ is bounded below and the extension, $A_E$, that we have chosen is also bounded below, $A_E \geq -C$ (as always will be the case for the Friedrichs extension). Then, for any initial data $(\phi_0, \dot{\phi}_0)$ in $L^2(\Sigma, N^{-1} d \Sigma) \times L^2(\Sigma, N^{-1} d \Sigma)$, we can define
\begin {align}
\phi_t = \cos(A_E^{1/2} t)\phi_0 + A_E^{-1/2} \sin(A_E^{1/2} t) \dot{\phi}_0 .
\label{opsol}
\end {align}
The right side of this equation is well defined, since the functions $\cos(x^{1/2} t)$ and $x^{-1/2} \sin(x^{1/2} t)$ are bounded for $x \geq - C$, and any bounded function of a (possibly unbounded) self-adjoint operator is well defined as a bounded operator. Furthermore, the arguments of \cite{Wald1979} show that for smooth initial data, $(\phi_0, \dot{\phi}_0)$, of compact support, \eqref{opsol} yields a smooth solution to \eqref{sceq} and, for a globally hyperbolic spacetime, it agrees with the unique solution to \eqref{sceq} determined by this given initial data.

If $A$ is essentially self-adjoint, then \eqref{opsol} provides a unique prescription for defining dynamics. However, when $A$ is not essentially self-adjoint, we will obtain different dynamics depending on the choice of $A_E$. This corresponds to different choices of boundary conditions for $\phi$. 

The dynamics defined by \eqref{opsol} has a conserved energy for initial data in the domain of $A_E$ given by
\begin {align}
E_{IW} (\phi) = (\phi, A_E \phi) + ||\partial{\phi}/\partial t||^2
\label{IWen}
\end {align}
where the inner product and norm are taken in $L^2(\Sigma, N^{-1} d \Sigma)$. As already mentioned in the introduction, it was shown in \cite{IW2003} that any prescription giving well-defined dynamics with a conserved energy satisfying the properties listed in that reference must correspond to choosing a self-adjoint extension of $A$ and defining dynamics by \eqref{opsol}. Thus, the determination of all self-adjoint extensions of $A$ should yield all of the viable conservative boundary conditions for solutions to \eqref{sceq}. IW applied this general approach to scalar fields, electromagnetic fields, and linearized gravitational fields in AdS$_{1, d}$. We now review how IW reduced all of these cases to solving equations of the form \eqref{hilbev}.

\subsection {Decomposition of fields in AdS}
\label{sec:oview1}

\subsubsection{Scalar field}
\label{subscalar}

Consider a general, free scalar field in AdS$_{1, d}$ with Lagrangian
\begin {align}
\mathcal{L}_\phi = - \frac{1}{2} \bigg( g^{\alpha \beta} \partial_\alpha \phi \partial_\beta \phi + \mu^2 \phi^2 \bigg)
\label{eq:scalaract}
\end {align}
where $\mu^2$ denotes the effective squared mass\footnote {The curvature of AdS$_{1, d}$ is $ R = -d (d+1)/L^2$. Minimal coupling means $\xi = 0$; conformal coupling means $\xi = (d-1)/(4d)$ and $m=0$, in which the action is invariant under bulk Weyl transformations.} 
\begin{equation}
\mu^2 = m^2  +\xi R .
\label{effm}
\end{equation}
The equation of motion is the usual Klein-Gordon equation 
\begin {equation}
\nabla^2 \phi  - \mu^2 \phi = 0 .
\label{eq:KG}
\end {equation}
This equation is already of the form \eqref{sceq}, but we can reduce it to a $(1+1)$-dimensional equation by expanding in spherical harmonics
\begin {align}
\phi (t, r, z^i) = r^{-\frac{(d-1)}{2}} \sum_{\textbf{k}} \Phi_{\textbf{k}} (t, x) \mathbb{S}_{\textbf{k}} (z^i)
\label{eq:resexp}
\end {align}
where $z^i$ denotes the angular coordinates (with $i =1, 2, \cdots, d-1$) and $\textbf{k}$ denotes the spherical harmonic mode\footnote{In four-dimensional spacetime, the spheres are two dimensional and $\mathbb{S}_{\textbf{k}}$ would normally be written as $Y_{l m}$.}. Substituting this expansion into \eqref{eq:KG}, we obtain
\begin {align}
\frac{\partial^2 \Phi_{\textbf{k}}}{\partial t^2} = \bigg(\partial_x^2 - \frac{\nu^2 - 1/4} {\sin^2 x} - \frac{\sigma^2 - 1/4}{\cos^2 x} \bigg) \Phi_{\textbf{k}} 
\label{eq:waveeq}
\end {align}
where
\begin {align}
\nu^2 = \frac{d^2}{4} + \mu^2 L^2
\label{eq:scalarnu}
\end {align}
and
\begin{align}
\sigma = \frac{d-2}{2}+l  
\label{eq:sigma}
\end{align}
where $l = 0, 1, 2, \dots$ labels the eigenvalues of the sphere Laplacian, given by $-l(l+d -2)$. Equation \eqref{eq:waveeq} is of the form \eqref{hilbev} with the operator $A$ given by
\begin {align}
A = - \bigg(\partial_x^2 - \frac{\nu^2 - 1/4} {\sin^2 x} - \frac{\sigma^2 - 1/4}{\cos^2 x} \bigg)  
\label{eq:Awaveeq}
\end {align}
and the Hilbert space on which $A$ is defined as a symmetric operator is simply $\mathcal{H} = L^2 \big([0, \pi/2], dx \big)$. 
 
\subsubsection{Electromagnetic field}

The Lagrangian for the Maxwell field $A_\mu$ is given by
\begin {align}
\mathcal{L}_M = - \frac{1}{4} F_{\mu \nu} F^{\mu \nu}
\end {align}
Following IW we decompose $A_\mu$ into its scalar and vector parts relative to spheres
\begin {align}
A_\mu (t, r, z^i) =A_\mu^S (t, r, z^i)  + A_\mu^V (t, r, z^i) 
\end {align}
The vector part is gauge invariant and only has components tangent to the spheres. We expand it in vector spherical harmonics $(\mathbb{V}_{\textbf{k}})_i $ as
\begin {align}
A_i^V  & = r^{-\frac{(d-3)}{2}} \sum_{\textbf{k}} \Phi_{V{\textbf{k}}}^{\rm EM} (\mathbb{V}_{\textbf{k}})_i
\end {align}
We expand the scalar part in scalar spherical harmonics as
\begin {align}
A_\mu^S dx^\mu & = \sum_{\textbf{k}} \big(A_{\textbf{k} a} \mathbb{S}_{\textbf{k}} dy^a + A_{\textbf{k}} \hat{D}_i \mathbb{S}_{\textbf{k}} dz^i \big)
\end {align}
where $y^a = (t,r)$. We then introduce the gauge invariant potential for the scalar part via\footnote{Here we correct a typo in eq.(62) in the original IW paper \cite{IW2004}.}
\begin{align}
\hat{\nabla}_a \bigg( r^{\frac{(d-3)}{2}} \Phi_{S{\textbf{k}}}^{\rm EM} \bigg) & = \epsilon_{ab} r^{d-3} \big( \hat{\nabla}^b A_{\textbf{k}} - A_{\textbf{k}}^b)   
\label{phisem}
\end{align}
It then follows that $\Phi_{V{\textbf{k}}}^{\rm EM}$ and $\Phi_{S{\textbf{k}}}^{\rm EM}$ satisfy \eqref{eq:waveeq} with $\sigma$ given by \eqref{eq:sigma} and $\nu^2$ given by 
\begin {align}
\nu_S^2 = \frac{(d-4)^2}{4}, \quad \nu_V^2 = \frac{(d-2)^2}{4}. 
\label{eq:EMnu}
\end {align}

\subsubsection{Linearized gravitational field}

The Lagrangian for the linearized gravitational field off of AdS$_{1, d}$ spacetime (with metric $g_{\mu \nu}^0$ given by \eqref{eq:AdSmetric1}) is obtained by writing $g_{\mu \nu} = g_{\mu \nu}^0 + h_{\mu \nu}$ and expanding the Einstein-Hilbert Lagrangian
\begin {align}
\mathcal{L}_{\rm EH} = \frac{1}{16 \pi G_{d+1}} \big(R [g] - 2 \Lambda \big)
\end {align}
to second order in $h_{\mu \nu}$. We decompose $h_{\mu \nu}$ into its scalar, vector, and tensor parts with respect to spheres by
\begin {align}
h_{ab} = & h_{ab}^S \\
h_{ai} = &  h^{V}_{ai}  + \hat{D}_i h_a^S \\
h_{ij}  = & h_{ij}^T + 2 \hat{D}_{(i} h_{P  j)}^V \nonumber \\
& + \bigg[h_L^S \gamma_{ij} + \bigg(\hat{D}_i \hat{D}_j - \frac{1}{d-1} \gamma_{ij} \hat{D}^m \hat{D}_m \bigg) h_P^{S} \bigg]
\end {align}
Here, the tensor part is $h_{ij}^T$, the vector parts are $h^{V}_{ai}$ and $h_{P j}^V$, and the scalar parts are $h_{ab}^S$, $h_a^S$, $h_L^S$, and $h_P^{S}$.

The tensor part, $h_{ij}^T$, is gauge-invariant and can be expanded in tensor spherical harmonics, $(\mathbb{T}_{\textbf{k}})_{ij}$, as
\begin {align}
h_{ij}^T = r^{-\frac{(d-1)}{2}+2} \sum_{\textbf{k}} \Phi^G_{T {\textbf{k}}} \cdot (\mathbb{T}_{\textbf{k}})_{ij} 
\end {align}
where the dot is a regular functional multiplication and we keep it explicit for the clarity of expression. 

The vector parts, $h^V_{ai}$ and $h^V_{P i}$, are not gauge invariant. They can be expanded in vector spherical harmonics as
\begin {align}
h^V_{ai} & = \sum_{\textbf{k}_V} h^V_{\textbf{k} a} \cdot (\mathbb{V}_{\textbf{k}})_i, \quad h^V_{Pi} = \sum_{\textbf{k}_V} h_{P\textbf{k}}^V \cdot (\mathbb{V}_{\textbf{k}})_i.
\end {align}
As shown in \cite{IW2004}, one can introduce a gauge-invariant potential, $\Phi_{V{\textbf{k}}}^{\rm G}$, characterizing the vector parts of the metric perturbation via
\begin {align}
\hat{\nabla}_a \left(r^{\frac{(d-1)}{2}} \Phi_{V{\textbf{k}}}^{\rm G} \right) = r^{d-3}
\epsilon_{ab} \left[ h^{V b}_{\textbf{k}} - r^2 \hat{\nabla}^b (h_{P\textbf{k}}^V/r^2) \right]
\label{phivg}
\end {align}

Finally, we expand the scalar parts in scalar spherical harmonics
\begin {align}
h_{ab}^S & = \sum_{\textbf{k}_S} h_{\textbf{k} \ ab}^S \cdot \mathbb{S}_{\textbf{k}}, \quad h_a^S = \sum_{\textbf{k}_S} h_{\textbf{k} \ a}^S \cdot \mathbb{S}_{\textbf{k}} \\
h_{L}^S & =  \sum_{\textbf{k}_S}  h_{L \textbf{k}}^S \cdot \mathbb{S}_{\textbf{k}}, \quad h_{P}^S =  \sum_{\textbf{k}_S} h_{P \textbf{k}}^S \cdot \mathbb{S}_{\textbf{k}}. 
\end {align}
We define the gauge-invariant quantity $Z_{ab}$ by 
\begin {align}
Z_{ab} = & r^{d-3} \big(h_{\textbf{k} \ ab}^S + \hat{\nabla}_a X_b + \hat{\nabla}_b X_a \big) + \frac{(d-2)}{(d-1)} Z g_{ab}
\end {align}
where
\begin {align}
Z = & (d-1) r^{d-5} \bigg[h_{L \textbf{k}}^S + \frac{l (l+d-2)}{(d-1)} h_{P \textbf{k}}^S + 2 r \big(\hat{\nabla}^a r \big) X_a \bigg] 
\end {align}
and
\begin {align}
X_a = - h_{\textbf{k} \ a}^S + \frac{1}{2} r^2 \hat{\nabla}_a \bigg(\frac{h_{P \textbf{k}}^S} {r^2} \bigg)
\end {align}
(The linearized Einstein equation implies that $Z = Z^a{}_a$.) Finally, as shown in \cite{IW2004}, one can introduce a gauge invariant potential, $\Phi_{S{\textbf{k}}}^{\rm G}$, characterizing the scalar parts of the metric perturbation via
\begin {align}
Z_{ab} = \bigg(\hat{\nabla}_a \hat{\nabla}_b - \frac{1}{L^2} g_{ab} \bigg) \bigg(r^{\frac{(d-1)}{2}} \Phi_{S{\textbf{k}}}^{\rm G} \bigg).
\label{phisg}
\end {align}

Each of the quantities $\Phi_{S{\textbf{k}}}^{\rm G}$, $\Phi_{V{\textbf{k}}}^{\rm G}$, and $\Phi^G_{T {\textbf{k}}}$ satisfies an equation of the form \eqref{eq:waveeq} with $\sigma$ given by \eqref{eq:sigma}. The values of $\nu^2$ for these quantities are, respectively,
\begin {align}
\nu_S^2 = \frac{(d-4)^2}{4}, \quad \nu^2_V = \frac{(d-2)^2}{4}, \quad \nu_T^2 = \frac{d^2}{4}
\label{eq:rescaledGT}
\end {align}

\subsection {Field asymptotics and boundary conditions}
\label{sec:oview2}

In the previous subsection, we reviewed how IW reduced the problem of solving the scalar, electromagnetic, and linearized gravitational equations in AdS$_{1, d}$ to solving an equation of the form \eqref{eq:waveeq}. In this subsection, we give the leading terms in an asymptotic expansion of solutions to \eqref{eq:waveeq} near infinity. The IW boundary conditions can then be expressed as relations between the coefficients appearing in this asymptotic expansion. For $\nu^2 < 0$, the operator $A$ given by \eqref{eq:Awaveeq} is unbounded below and the dynamics defined by \eqref{eq:waveeq} is highly unstable (with solutions with arbitrarily large exponential growth rates), so we will restrict attention in the following to the case $\nu^2 \geq 0$, in which case $\nu$ is real, and we take $\nu \geq 0$.

Consider, first, solutions, $\Phi$, to \eqref{eq:waveeq} that oscillate with a fixed frequency, $\omega$. Then \eqref{eq:waveeq} reduces to a second order ordinary differential equation, so there are two, linearly independent solutions, i.e., the general solution can be written as \footnote{As discussed in \cite{IW2004}, the solution $\Phi_2$ does not have acceptable behavior at the origin ($x=\pi/2$), so if we were seeking globally acceptable solutions, we would set $B_2=0$. However, we are seeking here to obtain the general asymptotic behavior of solutions near infinity ($x=0$), in terms of which the IW boundary conditions can be formulated. The eigenvectors of the extension of $A$ corresponding to a choice of IW boundary conditions at infinity would then be obtained by imposing these boundary conditions at infinity together with $B_2 = 0$, as would be possible, in general, only for discrete values of $\omega$.}
\begin {equation}
\Phi (t, x) = e^{-i \omega t} \big[B_1 \Phi_1 (x) + B_2 \Phi_2 (x) \big] 
\end {equation}
where $B_1, B_2$ are arbitrary constants. The basis solutions, $\Phi_1$ and $\Phi_2$, can be expressed in terms of hypergeometric functions $F(a,b;c;z)$ as
\begin {align}
\Phi_1 &= (\cos x)^{\sigma +\frac{1}{2}} (\sin  x)^{\nu+ \frac{1}{2}} F (\zeta^\omega_{\nu, \sigma}, \zeta^{-\omega}_{\nu, \sigma}; 1 + \sigma; \cos^2 x) \nonumber \\
\Phi_2 &= (\cos x)^{-\sigma +\frac{1}{2}} (\sin  x)^{\nu+ \frac{1}{2}} \tilde{F}(\cos^2 x) \label{eq:Phi12} 
\end {align}
where $\sigma$ was defined by \eqref{eq:sigma} and
\begin {align}
 \zeta^\omega_{\nu, \sigma} \equiv \frac{\nu + \sigma + 1 + \omega}{2} 
\label{eq:zeta}
\end {align}
The functional form of $\tilde{F}$ in the second solution $\Phi_2$ varies by spacetime dimensions. For even-dimensional AdS spacetime (i.e., for $d$ odd), the parameter $\sigma $ is non-integer and we have
\begin {align}
\tilde{F}(\cos^2 x) = F (\zeta^\omega_{\nu, -\sigma}, \zeta^{-\omega}_{\nu, -\sigma}, 1 - \sigma; \cos^2 x)
\end {align}
On the other hand, for odd-dimensional AdS spacetime (i.e., $d$ even), the parameter $\sigma$ is an integer and we have\footnote{Here we corrected a typo in eq.(147) of the original IW paper \cite{IW2004}.}
\begin {multline}
\tilde{F}(\cos^2 x) = F (\zeta^\omega_{\nu, \sigma}, \zeta^{-\omega}_{\nu, \sigma}, 1 + \sigma; \cos^2 x) \cdot \text{log} (\cos^2 x)  (\cos x)^{2 \sigma} \\
 +  (\cos x)^{2 \sigma}  \sum_{k=1}^\infty \frac{(\zeta^{\omega}_{\nu, \sigma} \big)_k (\zeta^{-\omega}_{\nu, \sigma} \big)_k}{(1 + \sigma)_k \ k!} \cdot \{h(k) - h(0) \} \cdot (\cos x)^{2k} \\
 -  (\cos x)^{2 \sigma}  \sum_{k=1}^\sigma \frac{(k-1)! (-\sigma)_k} {(\zeta^{\omega}_{-\nu, -\sigma} \big)_k (\zeta^{-\omega}_{- \nu, - \sigma} \big)_k} \cdot (\cos x)^{-2k}
\end {multline}
where the last term should be ignored for the case of $\sigma=0$. Here, the notation $(\lambda)_k$ denotes
\begin {equation}
\big(\lambda \big)_k  \equiv \frac {\Gamma (\lambda + k)}{\Gamma (\lambda)}
\end {equation}
and $h(k)$ is defined by
\begin {align}
h (k) & \equiv \psi (\zeta^\omega_{\nu, \sigma} + k) + \psi (\zeta^{-\omega}_{\nu, \sigma} + k ) \nonumber \\
& \hspace{20 pt} - \psi (1 + \sigma + k) - \psi (k+1)
\end {align}
with
\begin {align}
\psi (z) & \equiv \frac{d}{dz} \text{log} \Gamma (z)
\end {align}

The asymptotic behavior near infinity of solutions to \eqref{eq:waveeq} with definite frequency $\omega$ can now be obtained by expanding the above explicit solutions about $x=0$. We obtain the following results:

\medskip

\noindent
\textit{Non-integer} $\nu >0$
\begin {multline}
\Phi_\nu = x^{\frac{1}{2} - \nu} \big(a_\nu + a_{\nu}^{(2)} x^2 + \cdots \big) \\
+ x^{\frac{1}{2} + \nu} \big(b_\nu + b_\nu^{(2)} x^2 + \cdots \big) .
\label{eq:nonintnu}
\end {multline}

\medskip
\noindent
\textit{Integer} $\nu \geq 1$
\begin {multline}
\Phi_\nu = x^{\frac{1}{2} - \nu} \big[a_\nu +  a_{\nu}^{(2)} x^2 + \cdots + a_{\nu}^{(2\nu-2)}  x^{2 \nu-2} \\
 + a_\nu x^{2\nu} \lambda_\nu \text{log}(x^2) + \cdots \big] \\
+ x^{\frac{1}{2} + \nu} \big(  b_\nu + b_\nu^{(2)} x^2 + \cdots \big) 
\end {multline}
where $\lambda_\nu$ is a constant determined by $\nu, \omega$ and $\sigma$.

\medskip
\noindent
$\nu = 0$
\begin {multline}
\Phi_{\nu=0} = x^{\frac{1}{2}} \big[ a_0 \text{log}(x^2) + a_{0, \rm log}^{(2)} x^2 \text{log} (x^2) +  a_{0}^{(2)} x^2 + \cdots \big] \\
+ x^{\frac{1}{2} } \big(  b_0 + b_0^{(2)} x^2 + \cdots \big).
\label{eq:zeronu}
\end {multline}
In all cases, the coefficients $a_\nu$ and $b_\nu$ can be chosen arbitrarily, whereas the higher order coefficients $a_{\nu}^{(i)}$ are determined by $a_\nu$ and the higher order coefficients $b_{\nu}^{(i)}$ are determined by $b_\nu$. The higher order coefficients also depend on $\nu, \omega$ and $\sigma$.

Equations \eqref{eq:nonintnu}-\eqref{eq:zeronu} give the asymptotic behavior of solutions at a definite frequency $\omega$. However, by taking superpositions of such solutions, we get the asymptotic behavior of general solutions. The upshot is that the asymptotic behavior of general solutions to \eqref{eq:waveeq} is given by \eqref{eq:nonintnu}-\eqref{eq:zeronu}, where $a_\nu$ and $b_\nu$ are now arbitrary functions of time, $t$.

The IW results \cite{IW2004} on allowed boundary conditions may now be stated as follows. For $\nu^2 \geq 1$, the operator $A$ is essentially self-adjoint, and the unique allowed boundary condition\footnote{Alternative boundary conditions for the case $1 < \nu <2$ have been considered by Andrade and Marolf \cite{AnMa2011}. These alternative boundary conditions would fail to have a conserved, positive energy corresponding to the natural energy defined for data of compact support (see \cite{IW2003}). Note also that for $\nu^2 \geq 1$, solutions with $a_\nu \neq 0$ fail to be square integrable.} is $a_\nu = 0$. The operator $A$ is positive in this case, so the dynamics is stable. 

For $0 < \nu^2 < 1$, there is a one-parameter family of self-adjoint extensions of $A$, corresponding to a choice of parameter $q_\nu$ defined by
\begin {align}
q_\nu = \frac{b_\nu} {a_\nu}
\label{eq:nonlocalq}
\end {align}
where $q_\nu$ can take any real value (including $q_\nu = \infty$). We refer to the choice $q_\nu = \infty$ (i.e., $a_\nu = 0$) as Dirichlet boundary conditions; we refer to the choice $q_\nu = 0$ as Neumann boundary conditions; and we refer to any finite, nonzero choice of $q_\nu$ as Robin boundary conditions. The self-adjoint extension of $A$ is positive for Dirichlet and Neumann boundary conditions but fails to be positive for Robin boundary conditions with sufficiently negative $q_\nu$, as given by eq.(169) of \cite{IW2004}.

For $\nu = 0$, there is also a one-parameter family of self-adjoint extensions of $A$, corresponding to a choice of $q_0 = b_0/a_0$ taking any real value (including $\infty$). In this case, the range of $q_0$ for which the self-adjoint extension of $A$ is positive is given by eq.(171) of \cite{IW2004}.

\section {Local and AdS-invariant Boundary Conditions}
\label{sec:BC}

In the previous section, we reviewed how IW \cite{IW2004} reduced the problem of scalar, electromagnetic, and linearized gravitational fields in AdS$_{1, d}$ to solving the $(1+1)$-dimensional wave equation \eqref{eq:waveeq}, and we reviewed the IW results for the boundary conditions for this equation. However, the reduction to \eqref{eq:waveeq} involved an expansion in spherical harmonics. The boundary conditions given at the end of subsection \ref{sec:oview2} can be applied independently to each spherical harmonic mode of the field, but if one were to do so, the resulting boundary conditions would not, in general, be local in the fields and, thereby, would not be of physical interest. Thus, we wish to determine the subset of the IW boundary conditions that correspond to local boundary conditions for the fields. For such local boundary conditions, we also wish to determine which are AdS-invariant, i.e., which boundary conditions have the property that AdS isometries map solutions satisfying the boundary conditions to solutions satisfying the same boundary conditions. It is relatively straightforward to perform the local and AdS-invariant analyses for scalar fields, but it is considerably less straightforward in the electromagnetic and linearized gravitational cases, since the potentials $\Phi_{S{\textbf{k}}}^{\rm EM}$, $\Phi_{V{\textbf{k}}}^{\rm G}$, and $\Phi_{S{\textbf{k}}}^{\rm G}$ are nonlocally defined in terms of the fields (see eqs. \eqref{phisem}, \eqref{phivg}, and \eqref{phisg}). In AdS$_{1,3}$, there is a further complication in the electromagnetic and gravitational cases, in that there is a duality rotation invariance under which $\Phi_S$ and $\Phi_V$ transform into each other. In particular, $\Phi_S$ and $\Phi_V$ satisfy the same equation. Consequently, the IW formalism should be generalized to allow boundary conditions on linear combinations of $\Phi_S$ and $\Phi_V$ rather than requiring---as IW did---that the boundary conditions be applied separately to $\Phi_S$ and $\Phi_V$. 

\subsection {Scalar fields in AdS$_{1,d}$ }
\label{sec:scalarBC}

Consider a real scalar field, $\phi$, in AdS$_{1, d}$, as discussed in subsection \ref{subscalar}. As seen from \eqref{effm} and \eqref{eq:scalarnu}, any value of $\nu^2$ is possible, depending on the mass of the scalar field and its coupling to curvature. The case $\nu^2 < 0$ corresponds to being below the Breitenlohner and Freedman bound \cite {BF1982}, in which case the dynamics is highly unstable. We will not consider this case further here. For $\nu^2 \geq 1$, the unique IW boundary condition for each mode is $a_\nu = 0$. Taking eqs. \eqref{eq:nonintnu} and \eqref{eq:resexp} into account, we see that this corresponds to the Dirichlet boundary condition on $\phi$
\begin {align}
\big( r^{\frac{d}{2} - \nu} \phi \big) \big|_{\mathcal{I}} = 0
\label{nu2g1}
\end {align}
where $\mathcal{I}$ denotes the AdS timelike conformal boundary at infinity. This boundary condition is manifestly local in $\phi$, and it is not difficult to see that it is AdS-invariant.

For $0 < \nu < 1$, we have a choice of the parameter $q_\nu$ given by \eqref{eq:nonlocalq} for each spherical harmonic mode. However, it is not difficult to see that such choices will give rise to a boundary condition that is local in $\phi$ if and only if the same value of $q_\nu$ is chosen for each spherical harmonic mode. Thus, we obtain a single, one-parameter family of local boundary conditions, labeled by the parameter $q_\nu$. According to eqs. \eqref{eq:nonintnu} and \eqref{eq:resexp}, the asymptotic behavior of $\phi$ is given by
\begin {align}
\phi (t, r, z^i) = L^{\frac{1}{2} - \nu} \bigg( \frac{a_{\nu}}{r^{\frac{d}{2} - \nu}} + \frac{L^{2\nu}  b_{\nu}}{r^{\frac{d}{2} + \nu}} + \cdots \bigg)
\label{eq:gsfall}
\end {align}
Thus, the boundary condition on $\phi$ corresponding to a choice of $q_\nu$ can be expressed as
\begin {align}
\bigg[r^{2 \nu-1} n^\mu \nabla_\mu \big(r^{\frac{d}{2}-\nu} \phi \big) + 2 \nu L^{2\nu-1} q_{\nu}  \big( r^{\frac{d}{2} - \nu} \phi \big) \bigg] \bigg|_{\mathcal{I}} = 0
\label{eq:gsBC}
\end {align}
where we have defined
\begin {align}
n^\mu = \frac{r^2}{L} \bigg(\frac{\partial}{\partial r} \bigg)^\mu 
\label{ndef}
\end {align}
so that $n^\mu$ is the unit normal to $\mathcal{I}$ in the unphysical metric 
\begin {align}
\tilde{g}_{\mu \nu} = \Omega^2 g_{\mu \nu}
\end {align}
with $\Omega = 1/r$. The boundary condition \eqref{eq:gsBC} will give rise to stable dynamics (positive $A$) if and only if $q_\nu$ satisfies eq.(169) of \cite{IW2004} for all $l$, as will be the case if it satisfies that equation for $l=0$. 

The boundary condition \eqref{eq:gsBC} is manifestly local in $\phi$ for all values of $q_\nu$. It also is easily seen to be invariant under time translations and spatial rotations for all $q_\nu$. To determine invariance under more general AdS isometries, we consider an isometry corresponding to a Lorentz boost in the embedding space. For simplicity and definiteness, we will restrict our calculation here to AdS$_{1, 3}$, but the generalization of the analysis to other dimensions is straightforward. The particular isometry $(t, r, \theta, \phi) \to (t', r', \theta', \phi')$ that we shall consider is \footnote{The AdS$_{1, 3}$ spacetime is the covering space of the hyperboloid
\begin {align}
- (Y^0)^2 - (Y^1)^2 + (Y^2)^2 +(Y^3)^2 + (Y^4)^2  = - L^2 \nonumber
\end{align}
in $\mathbb{R}^{2, 3}$.
The coordinates $(t, r, \theta, \phi)$ are related to the global inertial coordinates of $\mathbb{R}^{2, 3}$ by
\begin {align}
& Y^0 = \sqrt{L^2 + r^2} \sin (t), \ Y^1 = \sqrt{L^2 + r^2} \cos (t) \nonumber \\
& Y^2 = r \sin (\theta) \cos (\phi), \ Y^3 =  r \sin (\theta) \sin (\phi), \ Y^4 =  r \cos (\theta). \nonumber 
\end {align}
The isometry \eqref{eq:r} corresponds to a Lorentz boost in the $Y^0$-$Y^4$ plane.}
\begin {align}
& r' = \sqrt{r^2 \sin^2 \theta - \frac{(\sqrt{r^2 +L^2} \beta \sin(t) - r \cos\theta)^2}{\beta^2-1}} \nonumber \\
& \cos (t') = \frac{\sqrt{r^2 + L^2} \cos(t)}{\sqrt{r^{'2} + L^2}}, \quad \sin (\theta')  = \frac{r \sin (\theta)}{r'}, \quad \phi' = \phi. 
\label{eq:r} 
\end {align}
where $\beta$ is the boost parameter. A straightforward but lengthy calculation shows that, under this isometry, the transformed scalar field, $\phi'$, satisfies a boundary condition of the general form of \eqref{eq:gsBC} but with a new value of $q_\nu$ given by
\begin {align}
q_\nu' = \frac{b_{\nu}'}{a_{\nu}'}  = \lambda^{2\nu} q_\nu 
\end {align}
where
\begin {align}
\lambda =  \sqrt{\sin^2 \theta - \frac{( \beta \sin t - \cos\theta)^2}{\beta^2-1}} .
\end {align}
Thus, we see that the general Robin boundary condition \eqref{eq:gsBC} fails to be AdS-invariant except in the special cases of Dirichlet conditions ($q_\nu = \infty$) and Neumann conditions ($q_\nu = 0$). This corresponds to the two Breitenlohner-Freedman \cite{BF1982} boundary conditions invariant under supersymmetry transformations.

For $\nu = 0$, the local IW boundary conditions analogous to \eqref{eq:gsBC} are
\begin {align}
\bigg[ r \big(\text{log} (r/L) - q_0/2 \big) \partial_r \big( r^{\frac{d}{2}} \phi \big) - r^{\frac{d}{2}} \phi \bigg] \bigg|_{\mathcal{I}} = 0
\label{eq:bfsBC}
\end {align}
This boundary condition will give rise to stable dynamics (positive $A$) if and only if $q_0$ satisfies eq.(171) of \cite{IW2004} for all $l$, as will be the case if it satisfies that equation for $l=0$. On account of the logarithmic term in \eqref{eq:bfsBC}, the boundary condition $q_0 = 0$ does not remain invariant under a Lorentz boost of the embedding space. Thus, for $\nu = 0$, the only local boundary condition \eqref{eq:bfsBC} that is AdS-invariant is $q_0 = \infty$ (i.e., $r \partial_r(r^{d/2} \phi) |_\mathcal{I} = 0$), which may be viewed as a generalized Dirichlet-Neumann condition. 

\medskip

{\em In summary, for the scalar field $\phi$ in AdS$_{1, d}$ with $\nu^2 \geq 1$, the general IW boundary condition that is local in $\phi$ is \eqref{nu2g1} and it is AdS-invariant; for $0 < \nu < 1$, the general local IW boundary condition is \eqref{eq:gsBC} but only the Dirichlet ($q_\nu = \infty$) and Neumann ($q_\nu = 0$) cases are AdS-invariant; and for $\nu = 0$, the general IW local boundary condition is \eqref{eq:bfsBC} but only the Dirichlet-Neumann case $q_0 = \infty$ is AdS-invariant.}

\subsection {Maxwell theory in AdS$_{1, 3}$}
\label{sec:EMBC}

As noted at the beginning of the previous subsection, the scalar field in AdS in any dimension can take any value of $\nu^2$, depending on its mass and coupling to curvature. However, the value of $\nu^2$ for the potentials describing the Maxwell field is fixed by \eqref{eq:EMnu}. Taking account of the fact that the vector part is present only for $d \geq 3$, we see that the IW formalism allows nontrivial boundary conditions (i.e., $0 \leq \nu < 1$) only in AdS$_{1, 3}$, AdS$_{1, 4}$, and AdS$_{1, 5}$. Furthermore, in AdS$_{1, 4}$ and AdS$_{1, 5}$, the boundary conditions are nontrivial only in the scalar sector. We will defer discussion of AdS$_{1, 4}$ and AdS$_{1, 5}$ to appendix \ref{sec:apphigher} and will consider only AdS$_{1, 3}$ in this section. 

In AdS$_{1, 3}$, we have\footnote{Note that this is the same value of $\nu$ as for a conformally invariant scalar field ($m^2 =0$, $\xi = (d-1)/4d$).} $\nu_S = \nu_V = 1/2$. Thus, $\Phi_{V{\textbf{k}}}^{\rm EM}$ and $\Phi_{S{\textbf{k}}}^{\rm EM}$ satisfy exactly the same equation. This is not an accident: In 4 spacetime dimensions, Maxwell's equations are invariant under a duality transformation
\begin {align}
F_{\mu \nu} \to {}^*F_{\mu \nu} \equiv \frac{1}{2} \epsilon_{\mu \nu \lambda \sigma} F^{\lambda \sigma}, \quad {}^*F_{\mu \nu} \to - F_{\mu \nu}
\end {align}
under which we have
\begin {equation}
\Phi_{S{\textbf{k}}}^{\rm EM} \rightarrow \Phi_{V{\textbf{k}}}^{\rm EM}, \quad \Phi_{V{\textbf{k}}}^{\rm EM} \rightarrow - \Phi_{S{\textbf{k}}}^{\rm EM} . 
\label{dualtrnem}
\end {equation}
Since $\Phi_{V{\textbf{k}}}^{\rm EM}$ and $\Phi_{S{\textbf{k}}}^{\rm EM}$ satisfy the same equation, we have additional freedom---not considered in the IW analysis---of applying boundary conditions to linear combinations of these quantities. We will make use of this additional freedom below.

The IW formalism allows one to choose boundary conditions independently for each spherical harmonic component. However, as in the case of the scalar field, in order to get local boundary conditions, we restrict to the same value of $q_\frac{1}{2}$ for each spherical harmonic component. Therefore, the local boundary conditions will be much more conveniently expressed in terms of the quantities $\Phi_S^{\rm EM}$ and $\Phi_V^{\rm EM}$ defined by
\begin {align}
\Phi_S^{\rm EM} = \sum_{\textbf{k}} \Phi_{S{\textbf{k}}}^{\rm EM} \mathbb{S}_{\textbf{k}} , \quad \epsilon_{ij} \hat{D}^j \Phi_{V}^{\rm EM} = \sum_{\textbf{k}} \Phi_{V{\textbf{k}}}^{\rm EM} \mathbb{V}_{\textbf{k} i} 
\end {align}
where we have used the fact that $\mathbb{V}_{\textbf{k}i} = \epsilon_{ij} \hat{D}^j \mathbb{S}_{\textbf{k}}$. Equation \eqref{eq:nonintnu} with $\nu = 1/2$ yields simply 
\begin {align}
\Phi_S^{\rm EM} (t, r, \theta, \phi) & = \bigg[ a_{\frac{1}{2}}^S (t, \theta, \phi) + \frac{L}{r} b_{\frac{1}{2}}^S (t, \theta, \phi)  + \cdots \bigg]  \nonumber \\
\Phi_V^{\rm EM} (t, r, \theta, \phi) & = \bigg[ a_{\frac{1}{2}}^V (t, \theta, \phi) + \frac{L}{r} b_{\frac{1}{2}}^V (t, \theta, \phi)  + \cdots \bigg]
\label{abEMfall}
\end {align}
It is then easily seen that the IW boundary conditions that are local in $\Phi_S^{\rm EM}$ and $\Phi_V^{\rm EM}$ are labeled by two real parameters, $q_{\frac{1}{2}}^S$ and $q_{\frac{1}{2}}^V$ (which are allowed to take the value $\infty$), and take the form,
\begin {align}
& \big[ n^\mu \partial_\mu \Phi_S^{\rm EM} + q_{\frac{1}{2}}^S \Phi_S^{\rm EM} \big] \big|_{\mathcal{I}} = 0 \label{eq:EMphiqS} \\
& \big[ n^\mu \partial_\mu \Phi_V^{\rm EM} + q_{\frac{1}{2}}^V \Phi_V^{\rm EM} \big] \big|_{\mathcal{I}} = 0  
\label{eq:EMphiqV}
\end {align}
where $n^\mu$ was defined by \eqref{ndef}.
The dynamics defined by these boundary conditions will give rise to stable dynamics (positive $A$) if and only if $q_{\frac{1}{2}}^S$ and $q_{\frac{1}{2}}^V$ satisfy eq.(169) of \cite{IW2004} for all $l$, as will be the case if they satisfy that equation for $l=1$.
In accord with the comments of the previous paragraph, these boundary conditions can be generalized by applying them to linear combinations of $\Phi_S^{\rm EM}$ and $\Phi_V^{\rm EM}$.

Although \eqref{eq:EMphiqS} and \eqref{eq:EMphiqV} are local in $\Phi_S^{\rm EM}$ and $\Phi_V^{\rm EM}$, it remains to determine whether they are local in the electromagnetic field itself. It also remains to determine for which values of $q_{\frac{1}{2}}^S$ and $q_{\frac{1}{2}}^V$ the boundary conditions are AdS-invariant. With regard to the locality in the electromagnetic field, it can be verified that the field strength tensor $F_{\mu \nu}$ is given in terms of $\Phi_S^{\rm EM}$ and $\Phi_V^{\rm EM}$ in AdS$_{1,3}$ by
\begin {align}
F_{tr} & = - \frac{L}{r^2} \hat{D}^2 \Phi_S^{\rm EM} \label{eq:Ftrphi} \\
F_{ai} & = \epsilon_{ab} \hat{\nabla}^b \hat{D}_i \Phi_S^{\rm EM}  +  \epsilon_{ij} \hat{D}^j \hat{\nabla}_a \Phi_V^{\rm EM} \label{eq:Faiphi}  \\
F_{ij} & = - \hat{D}^2 \Phi_{V}^{\rm EM} \epsilon_{ij}
\label{eq:Fijphi} 
\end {align}
where our notational conventions on indices were stated at the end of section \ref{sec:intro}. The boundary conditions \eqref{eq:EMphiqS} and \eqref{eq:EMphiqV} are most conveniently reformulated by introducing the (rescaled) ``electric'' field $\tilde{E}_I$ and ``magnetic'' field $\tilde{B}_I$ at $\mathcal I$ by\footnote{Note that ${\epsilon^{\mu \nu}}_{\lambda \rho}$ is conformally invariant, so we may use either the physical $\epsilon$ (with indices raised by the physical metric) or the unphysical $\epsilon$ (with indices raised by the unphysical metric) in \eqref{EBdef}.}
\begin {align}
\tilde{E}_I \equiv F_{I \mu} n^\mu, \quad \tilde{B}_I \equiv {}^*F_{I \mu} n^\mu = \frac{1}{2} \epsilon^{JK}{}_{I\mu} F_{JK} n^\mu 
\label{EBdef}
\end {align}
We put quotes on ``electric'' and ``magnetic'' (dropped hereinafter) because the decomposition of $F_{\mu \nu}$ into electric and magnetic parts is normally defined relative to a choice of timelike vector, whereas our $n^\mu$ is spacelike. We referred to these fields as ``rescaled'' because $n^\mu$ is the unit normal to $\mathcal I$ in the unphysical metric; the corresponding quantities defined with respect to the unit normal $\Omega n^\mu = n^\mu/r$ with respect to the physical metric would be $E_I = \Omega \tilde{E}_I$ and $B_I = \Omega \tilde{B}_I$. 

Note that by \eqref{eq:Ftrphi} $\tilde{E}_t \propto F_{tr}$ depends only on $\Phi_S^{\rm EM}$, whereas by \eqref{eq:Fijphi} $\tilde{B}_t \propto F_{ij}$ depends only on $\Phi_V^{\rm EM}$. Using these facts, it can be seen that the boundary conditions \eqref{eq:EMphiqS} and \eqref{eq:EMphiqV} can be reformulated as local boundary conditions on $\tilde{E}_t$ and $\tilde{B}_t$ as
\begin {align}
&  \big[ n^\mu \partial_\mu \big(\tilde{E}_t \big) + q_{\frac{1}{2}}^S \big( \tilde{E}_t \big) \big] \big|_{\mathcal{I}} = 0  \label{eq:Etq} \\ 
&  \big[ n^\mu \partial_\mu \big( \tilde{B}_t \big)  + q_{\frac{1}{2}}^V \big( \tilde{B}_t \big) \big] \big|_{\mathcal{I}}  = 0  \label{eq:Btq}
\end {align}
Again, these boundary conditions can be generalized by applying them to linear combinations of $\tilde{E}_t$ and $\tilde{B}_t$.

It remains to determine which of the boundary conditions \eqref{eq:Etq} and \eqref{eq:Btq} are AdS-invariant. The IW boundary conditions are guaranteed, by construction, to be invariant under time translations, and it is easily seen that \eqref{eq:Etq} and \eqref{eq:Btq} are invariant under rotations. However, we cannot determine the behavior of \eqref{eq:Etq} and \eqref{eq:Btq} under the ``boost isometry'' \eqref{eq:r} by simply applying the scalar results to \eqref{eq:EMphiqS} and \eqref{eq:EMphiqV} because $\Phi_S^{\rm EM}$ and $\Phi_V^{\rm EM}$ mix with each other in a complicated way under this isometry. Nevertheless, we have performed the necessary computations and shown that the only local, AdS-invariant boundary conditions within the original IW framework are given by the choices ($q_{\frac{1}{2}}^S=0, \, q_{\frac{1}{2}}^V = \infty$) and ($q_{\frac{1}{2}}^S=\infty, \, q_{\frac{1}{2}}^V = 0$). The choice ($q_{\frac{1}{2}}^S=0, \, q_{\frac{1}{2}}^V = \infty$) requires that $n^\mu \partial_\mu \Phi_S^{\rm EM}=0$ and $\Phi_V^{\rm EM} = 0$. From \eqref{eq:Faiphi}, \eqref{eq:Fijphi}, and \eqref{EBdef}, it can be seen that this is equivalent to the condition
\begin {align}
 \tilde{B}_I \big|_{\mathcal{I}} = 0 \label{eq:EMB} .
\end{align}
Similarly, the choice ($q_{\frac{1}{2}}^S=\infty, \, q_{\frac{1}{2}}^V = 0$) is equivalent to 
\begin{align}
 \tilde{E}_I \big|_{\mathcal{I}}= 0 \label{eq:EME}
\end {align}
The freedom to apply the boundary conditions to linear combinations of $\Phi_S^{\rm EM}$ and $\Phi_V^{\rm EM}$ yields the more general local, AdS-invariant boundary condition
\begin {equation}
\big[ \tilde{E}_I + \alpha \tilde{B}_I \big]_{\mathcal{I}} = 0
\label{eq:EMdualrot}
\end {equation}
where $\alpha$ is an arbitrary real number (including infinity). The dynamics defined by \eqref{eq:EMdualrot} is well posed and stable for all values of $\alpha$. We conjecture that---independently of the IW prescription---the boundary conditions \eqref{eq:EMdualrot} are the most general, local, AdS-invariant boundary conditions on the electromagnetic field in AdS$_{1, 3}$ that admit a well-posed evolution and are ``conservative'' in the sense that it has a conserved positive energy of the type discussed in \cite{IW2003}. As we shall show in section \ref{sec:phase}, we will be able to define a covariant phase space structure---with a conserved symplectic form---on the space of solutions satisfying the boundary conditions \eqref{eq:EMdualrot}, which provides an additional important sense in which the dynamics are conservative.

\medskip

{\em In summary, for the Maxwell field in AdS$_{1, 3}$, if we follow the IW prescription of applying boundary conditions for $\Phi_S^{\rm EM}$ and $\Phi_V^{\rm EM}$ separately, then the most general local boundary conditions are \eqref{eq:Etq} and \eqref{eq:Btq} and the most general AdS-invariant boundary conditions are \eqref{eq:EMB} and \eqref{eq:EME}. If we generalize the IW prescription to boundary conditions to be applied to linear combinations of $\Phi_S^{\rm EM}$ and $\Phi_V^{\rm EM}$, then the most general, local, AdS-invariant boundary conditions are \eqref{eq:EMdualrot}. We believe that the boundary conditions \eqref{eq:EMdualrot} are the most general possible local, AdS-invariant, conservative boundary conditions for the electromagnetic field.}

\subsection {Yang-Mills Theory in AdS$_{1, 3}$}
\label{sec:YMBC}

The Yang-Mills Lagrangian is
\begin {align}
\mathcal{L}_{\rm YM} =  - \frac{1}{2 g_{\rm YM}^2} \big(F^a  \wedge {}^*F_a \big)
\label{YMlag}
\end {align}
where $a$ here denotes a Lie algebra index and 
\begin  {align}
F_{\mu \nu}^a = \partial_\mu A_\nu^a - \partial_\nu A_\mu^a +  c^a{}_{bc} A_\mu^b A_\nu^c
\end {align}
where $c^a{}_{bc}$ is the structure tensor of the (assumed to be compact and semi-simple) Lie algebra \footnote{The structure tensor is defined by $[T,S]^a = {c^a}_{bc} T^b S^c$, where $[,]$ denotes the Lie bracket and $T^a$ and $S^a$ are arbitrary elements of the Lie algebra. Lie algebra indices are raised and lowered with the positive-definite bilinear form $k_{ab}  = - c^c{}_{ad}c^d{}_{bc}$. When we switch to index free notation in section \ref{sec:YMphase}, we will write $k_{ab} X^a Y^b$ as $\text{Tr} (XY)$.}. The Yang-Mills equations arising from \eqref{YMlag} are nonlinear. Consequently, they do not fall within the framework of theories that can be analyzed using the IW methods. In particular, we cannot introduce analogs of the electromagnetic potentials $\Phi_S^{\rm EM}$ and $\Phi_V^{\rm EM}$. Nevertheless, the Yang-Mills analogs of the local boundary conditions \eqref{eq:Etq} and \eqref{eq:Btq}, i.e., 
\begin {align}
&  \big[ n^\mu \mathcal{D}_\mu \big(\tilde{E}^a_t \big) + q_{\frac{1}{2}}^S \big( \tilde{E}^a_t \big) \big] \big|_{\mathcal{I}} = 0  \label{eq:YMEtq} \\ 
&  \big[ n^\mu \mathcal{D}_\mu \big( \tilde{B}^a_t \big)  + q_{\frac{1}{2}}^V \big( \tilde{B}^a_t \big) \big] \big|_{\mathcal{I}}  = 0  \label{eq:YMBtq}
\end {align}
make sense in Yang-Mills theory, where we have promoted the partial derivatives to the covariant versions defined via $(\mathcal{D}_\mu X)^a = \partial_\mu X^a + c^a{}_{bc} A^b_\mu X^c $ with $X = X^a T_a$. We conjecture that they give rise to well-posed evolution. 

Furthermore, the Yang-Mills analog of the boundary conditions \eqref{eq:EMdualrot}, namely,
\begin {align}
\big[ \tilde{E}_I^a + \alpha \tilde{B}_I^a \big] \big|_{\mathcal{I}} = 0
\label{eq:YMdualrot}
\end {align}
are well defined and AdS-invariant. We conjecture that, for all $\alpha$ (including $\alpha = \infty$), the Yang-Mills equations with the boundary condition \eqref{eq:YMdualrot} yield well-posed, stable evolution in AdS$_{1, 3}$. As we shall show in section \ref{sec:phase}, we will be able to define a covariant phase space structure on the solutions to the Yang-Mills equations with boundary conditions \eqref{eq:YMdualrot}, and we conjecture that \eqref{eq:YMdualrot} are the most general AdS-invariant boundary conditions for which this is possible.

\medskip

{\em In summary, although the IW analysis does not apply to Yang-Mills theory, analogs of the electromagnetic boundary conditions found in the previous section make sense in Yang-Mills theory in AdS$_{1, 3}$ and we expect that they will give rise to well-posed evolution. We also conjecture that \eqref{eq:YMdualrot} are the most general AdS-invariant boundary conditions that admit a covariant phase space structure on solutions.}

\subsection {Linearized Einstein gravity in AdS$_{1, 3}$}
\label{sec:GRAVBC}

In AdS$_{1, d}$, the values of $\nu^2$ for the scalar, vector, and tensor parts of a linearized gravitational perturbation are given by \eqref{eq:rescaledGT}. Since the tensor part is present only when $d \geq 4$, we see that we always have $\nu_T^2 > 1$, and the IW boundary conditions for the tensor part are unique. The values of $\nu_S^2$ and $\nu_V^2$ are exactly the same as in the electromagnetic case. Consequently, as in the electromagnetic case, the IW boundary conditions are nontrivial only for AdS$_{1, 3}$, AdS$_{1, 4}$, and AdS$_{1, 5}$. We will defer discussion of AdS$_{1, 4}$ and AdS$_{1, 5}$ to appendix \ref{sec:apphigher} and will consider only AdS$_{1, 3}$ in this section.

As in the electromagnetic case, since we are interested in local boundary conditions, it will be convenient to work with the quantities
\begin {align}
\Phi_S^{\rm G} = \sum_{\textbf{k}} \Phi_{S{\textbf{k}}}^{\rm G} \mathbb{S}_{\textbf{k}} , \quad \epsilon_{ij} \hat{D}^j \Phi_{V}^{\rm G} = \sum_{\textbf{k}} \Phi_{V{\textbf{k}}}^{\rm G} \mathbb{V}_{\textbf{k} i} 
\label{gravpots}
\end {align}
rather than the individual spherical harmonic modes $\Phi_{V{\textbf{k}}}^{\rm G}$ and $\Phi_{S{\textbf{k}}}^{\rm G}$. In AdS$_{1, 3}$, we have $\nu_S = \nu_V = 1/2$, therefore both fields $\Phi_{V}^{\rm G}$ and $\Phi_{S}^{\rm G}$ satisfy exactly the same equation as each other (and the same equation as $\Phi_{V}^{\rm EM}$ and $\Phi_{S}^{\rm EM}$ in AdS$_{1, 3}$). There is no general notion of ``duality transformation'' in nonlinear or linearized gravity, so there is no reason, a priori, why $\Phi_V^{\rm G}$ and $\Phi_S^{\rm G}$ need satisfy the same equation. Nevertheless, the fact that they do shows the presence of such a symmetry. Indeed, it can be verified from the formulas of appendix \ref{sec:appfall} that the transformation
\begin {equation}
\Phi_S^G \rightarrow - 2 \Phi_V^G, \quad \Phi_V^G \rightarrow \frac{1}{2} \Phi_S^G
\end {equation}
corresponds to a duality transformation\footnote{The existence of such a duality transformation is undoubtedly a consequence of the vanishing of the Weyl tensor in the AdS background.} of the linearized Weyl tensor $C^{(1)}_{\mu \nu \lambda \rho}$, namely
\begin {align}
C^{(1)}_{\mu \nu \lambda \rho} \to {}^*C^{(1)}_{\mu \nu \lambda \rho}, \quad {}^*C^{(1)}_{\mu \nu \lambda \rho} \to - C^{(1)}_{\mu \nu \lambda \rho}
\end {align}
where
\begin {align}
{}^*C^{(1)}_{\mu \nu \lambda \rho} \equiv \frac{1}{2} \epsilon_{\mu \nu \tau \sigma} {C^{(1) \tau \sigma}}_{\lambda \rho}
\end {align}
In any case, the fact that $\Phi_V^{\rm G}$ and $\Phi_S^{\rm G}$ satisfy the same equation allows us to generalize the IW framework by applying their boundary conditions to linear combinations of $\Phi_V^{\rm G}$ and $\Phi_S^{\rm G}$.

Since $\Phi_V^{\rm G}$ and $\Phi_S^{\rm G}$ satisfy the same equations as their electromagnetic counterparts, we can immediately write down the general IW boundary conditions that are local in $\Phi_S^{\rm G}$ and $\Phi_V^{\rm G}$, namely
\begin {align}
& \big[ n^\mu \partial_\mu \Phi_S^{\rm G} + q_{\frac{1}{2}}^S \Phi_S^{\rm G} \big] \big|_{\mathcal{I}} = 0 \label{eq:GphiqS} \\
& \big[ n^\mu \partial_\mu \Phi_V^{\rm G} + q_{\frac{1}{2}}^V \Phi_V^{\rm G} \big] \big|_{\mathcal{I}} = 0  
\label{eq:GphiqV}
\end {align}
(see \eqref{eq:EMphiqS} and \eqref{eq:EMphiqV}). The dynamics defined by these boundary conditions will give rise to stable dynamics if and only if $q_{\frac{1}{2}}^S$ and $q_{\frac{1}{2}}^V$ satisfy eq.(169) of \cite{IW2004} for all $l$, as will be the case if they satisfy that equation for $l=2$. These boundary conditions can be generalized by applying them to linear combinations of $\Phi_V^{\rm G}$ and $\Phi_S^{\rm G}$.

In order to express these conditions as local expressions in terms of the perturbed metric, we need formulas expressing the linearized Weyl tensor in terms of $\Phi_S^{\rm G}$ and $\Phi_V^{\rm G}$. These formulas are considerably more complicated than the corresponding formulas \eqref{eq:Ftrphi} - \eqref{eq:Fijphi} in the electromagnetic case; they are given in appendix \ref{sec:appfall}. The boundary conditions are most conveniently expressed in terms of the rescaled electric and magnetic parts of the linearized Weyl tensor, defined by
\begin {align}
\tilde{E}_{IJ}^{(1)} & \equiv \Omega \, C^{(1)}_{I \mu J \nu} n^\mu n^\nu \label{eq:EIJ} \\
\tilde{B}_{IJ}^{(1)} & \equiv \Omega \, {}^*C^{(1)}_{I \mu J \nu} n^\mu n^\nu  \label{eq:BIJ}
\end {align}
where $n^\mu$ was defined by \eqref{ndef} and $C^{(1)}_{\mu \nu \lambda \rho}$ is the physical Weyl tensor, i.e., with its first index lowered by the physical metric. (In terms of the unphysical Weyl tensor---i.e., with its first index lowered by the unphysical metric---the corresponding formula for $\tilde{E}_{IJ}$ would be $\tilde{E}_{IJ} = \Omega^{-1} \, \tilde{C}^{(1)}_{I \mu J \nu} n^\mu n^\nu$.) In parallel with the electromagnetic case, $\tilde{E}_{tt}$ depends only on $\Phi_S^{\rm G}$, whereas $\tilde{B}_{tt}$ depends only on $\Phi_V^{\rm G}$. The boundary condition \eqref{eq:GphiqS} can thereby be rewritten as a boundary condition on $\tilde{E}_{tt}$ and the boundary condition \eqref{eq:GphiqV} can thereby be rewritten as a boundary condition on $\tilde{B}_{tt}$. Using the formulas from appendix \ref{sec:appfall}, we have shown that \eqref{eq:GphiqS} and \eqref{eq:GphiqV} can be reformulated as 
\begin {align}
& \big[ n^\mu \partial_\mu \big(\tilde{E}_{tt}^{(1)} \big) + q_{\frac{1}{2}}^S  \tilde{E}_{tt}^{(1)} \big]\big|_{\mathcal{I}}  = 0 \label{eq:LGravtt1} \\ 
& \big[ n^\mu \partial_\mu \big( \tilde{B}_{tt}^{(1)} \big)  + q_{\frac{1}{2}}^V  \tilde{B}_{tt}^{(1)} \big]\big|_{\mathcal{I}} = 0 
\label{eq:LGravtt2}
\end {align}
Again, these boundary conditions can be generalized by applying them to linear combinations of $\tilde{E}_{tt}^{(1)}$ and $\tilde{B}_{tt}^{(1)}$.

In close parallel with the electromagnetic case, the boundary conditions in the above family that are AdS-invariant are ($q_{\frac{1}{2}}^S=0, \, q_{\frac{1}{2}}^V = \infty$), corresponding to 
\begin {align}
 \tilde{B}_{IJ}^{(1)}\big|_{\mathcal{I}} = 0 \label{eq:LGravB}
\end {align}
and ($q_{\frac{1}{2}}^S= \infty, \, q_{\frac{1}{2}}^V = 0$), corresponding to 
\begin {align}
 \tilde{E}_{IJ}^{(1)}\big|_{\mathcal{I}} = 0 \label{eq:LGravE}
\end {align}
Again, the freedom to apply the boundary conditions to linear combinations of $\Phi_S^{\rm G}$ and $\Phi_V^{\rm G}$ yields the more general local, AdS-invariant boundary condition
\begin {align}
\big[ \tilde{E}_{IJ}^{(1)} + \alpha \tilde{B}_{IJ}^{(1)} \big]_{\mathcal{I}} = 0 
 \label{eq:LGravdualrot}
\end {align}
where $\alpha$ is a real number (including infinity). The dynamics defined by \eqref{eq:LGravdualrot} is well posed and stable for all values of $\alpha$.

\medskip

{\em In summary, for the linearized gravitational field in AdS$_{1, 3}$, if we follow the IW prescription of applying boundary conditions for $\Phi_S^{\rm G}$ and $\Phi_V^{\rm G}$ separately, then the most general local boundary conditions are \eqref{eq:LGravtt1} and \eqref{eq:LGravtt2} and the most general AdS-invariant boundary conditions are \eqref{eq:LGravB} and \eqref{eq:LGravE}. If we generalize the IW prescription to boundary conditions to be applied to linear combinations of $\Phi_S^{\rm G}$ and $\Phi_V^{\rm G}$, then the most general, local, AdS-invariant boundary conditions are \eqref{eq:LGravdualrot}. We believe that the boundary conditions \eqref{eq:LGravdualrot} are the most general possible local, AdS-invariant, conservative boundary conditions for the linearized gravitational field.}

\subsection {Nonlinear Einstein gravity in 4-dimensional Fefferman-Graham (FG) Spacetimes}
\label{sec:NONLINGRAVBC}

If we wish to consider fully nonlinear gravity, then, of course, we no longer have a background AdS spacetime. A natural class of ``AdS-like'' spacetimes is the Fefferman-Graham (FG) class, consisting of solutions to Einstein's equation with negative cosmological constant with metric of the form
\begin {multline}
ds^2 =  \frac{L^2}{z^2} dz^2 + \frac{L^2}{z^2} \big[g_{(0)IJ} \\
+  z^2 g_{(2)IJ} +  z^3 g_{(3)IJ} + \cdots \big] dx^I dx^J 
\label{fgform}
\end {multline}
where the conformal boundary at infinity is $z=0$. 
The AdS$_{1, 3}$ metric itself \eqref{eq:AdSmetric1} takes this form, namely,
\begin {align}
ds^2 = \frac{L^2}{z^2} dz^2 + \frac{L^2}{z^2} \bigg[ - \frac{(1+z^2)^2}{4} dt^2 + \frac{(1-z^2)^2}{4} d\Omega_2^2  \bigg]
\label{fgexp}
\end {align}
with $z$ defined by
\begin {align}
r = \frac{L (1 - z^2)}{2 z}.
\label{rads}
\end {align}

Note the absence of a $g_{(1)IJ}$ term in the Fefferman-Graham (FG) expansion \eqref{fgform}, which is a consequence of Einstein's equation. An additional important consequence of Einstein's equation is that $g_{(2)IJ}$ is given by the Schouten tensor of $g_{(0)IJ}$ \footnote{In addition, $g_{(3)IJ}$ is trace-free and divergence-free with respect to $g_{(0)IJ}$.}
\begin {align}
 g_{(2) IJ} = - S_{IJ}[ g_{(0)}]
\label{g2schouten}
\end {align}
where, for any metric $\gamma_{IJ}$ in $3$ dimensions, the Schouten tensor is given in terms of the Ricci curvature of $\gamma_{IJ}$ by
\begin{align}
S_{IJ} [\gamma] =  R_{IJ}[\gamma] - \frac{1}{4} R[\gamma] \gamma_{IJ} .
\end{align}

The IW analysis applies only to linear fields in AdS spacetime, so, of course, it does not apply to nonlinear gravity with FG asymptotics. Furthermore, if we remove the ``$(1)$'' superscripts from the local, non-AdS-invariant boundary conditions \eqref{eq:LGravtt1} and \eqref{eq:LGravtt2}, the boundary conditions become highly non-covariant in a general FG setting. However, since for a general FG metric $\partial/\partial t$ need not be a Killing field, we see no reason to expect that the resulting boundary conditions will lead to well-defined dynamics. On the other hand, the general AdS-invariant boundary conditions \eqref{eq:LGravdualrot} with the ``$(1)$'' superscripts removed do make sense in a general FG setting. More precisely, it can be shown that for a general FG metric, the Weyl tensor $\tilde{C}^\mu{}_{\nu \lambda \rho}$ of the unphysical metric
\begin{align}
\widetilde{ds}^2 = & \Omega^2 ds^2 \nonumber \\
= & dz^2 + \big[g_{(0)IJ} +  z^2 g_{(2)IJ} +  z^3 g_{(3)IJ} + \cdots \big] dx^I dx^J
\label{FGunphys}
\end{align}
with $\Omega = z/L$ vanishes on $\mathcal I$ (i.e., $z=0$), so the quantities
\begin {align}
\tilde{E}_{IJ} & \equiv \Omega^{-1} \, \tilde{C}_{I \mu J \nu} n^\mu n^\nu \label{eq:FGEIJ} \\
\tilde{B}_{IJ} & \equiv \Omega^{-1} \, {}^*\tilde{C}_{I \mu J \nu} n^\mu n^\nu  \label{eq:FGBIJ}
\end {align}
are well defined on $\mathcal I$, where the first index of $\tilde{C}^\mu{}_{\nu \lambda \rho}$ is lowered with the unphysical metric \eqref{FGunphys} and the outward normal $n^\mu = - (\partial/\partial z)^\mu$.
Thus, the boundary condition
\begin {align}
\big[ \tilde{E}_{IJ} + \alpha \tilde{B}_{IJ} \big]_{\mathcal{I}} = 0 
 \label{eq:NLGravdualrot}
\end {align}
is well defined and covariant in a general FG setting. 
 As we shall explain below, it follows from the work of Friedrich \cite{Friedrich1995} and Sou\^{e}tre \cite{Souetre2025} that Einstein's equation with the boundary conditions \eqref{eq:NLGravdualrot} has a well-posed initial value formulation{\footnote{As shown in \cite{FAS2025}, the boundary conditions \eqref{eq:NLGravdualrot} also imply a vanishing Bel-Robinson flux through the conformal boundary.}.

First, however, we relate $\tilde{E}_{IJ}$ and $\tilde{B}_{IJ}$ on $\mathcal I$ to other quantities, as will be needed to relate the boundary conditions \eqref{eq:NLGravdualrot} to those of Sou\^{e}tre \cite{Souetre2025} and others. A direct calculation shows that $\tilde{E}_{IJ}$ is given by the third order term in the FG expansion
\begin{equation}
 \tilde{E}_{IJ}\big|_{\mathcal I} =  - \frac{3 L}{2} g_{(3) IJ} .
  \label{eq:EIJgIJ}
\end{equation}
Furthermore, $\tilde{E}_{IJ}$ also can be directly related to the rescaled ``boundary stress-energy tensor'' $\tilde{T}_{IJ}$, which is defined as follows. In the physical spacetime on a surface, $\mathcal{I}_z$, of constant $z > 0$, the Brown-York stress-energy tensor $T_{IJ}^{\rm BY}$ is defined by
\begin {align}
T_{IJ}^{\rm BY} [\gamma]  = - \frac{2}{\sqrt{- \gamma}} \frac{\delta (\mathcal{S}_{\rm EH} + \mathcal{S}_{\rm GHY}\big)_{\rm on-shell}}{\delta \gamma^{IJ}}  
\end{align}
where $S_{\rm EH}$ is the Einstein-Hilbert action and the Gibbons-Hawking-York action is
\begin{equation}
\mathcal{S}_{\rm GHY} = \frac{1}{8 \pi G_4} \int d^{3}x \sqrt{-\gamma} K .
\label{eq:GHY}
\end{equation}
Here $\gamma_{IJ}$ is the induced physical metric on $\mathcal{I}_z$, and $K$ is the trace of the extrinsic curvature of $\mathcal{I}_z$, calculated with respect to the physical metric. The ``counterterm stress-energy'' $T_{IJ}^{\rm CT}$ of Balasubramanian and Kraus \cite{BK1999} is similarly defined using the gravitational counterterm action \footnote{See also Henningson and Skenderis \cite{HeSk1998} for the subtraction of divergences in the on-shell gravitational action using local covariant counterterms. The gravitational counterterm action was generalized to arbitrary dimensions by de Haro, Skenderis, and Solodukhin\cite{DSS2000}.}
\begin{equation}
\mathcal{S}_{\rm CT} = - \frac{L}{16 \pi G_4} \int d^3x \sqrt{-\gamma} \bigg( \frac{4}{L^2} +  {}^{(3)}R \bigg)
 \label{eq:CT}
\end{equation}
where $^{(3)}R$ is the Ricci scalar of $\gamma_{IJ}$ on $\mathcal{I}_z$. We then have
\begin {align}
T_{IJ}^{\rm BY} (z) & =  - \frac{1}{8 \pi G_4} \big(K_{IJ} - K \gamma_{IJ} \big) \\
T_{IJ}^{\rm CT} (z) & =  - \frac{L}{8 \pi G_4} \bigg[\frac{2}{L^2} \gamma_{IJ} - G_{IJ} \bigg] 
\end {align}
where $G_{IJ}$ is the Einstein tensor of $\gamma_{IJ}$.
The rescaled boundary stress-energy tensor is then defined by
\begin {align}
\tilde{T}_{IJ} \equiv \lim_{z \to 0} \Omega^{-1} \left( T_{IJ}^{\rm BY} + T_{IJ}^{\rm CT} \right)
\end {align}
It can then be shown that 
\begin{equation}
 \tilde{E}_{IJ} \big|_{\mathcal I} = - \frac{8 \pi G_4}{L} \tilde{T}_{IJ} 
 \label{eq:EIJTIJ}
\end{equation}

On the other hand, $\tilde{B}_{IJ} \big|_{\mathcal I}$ can be shown to be related to $g_{(0) IJ}$ as follows. On the surface, $\mathcal{I}_z$, of constant $z > 0$ in the physical spacetime, consider the magnetic part of the Weyl tensor, $B_{IJ}$, given by
\begin {align}
B_{IJ} & = {}^* C_{I \mu J \nu} \eta^\mu \eta^\nu . \label{eq:physFGBIJ}
\end {align}
Here all quantities are defined relative to the physical metric and $\eta^\mu$ is the outward unit normal in the physical metric. The Einstein constraint equations imply that \cite{JM2019}
\begin {align}
 B_{IJ}  =  - \epsilon_{I}{}^{KL} \nabla_K K_{LJ} 
\end {align}
where $K_{LJ}$ is the extrinsic curvature of $\mathcal{I}_z$ and the volume element and derivative operator are defined by the induced physical metric, $\gamma_{IJ}$ on $\mathcal{I}_z$. From the FG form of the metric \eqref{fgform}, it can be seen that the outward unit normal in the physical spacetime is $\eta^\mu = - z/L (\partial/\partial z)^\mu$ and that the extrinsic curvature of $\mathcal{I}_z$ is
\begin {align}
K_{IJ} = \frac{1}{2} \mathcal{L}_{\eta} \gamma_{IJ} = \frac{L}{z^2} g_{(0)IJ} +O(z)
\end {align}
Since $\nabla_K \gamma_{IJ} = 0$, it follows that 
\begin {align}
\nabla_K \left(\frac{L}{z^2} g_{{(0)}IJ} \right) = & - L \nabla_K g_{{(2)}IJ} + O(z) \nonumber \\
= & - L \nabla^{(0)}_K g_{{(2)}IJ} + O(z) 
\end {align}
where $\nabla^{(0)}_K$ denotes the derivative operator associated with $g_{{(0)}IJ}$. Thus we have
\begin {align}
 B_{IJ}  = z \epsilon^{(0)}_{I}{}^{KL} \nabla^{(0)}_K g_{{(2)}LJ} + O(z^2)
\end {align}
where $\epsilon^{(0)}_{IKL}$ is the volume element (with outward-induced orientation) associated with $g_{{(0)}IJ}$ and indices are raised and lowered with this metric. Thus, taking \eqref{g2schouten} into account, we obtain
\begin {align}
 \tilde{B}_{IJ} \big|_{\mathcal I}  = & - L \epsilon^{(0)}_{I}{}^{KL} \nabla^{(0)}_K S_{LJ}[ g_{(0)}] \nonumber \\
 = & - L Y_{IJ} [g_{(0)}]
\label{BCY}
\end {align}
where $Y_{IJ}$ is the Cotton-York tensor, defined in terms of $S_{IJ}$ for any 3-dimensional metric $\gamma_{IJ}$ by
\begin{align}
Y_{IJ} =  \epsilon_{I}{}^{KL} \nabla_K S_{LJ} .
\label{cydef}
\end{align}
The Cotton-York tensor is automatically symmetric and trace-free, and its vanishing is necessary and sufficient for $\gamma_{IJ}$ to be locally conformally flat. Thus, \eqref{BCY} shows that the boundary metric is conformally flat if and only if $\tilde{B}_{IJ}|_{\mathcal I} = 0$.

It is worth pointing out that if we assume that $\mathcal I$ has the structure of $S^2 \times R$ with $S^2$ spacelike, then the local conformal flatness implies that $\mathcal I$ is locally conformal to a portion of the Einstein cylinder. If we further assume that $\mathcal I$ is in the conformal class of the full Einstein cylinder (i.e., the same conformal class as $\mathcal I$ of AdS$_{1, 3}$ spacetime), then by making use of the boundary diffeomorphism freedom and the freedom to redefine $z$, we may assume that the leading order $g_{(0)IJ}$ in the FG metric takes its AdS$_{1, 3}$ background value. By \eqref{g2schouten}, the subleading order $g_{(2)IJ}$ will also take its AdS$_{1, 3}$ value, and therefore the leading order deviation from the AdS$_{1, 3}$ background will arise at order $g_{(3)IJ}$. If we take an FG metric \eqref{fgform} with these properties and perform the coordinate transformation $z \to r$ given by \eqref{rads}, we get precisely the Henneaux and Teitelboim \cite {HT1985} metric fall-off conditions, with the additional restriction of being in the radial gauge (i.e., with $g_{rr}$ given by its AdS$_{1, 3}$ value and $g_{rI} =0$). Thus, under these assumptions, we see that the boundary condition $\tilde{B}_{IJ}|_{\mathcal I} = 0$ implies the Henneaux and Teitelboim \cite {HT1985} metric fall-off conditions; conversely, the Henneaux and Teitelboim fall-off conditions imply that $\tilde{B}_{IJ}|_{\mathcal I} = 0$.

The upshot of the above calculations yielding \eqref{eq:EIJgIJ}, \eqref{eq:EIJTIJ}, and \eqref{BCY} is that in our boundary conditions, we may replace the rescaled electric Weyl tensor $\tilde{E}_{IJ}|_{\mathcal I}$ by constants times either $g_{(3) IJ}$ or $\tilde{T}_{IJ}$ and we may replace $\tilde{B}_{IJ}|_{\mathcal I}$ by $-L Y_{IJ} [g_{(0)}]$. Thus, for example, the boundary conditions $\tilde{T}_{IJ} = 0$ considered by Comp\`ere and Marolf \cite{CM2008} correspond to the boundary conditions \eqref{eq:NLGravdualrot} with $\alpha = 0$. Most importantly, the boundary conditions 
\begin {align}
g_{(3) IJ} - \mu Y_{IJ}[g_{(0)}] \big|_{\mathcal{I}} = 0
\label{soubc}
\end {align}
for real $\mu = - 2\alpha/3$ are equivalent to the boundary conditions \eqref{eq:NLGravdualrot}. But the boundary conditions \eqref{soubc} are precisely the boundary conditions considered by Sou\^{e}tre \cite{Souetre2025}, who proved existence of a well-posed initial value formulation with these boundary conditions. Thus, the boundary conditions \eqref{eq:NLGravdualrot} give rise to a well-posed initial value formulation. It will be shown in appendix \ref{sec:appInh} that special cases of earlier results of Friedrich \cite{Friedrich1995} also show well-posedness with the boundary conditions \eqref{eq:NLGravdualrot}. 

As we shall show in section \ref{sec:phase}, we will be able to define a covariant phase space structure on the solutions to Einstein's equation with boundary conditions \eqref{eq:NLGravdualrot}, and we conjecture that \eqref{eq:NLGravdualrot} are the most general covariant boundary conditions for which this is possible.

\medskip

{\em In summary, the IW analysis does not apply to the nonlinear Einstein equation, and analogs of the linearized gravitational boundary conditions \eqref{eq:LGravtt1} and \eqref{eq:LGravtt2} do not appear to make sense in a general FG setting. However, \eqref{eq:NLGravdualrot}---which is the nonlinear analog of \eqref{eq:LGravdualrot}---does make sense in a general FG context and has previously been proven by Friedrich \cite{Friedrich1995} and Sou\^{e}tre \cite{Souetre2025} to give rise to well-posed dynamics. We conjecture that \eqref{eq:NLGravdualrot} are the most general covariant boundary conditions in a general FG setting that admit a covariant phase space structure on solutions.}

\section {Construction of Phase Space}
\label{sec:phase}

\subsection {General considerations on the construction of phase space and variational principles}
\label{gencon}

In the previous section, we obtained local boundary conditions for the scalar field in AdS$_{1, d}$, the electromagnetic, Yang-Mills, and linearized gravitational field in AdS$_{1, 3}$, and the nonlinear gravitational field in a general 4-dimensional Fefferman-Graham setting. These boundary conditions are known to give rise to well-posed dynamical evolution (except in the Yang-Mills case, where this was conjectured). However, well-posed evolution is not the only thing one might want: One also would like to have a phase space structure on the space of solutions. At the classical level, such a phase space structure would give rise to useful notions---such as a Hamiltonian and other conserved charges. In the case of linear fields, an appropriate phase space structure would allow one to formulate a quantum field theory corresponding to the classical field by the same procedures as used in quantum field theory in curved spacetime \cite{QFTCS}.

In this section, we employ the methods developed in \cite{LeeWald1990, BurnWald1990, IyerWald1994} to show that a phase space structure can be defined in the following cases: (i) the scalar field in AdS$_{1, d}$ with the general local boundary conditions \eqref{eq:gsBC} or \eqref{eq:bfsBC}; (ii) the electromagnetic and Yang-Mills fields in AdS$_{1, 3}$ with the general AdS-invariant boundary conditions \eqref{eq:EMdualrot} and \eqref{eq:YMdualrot}, respectively; and (iii) the nonlinear gravitational field in a general 4-dimensional FG setting with boundary conditions \eqref{eq:NLGravdualrot}. We will not consider the non-AdS-invariant boundary conditions \eqref{eq:Etq} and \eqref{eq:Btq} in the electromagnetic case nor the analogous non-AdS-invariant boundary conditions \eqref{eq:YMEtq} and \eqref{eq:YMBtq} in the Yang-Mills case, although we expect that a phase space structure can be put on the space of solutions in these cases as well. We also will not consider linearized gravity in AdS$_{1, 3}$ with boundary conditions \eqref{eq:LGravdualrot}, since the phase space in that case can be obtained as a special case of the general nonlinear phase space construction. We should emphasize that for the nonlinear cases of Yang-Mills theory and general relativity, when we say ``construct a phase space'' we merely mean we define a suitable symplectic product on linearized perturbations about solutions. We shall not consider any issues related to defining a topology or manifold structure on the space of solutions.

The general construction of phase space of \cite{LeeWald1990} can be described as follows. Consider a classical field theory in a $D$-dimensional spacetime defined by a Lagrangian $\boldsymbol{\mathcal{L}}$ constructed from dynamical fields $\varphi$ and possibly non-dynamical (``background'') fields $\psi$. We treat $\boldsymbol{\mathcal{L}}$ as a $D$-form and use boldface letters to denote differential forms when their indices are suppressed.
For the scalar, electromagnetic, and Yang-Mills cases, $\psi$ would be the background AdS metric; in the nonlinear gravitational case, no background fields would be allowed, i.e., the Lagrangian would be required to be diffeomorphism covariant. We assume that $\boldsymbol{\mathcal{L}}$
takes the general form
\begin {align}
\boldsymbol{\mathcal{L}} = \boldsymbol{\mathcal{L}} \big(\varphi, \nabla_\mu \varphi, \cdots, \nabla_{(\mu_1} \cdots \nabla_{\mu_k)} \varphi; \psi \big)
\end {align}
where, in this subsection, we will use boldface letters to denote differential forms on spacetime. 
If we vary the dynamical fields by $\delta \varphi$, the Lagrangian varies by 
\begin {align}
\delta \boldsymbol{\mathcal{L}} = \boldsymbol{\mathcal{E}} \cdot \delta \varphi + d \boldsymbol{\theta} (\varphi, \delta \varphi)
\label{eq:variedL}
\end {align}
This equation is commonly used to define the Euler-Lagrange equations of motion $\boldsymbol{\mathcal{E}} = 0$, but our main interest here is in the ``boundary term,'' which defines the symplectic potential $(D-1)$-form $\boldsymbol{\theta}$. It is clear from \eqref{eq:variedL} that $\boldsymbol{\theta}$ has the ambiguity 
\begin {align}
\boldsymbol{\theta} \to \boldsymbol{\theta} + d \boldsymbol{C} 
\label{thetaam}
\end {align}
where $\boldsymbol{C} (\varphi, \delta \varphi)$ is a $(D-2)$-form, sometimes referred to as a {\em corner term}.
We also have the freedom to change the Lagrangian by addition of an exact term (since this does not affect the equations of motion)
\begin {align}
\boldsymbol{\mathcal{L}} \to \boldsymbol{\mathcal{L}} + d \boldsymbol{\mu}
\label{lagfree}
\end {align}
in which case
\begin {align}
\boldsymbol{\theta} \to \boldsymbol{\theta} + \delta \boldsymbol{\mu}
\label{thetaam2}
\end {align}

The symplectic current, $\boldsymbol{\omega}$, is a $(D-1)$-form on spacetime defined by taking a second, antisymmetrized variation\footnote{Here, we assume that the variations defined by $\delta_1 \varphi$ and $\delta_2 \varphi$ commute. If they do not, the additional term $- \boldsymbol{\theta} (\varphi, [\delta_1 \varphi, \delta_2 \varphi])$ would have to be added to the right side of \eqref{symcur}, where $[\delta_1 \varphi, \delta_2 \varphi]$ denotes the commutator of $\delta_1 \varphi$ and $\delta_2 \varphi$.} of $\boldsymbol{\theta}$
\begin {align}
\boldsymbol{\omega}(\varphi; \delta_1 \varphi, \delta_2 \varphi) = \delta_1 \left[ \boldsymbol{\theta} (\varphi, \delta_2 \varphi) \right] - \delta_2 \left[ \boldsymbol{\theta} (\varphi, \delta_1 \varphi) \right]
\label{symcur}
\end {align}
The freedom to change $\boldsymbol{\theta}$ by \eqref{thetaam} gives rise to the freedom to modify $\boldsymbol{\omega}$ by 
\begin {align}
\boldsymbol{\omega} \to \boldsymbol{\omega} + d \overleftrightarrow{\boldsymbol{C}}  .
\label{omfree}
\end {align}
where we have introduced the notation
\begin {align}
\overleftrightarrow{\boldsymbol{C}} \equiv \delta_1 \left[ \boldsymbol{C} (\varphi, \delta_2 \varphi) \right] - \delta_2 \left[ \boldsymbol{C} (\varphi, \delta_1 \varphi) \right]
\end {align}
The freedom \eqref{thetaam2} does not affect $\boldsymbol{\omega}$.

We would like to define a symplectic structure, $\Omega$, on solutions by integrating $\boldsymbol{\omega}$ over a suitable hypersurface, $\Sigma$,
\begin {align}
\Omega (\varphi; \delta_1 \varphi, \delta_2 \varphi) = \int_\Sigma \boldsymbol{\omega}(\varphi; \delta_1 \varphi, \delta_2 \varphi)
\label{Omdef}
\end {align}
For a globally hyperbolic spacetime, we would take $\Sigma$ to be a Cauchy surface. In the case of interest here---i.e., AdS spacetime or, more generally, an FG spacetime---we take $\Sigma$ to be a complete, spacelike hypersurface that terminates on a cross-section of conformal infinity, $\mathcal I$. If $\Omega$ is well defined by \eqref{Omdef}, then we can view it as a two-form on field space, since it is bilinear and antisymmetric in $\delta_1 \varphi$ and $\delta_2 \varphi$. It is also exact (and, therefore, closed) as a 2-form in field space, since it was defined as a second antisymmetrized variation (i.e., a ``$d$'' operation in field space). Thus, after factoring by degeneracies of $\Omega$, \cite{LeeWald1990}, it will yield a symplectic form on the space of solutions, thereby providing the desired phase space structure. 

However, in order for $\Omega$ to be well-defined in the case of AdS or FG spacetimes, the following two properties must be satisfied:
\begin{enumerate}
\item \label{proper1} The integral in \eqref{Omdef} must be convergent.
\item \label{proper2} The pullback of the symplectic current $\boldsymbol{\omega}$ to conformal infinity, $\mathcal I$, must vanish. 
\end{enumerate}
The necessity of property \ref{proper1} for $\Omega$ to be well-defined is obvious. Property \ref{proper2} is necessary (and sufficient) for $\Omega$ to be independent of choice of $\Sigma$. 

Thus, our task in the following subsections is to compute $\boldsymbol{\omega}$ for the theories and boundary conditions under consideration and see if properties \ref{proper1} and \ref{proper2} hold. If they do hold for our initial choice of $\boldsymbol{\omega}$, then we are done. If not, we will then attempt to make use of the freedom \eqref{omfree} to find a ``corner term'' $\boldsymbol{C}$ such that the addition of $d \overleftrightarrow{\boldsymbol{C}}$ to $\boldsymbol{\omega}$ results in a symplectic current that satisfies properties 1 and 2.

We conclude this subsection with some remarks relating the criteria described above to criteria for a variational principle. For a theory in which a boundary, $B$, is present (such as $B = \mathcal{I}$ in our case), in most references the focus would be on having a variational principle for the action
\begin {align}
\mathcal{S} = \int \boldsymbol{\mathcal{L}}
\label{Slag}
\end {align}
such that $\mathcal S$ will be an extremum under variations of $\varphi$ that satisfy the boundary conditions on $B$ if and only if $\varphi$ satisfies the equations of motion. From \eqref{eq:variedL}, we see that when the equations of motion hold, we have
\begin {align}
\delta \mathcal{S} = \int_B \boldsymbol{\theta}
\end {align}
Thus, in order to have a ``good variational principle,'' one would want (i) the action integral \eqref{Slag} (over a finite time interval) to be finite for all admissible fields and (ii) the pullback, $\boldsymbol{\theta}|_B$, of $\boldsymbol{\theta}$ to $B$ to vanish for all admissible variations about solutions. The finiteness of the action integral will not be relevant to our considerations for the construction of phase space. However, the vanishing of $\boldsymbol{\theta}|_B$ is highly relevant, since it implies the vanishing of $\boldsymbol{\omega}|_B$ (i.e., our property \ref{proper2} above)---although it is a significantly stronger condition. 

If $\boldsymbol{\theta}|_B$ does not vanish, the usual procedure for obtaining a good variational principle is to try to find a boundary term to add to the action such that its variation cancels $\boldsymbol{\theta}|_B$ modulo an exact term. More precisely, suppose one can find a ``boundary Lagrangian'' $(D-1)$-form $\boldsymbol{\mathcal{L}}_B$ on $B$ such that $\boldsymbol{\theta}|_B$ agrees with $\delta \boldsymbol{\mathcal{L}}_B$ up to an exact term, i.e., on $B$ we have
\begin {align}
\boldsymbol{\theta}|_B = \delta \boldsymbol{\mathcal{L}}_B - d \boldsymbol{C}_B
\label{corner}
\end {align}
for some $(D-2)$-form $\boldsymbol{C}_B$ on $B$.
If we suitably extend $\boldsymbol{\mathcal{L}}_B$ to the spacetime, then the modified Lagrangian 
\begin {align}
\boldsymbol{\mathcal{L}}^{\rm mod} = \boldsymbol{\mathcal{L}} - d \boldsymbol{\mathcal{L}}_B
\end {align}
will have symplectic potential
\begin {align}
\boldsymbol{\theta}^{\rm mod} = \boldsymbol{\theta} - \delta \boldsymbol{\mathcal{L}}_B
\end {align}
When pulled back to the boundary, we have
\begin {align}
\boldsymbol{\theta}^{\rm mod}|_B = \boldsymbol{\theta}|_B - \delta \boldsymbol{\mathcal{L}}_B =  -d \boldsymbol{C}_B .
\label{modth}
\end {align}
If we assume that $B$ is compact without boundary (or if $\int_{\partial B} \boldsymbol{C}_B = 0$ by endpoint conditions), then $\int_B \boldsymbol{\theta}^{\rm mod} = 0$, as needed for a good variational principle. 

Now, suppose that one has succeeded in carrying out the above procedure of finding a boundary Lagrangian $\boldsymbol{\mathcal{L}}_B$ such that \eqref{corner} holds. Let $\boldsymbol{C}$ be a suitable extension of $\boldsymbol{C}_B$ to the spacetime. Then, using the freedom to redefine $\boldsymbol{\theta}$ via \eqref{thetaam}, we can take as our ``final'' choice of symplectic potential
\begin {align}
\boldsymbol{\theta}^{\rm fin} = \boldsymbol{\theta}^{\rm mod} + d \boldsymbol{C}
\end {align}
It is clear from \eqref{modth} that $\boldsymbol{\theta}^{\rm fin}|_B = 0$ and, hence, $\boldsymbol{\omega}^{\rm fin}|_B = 0$. Thus, if we can find a boundary Lagrangian such that \eqref{corner} holds, the associated ``corner term'' $\boldsymbol{C}_B$ in \eqref{corner} will give us the allowed modification to $\boldsymbol{\omega}$ that we need to make its flux through $B$ vanish. The ``final theory'' will then have
\begin {align}
\boldsymbol{\mathcal{L}}^{\rm fin} = \boldsymbol{\mathcal{L}}^{\rm mod} = \boldsymbol{\mathcal{L}} - d \boldsymbol{\mathcal{L}}_B
\label{Lfin}
\end {align}
\begin {align}
\boldsymbol{\theta}^{\rm fin} = \boldsymbol{\theta} - \delta \boldsymbol{\mathcal{L}}_B + d \boldsymbol{C}
\label{theta fin}
\end {align}
\begin {align}
\boldsymbol{\omega}^{\rm fin} = \boldsymbol{\omega} + d \overleftrightarrow{\boldsymbol{C}}
\label{omegafin}
\end {align}

A possible drawback of the above procedure is that it may not be possible to extend $\boldsymbol{\mathcal{L}}_B$ or $\boldsymbol{C}_B$ into the interior of the spacetime in a covariant way, in which case one would not have a globally well-defined covariant symplectic current $\boldsymbol{\omega}^{\rm fin}$. However, it turns out that when we follow this procedure in the following subsections, we will be led to a spacetime covariant $\boldsymbol{\mathcal{L}}^{\rm fin}$, $\boldsymbol{\theta}^{\rm fin}$, and $\boldsymbol{\omega}^{\rm fin}$.

\subsection {Phase space for the scalar field in AdS$_{1, d}$}
\label{sec:SCphase}

The construction of a phase space for the scalar field in AdS$_{1, d}$ for the general local boundary conditions \eqref{eq:gsBC} or \eqref{eq:bfsBC} can be carried out in an entirely straightforward manner, so it is useful to carry out this example before moving on to the electromagnetic, Yang-Mills, and gravitational cases.
The variation of the scalar field Lagrangian \eqref{eq:scalaract} (with the volume element $\epsilon_{\mu_1 \cdots \mu_{d+1}}$ inserted so as to make it a $(d+1)$-form) is
\begin {multline}
\delta \mathcal{L}_\phi = \delta \phi \big(\nabla^2 \phi  -  \mu^2 \phi \big) \epsilon_{\mu_1 \cdots \mu_{d+1}} \\
- d \big[ \delta \phi (\nabla^\nu \phi) \epsilon_{\nu \mu_1 \cdots \mu_{d}} \big]
\end {multline}
Thus, the symplectic potential $d$-form is
\begin {align}
\theta_{\mu_1 \cdots \mu_{d}}^{(\phi)} (\phi, \delta \phi) = - (\nabla^\nu \phi) (\delta \phi) \epsilon_{\nu \mu_1 \cdots \mu_{d}} \, 
\label{eq:theta4phi}
\end {align}
and the symplectic current $d$-form \eqref{symcur} is\footnote{Note that $\omega_{\mu_1 \cdots \mu_{d}}^{(\phi)}$ does not depend on the background $\phi$, as is the case for any linear theory.}
\begin {multline}
\omega_{\mu_1 \cdots \mu_{d}}^{(\phi)}  (\delta_1 \phi, \delta_2 \phi) \\
=   [\delta_1 \phi (\nabla^\nu \delta_2 \phi)  -  \delta_2 \phi (\nabla^\nu \delta_1 \phi)  ] \epsilon_{\nu \mu_1 \cdots \mu_{d}} \, . 
\label{eq:omega4phi}
\end {multline}

For $\nu \geq 1$, the required boundary condition is $a_\nu = 0$, and it is easily seen that the allowed solutions fall off sufficiently fast at infinity that properties \ref{proper1} and \ref{proper2} of subsection \ref{gencon} are satisfied. Thus, there is no difficulty in obtaining a phase space structure.

For $0 < \nu < 1$ with the boundary condition $b_\nu = q_\nu a_\nu$, it can be seen from the fall-off \eqref{eq:gsfall} that 
\begin {equation}
\omega_{r z^1 \cdots z^{d-1}}^{(\phi)} (\delta_1 \phi, \delta_2 \phi) = \mathcal{O} \bigg(\frac{1}{r^{3-2\nu}} \bigg) 
\label{eq:sfinite1}
\end {equation}
where $z^i$ denote the angular coordinates. Consequently, the $r$-integral in \eqref{Omdef} converges, and property \ref{proper1} of subsection \ref{gencon} holds. On the other hand, $\omega_{t z^1 \cdots z^{d-1}}^{(\phi)}$ contains individual terms that are $\mathcal{O}(r^{2\nu})$ and $\mathcal{O}(1)$. However, the $\mathcal{O}(r^{2\nu})$ terms involve products of $\delta_1 a_\nu$ and $\delta_2 a_\nu$ and vanish under anti-symmetrization. The $\mathcal{O}(1)$ terms are mixed terms in $\delta a_\nu$ and $\delta b_\nu$, but these also vanish under antisymmetrization when the same boundary condition $b_\nu = q_\nu a_\nu$ is imposed on both solutions. Thus, we find that the pullback of $\boldsymbol{\omega}^{(\phi)}$ to $\mathcal I$ vanishes,
\begin {align}
\boldsymbol{\omega}^{(\phi)} (\delta_1 \phi, \delta_2 \phi) \bigg|_{\mathcal{I}} =  0
\label{symfluxI1}
\end {align}
and property \ref{proper2} of subsection \ref{gencon} also is satisfied. Thus, there also is no difficulty in obtaining a phase space structure.

Finally, for $\nu = 0$ with the boundary condition $b_0 = q_0 a_0$ we find
\begin {align}
\omega_{r z^1 \cdots z^{d-1}}^{(\phi)}  (\delta_1 \phi, \delta_2 \phi) = \mathcal{O} \bigg(\frac{\text{log}^2 (r/L)}{r^3} \bigg)
\label{eq:sfinite2}
\end {align}
so property \ref{proper1} holds. Similarly, under antisymmetrization and the application of the same boundary condition to both solutions, we again find 
\begin {align}
\boldsymbol{\omega}^{(\phi)} (\delta_1 \phi, \delta_2 \phi) \bigg|_{\mathcal{I}} =  0
\label{symfluxI2}
\end {align}
so property \ref{proper2} of subsection \ref{gencon} holds and we have a well-defined phase space. 

We conclude this subsection with a remark on energy. For any choice of boundary conditions for the scalar field specified by $q_\nu$, the phase space structure together with the time translation symmetry of the AdS background automatically gives a well-defined, conserved
\textit{canonical energy} on solutions $\phi$ given by
\begin {align}
\mathcal{E} \ \equiv \ \Omega ( \phi, \mathcal{L}_t \phi) = \int_{\Sigma} \boldsymbol{\omega}^{(\phi)}  (\phi, \mathcal{L}_t \phi) 
\end {align}
where $\mathcal{L}_t$ denotes the Lie derivative with respect to the timelike Killing field $t^a$. This energy corresponds to the conserved energy \eqref{IWen} automatically provided by the IW approach \cite{IW2003}. However, the canonical energy current density ${\omega}^{(\phi)}_{\mu_1 \cdots \mu_{d}} (\phi, \mathcal{L}_t \phi)$ is {\em not} locally equal to the energy current density ${T^{\nu}}_\rho t^\rho \epsilon_{\nu \mu_1 \cdots \mu_{d}}$, where $T_{\mu \nu}$ is the stress-energy tensor of $\phi$. Consequently, although by \eqref{symfluxI1} or \eqref{symfluxI2} the canonical energy flux vanishes at each point of $\mathcal{I}$, the stress-energy flux need not vanish. This fact is undoubtedly related to the phenomena discussed in \cite{AIS1978} where the authors proposed boundary conditions where the energy carried by the stress-energy tensor is allowed to ``re-circulate" through the AdS timelike boundary and the energy is conserved only when a time average is taken.

\subsection {Phase space for Maxwell theory in AdS$_{1,3}$}
\label{sec:EMphase}

In this subsection, we provide a phase space structure for Maxwell theory in AdS$_{1,3}$ with the general AdS-invariant boundary conditions \eqref{eq:EMdualrot}. 

It will be convenient to use differential forms notation throughout this subsection, so the one-form vector potential will be denoted as $\boldsymbol{A}$ and the field strength 2-form will be denoted as $\boldsymbol{F} = d \boldsymbol{A}$. The Maxwell Lagrangian in this notation is
\begin {align}
\boldsymbol{ \mathcal{L}_{\text{M}} } = - \frac{1}{2} \boldsymbol{F} \wedge {}^*\boldsymbol{F}
\end {align}
where the Hodge dual, ${}^*\boldsymbol{F}$, of $\boldsymbol{F}$ is given by
\begin {align}
{}^*F_{\mu \nu} = \frac{1}{2} {\epsilon^{\lambda \rho}}_{\mu \nu} F_{\lambda \rho}
\end {align}
Variation of $\boldsymbol{ \mathcal{L}_{\text{M}} }$ yields
\begin {align}
\delta \boldsymbol{\mathcal{L}_{\text{M}}} = -  \delta \boldsymbol{A} \wedge d {}^*\boldsymbol{F}   - d \left[ \delta \boldsymbol{A} \wedge {}^*\boldsymbol{F} \right]
\end {align}
from which one can read off that Maxwell's equations are 
\begin {align}
d {}^*\boldsymbol{F} = 0
\end {align}
and that the symplectic potential $d$-form is 
\begin {align}
\boldsymbol{\theta}_{\text{M}} (\boldsymbol{A}, \delta \boldsymbol{A}) = - \delta \boldsymbol{A} \wedge {}^*\boldsymbol{F} .
\label{eq:EMtheta}
\end {align}
Taking an antisymmetrized second variation, we obtain the symplectic current
\begin {align}
\boldsymbol{\omega}_{\text{M}} (\delta_1 \boldsymbol{A}, \delta_2 \boldsymbol{A}) = \delta_1 \boldsymbol{A} \wedge \delta_2 {}^*\boldsymbol{F} - \delta_2 \boldsymbol{A} \wedge \delta_1 {}^*\boldsymbol{F} 
\label{eq:EMomega}
\end {align}

Maxwell theory in 4 spacetime dimensions is conformally invariant, and $\boldsymbol{A}$, $\boldsymbol{F}$, and ${}^*\boldsymbol{F}$ all have conformal weight zero, so the above formulas for $\boldsymbol{\theta}_{\text{M}}$ and $\boldsymbol{\omega}_{\text{M}}$ hold for the unphysical fields in the conformal spacetime as well. The asymptotic behavior of $\boldsymbol{F}$ prior to the imposition of any boundary conditions is determined by \eqref{abEMfall} and \eqref{eq:Ftrphi}-\eqref{eq:Fijphi}, which imply that $\boldsymbol{F}$ extends smoothly to $\mathcal I$. We require that the gauge of $\boldsymbol{A}$ is chosen so that it extends smoothly to $\mathcal I$ as well. In the unphysical spacetime $\Sigma$ is compact (with boundary at $\mathcal I$), so the integral in \eqref{Omdef} defining $\Omega$ is manifestly convergent. Thus, property \ref{proper1} of subsection \ref{gencon} holds for $\Omega$ defined using $\boldsymbol{\omega}_{\text{M}}$ for all boundary conditions. However, although the pullback of $\boldsymbol{\omega}_{\text{M}}$ to $\mathcal I$ is well-defined and finite, it is nonvanishing in general, so property \ref{proper2} of subsection \ref{gencon} does not hold in general. 

To investigate this further, we first consider the case $\alpha = \infty$, corresponding to the usual choice of boundary conditions $\tilde{B}_I|_\mathcal I = 0$. This condition can be rewritten as $F_{IJ}|_\mathcal I = 0$, i.e., the pullback of $\boldsymbol{F}$ to ${\mathcal{I}}$ vanishes. But, in this case, the pullback of $\boldsymbol{A}$ is pure gauge, so for the boundary conditions $\tilde{B}_I|_\mathcal I = 0$ we may impose the gauge condition that the pullback of $\boldsymbol{A}$ vanishes. With this gauge condition, it is clear from \eqref{eq:EMomega} that the pullback of $\boldsymbol{\omega}_{\text{M}}$ to $\mathcal I$ vanishes, so property \ref{proper2} holds for the boundary conditions $\tilde{B}_I|_\mathcal I = 0$, and there is no difficulty in the construction of phase space using the unmodified Maxwell symplectic current $\boldsymbol{\omega}_{\text{M}}$.

Since the boundary condition $\tilde{E}_I|_\mathcal I = 0$ (corresponding to $\alpha = 0$) can be re-expressed as the vanishing of the pullback of ${}^*\boldsymbol{F}$ to $\mathcal I$, it is clear from \eqref{eq:EMomega} that the pullback of $\boldsymbol{\omega}_{\text{M}}$ to $\mathcal I$ vanishes in this case, so property \ref{proper2} also holds using the unmodified Maxwell symplectic current when $\alpha = 0$. However, for $\alpha \neq 0, \infty$, the pullback of $\boldsymbol{\omega}_{\text{M}}$ to $\mathcal I$ will be nonvanishing, so, in order to satisfy property \ref{proper2}, we must modify $\boldsymbol{\theta}_{\text{M}}$ and $\boldsymbol{\omega}_{\text{M}}$ by a suitable ``corner term'' $\CT{}_\alpha$ as in \eqref{thetaam} and \eqref{omfree}. It is not difficult to see that a suitable such corner term is 
\begin {align}
\CT{}_\alpha  = \frac{\alpha}{2} \big( \boldsymbol{A} \wedge \delta \boldsymbol{A} \big) .
\label{eq:EMcorner}
\end {align}
With this corner term, the modified symplectic current given by \eqref{omfree} is
\begin {align}
\boldsymbol{\omega}_{\text{EM}} = & \boldsymbol{\omega}_{\text{M}} + d \overleftrightarrow{ \CT{}_{\alpha}} \nonumber \\
= & \delta_1 \boldsymbol{A} \wedge \delta_2 \left[ {}^*\boldsymbol{F} - \alpha \boldsymbol{F} \right] - \delta_2 \boldsymbol{A} \wedge \delta_1 \left[ {}^*\boldsymbol{F} - \alpha \boldsymbol{F} \right]
\end {align}
Since the boundary conditions $[ \tilde{E}_I + \alpha \tilde{B}_I]_{\mathcal{I}} = 0$ are equivalent to the vanishing of the pullback of $[ {}^*\boldsymbol{F} - \alpha \boldsymbol{F}]$ to $\mathcal{I}$, it is easily seen that property \ref{proper2} of subsection \ref{gencon} holds for $\boldsymbol{\omega}_{\text{EM}}$. Furthermore, the modification provided by the corner term \eqref{eq:EMcorner} does not affect property \ref{proper1}, which remains satisfied for $\boldsymbol{\omega}_{\text{EM}}$. Thus, taking $\boldsymbol{\omega}_{\text{EM}}$ as the symplectic current for Maxwell theory in AdS$_{1,3}$ with boundary conditions \eqref{eq:EMdualrot}, we are able to obtain a completely satisfactory phase space structure.

Although, as we have just seen, the corner term \eqref{eq:EMcorner} results in a symplectic current whose pullback to $\mathcal{I}$ vanishes, the modified symplectic potential is given by
\begin {align}
\boldsymbol{\theta}'_{\text{M}} = & \boldsymbol{\theta}_{\text{M}} + d \CT{}_\alpha \nonumber \\
= & - \delta \boldsymbol{A} \wedge \left[ {}^*\boldsymbol{F} - \alpha \boldsymbol{F} \right] - \frac{\alpha}{2}\left( \boldsymbol{A} \wedge \delta \boldsymbol{F} + \delta \boldsymbol{A} \wedge \boldsymbol{F} \right) . 
\label{modsp}
\end {align}
The pullback of $\boldsymbol{\theta}'_{\text{M}}$ to $\mathcal{I}$ does not vanish modulo exact terms, so we do not have a good variational principle in the sense discussed at the end of subsection \ref{gencon}.
It therefore is illuminating to follow the procedure of that subsection, as we now shall do. This will lead us to reformulate Maxwell theory with boundary conditions \eqref{eq:EMdualrot} as being naturally associated with a modified Lagrangian.

In order to implement the strategy discussed at the end of subsection \ref{gencon}, 
we seek to find $\boldsymbol{\mathcal{L}}_\mathcal{I}$ and $\boldsymbol{C}_\mathcal{I}$ on $\mathcal I$ so that $\boldsymbol{\theta}_{\text{M}}|_{\mathcal I}$ satisfies \eqref{corner}. By inspection, the following choices work:
\begin {align}
\boldsymbol{\mathcal{L}}_\mathcal{I} = - \frac{\alpha}{2} \boldsymbol{\mathcal{L}}_{\rm aCS} = - \frac{\alpha}{2} \boldsymbol{A} \wedge d\boldsymbol{A}
\end {align}
where $\boldsymbol{\mathcal{L}}_{\rm aCS} = \boldsymbol{A} \wedge d\boldsymbol{A}$ is the abelian Chern-Simons Lagrangian, together with
\begin {align}
\boldsymbol{C}_\mathcal{I} = \CT{}_\alpha  = \frac{\alpha}{2} \big( \boldsymbol{A} \wedge \delta \boldsymbol{A} \big) .
\end {align}
It is easily verified that 
\begin {align}
\boldsymbol{\theta}_{\text{M}} \big|_{\mathcal I} = \delta \boldsymbol{\mathcal{L}}_\mathcal{I} - d \boldsymbol{C}_\mathcal{I}.
\label{blagth}
\end {align}
as desired.  Note that 
\begin {align}
 d \boldsymbol{\mathcal{L}}_{\rm aCS} = \boldsymbol{F} \wedge \boldsymbol{F} =\boldsymbol{\mathcal{L}}_{\rm aP} 
\end {align}
where $\boldsymbol{\mathcal{L}}_{\rm aP}$ is the abelian Pontryagin Lagrangian. 

Both $\boldsymbol{\mathcal{L}}_\mathcal{I}$ and $\boldsymbol{C}_\mathcal{I}$ extend to the bulk in a natural way, so we get a ``final theory'' by the prescription \eqref{Lfin} - \eqref{omegafin}, namely
\begin {align}
\boldsymbol{\mathcal{L}}_{\rm EM} = & \boldsymbol{\mathcal{L}}_{\rm M} + 
\frac{\alpha}{2} \boldsymbol{\mathcal{L}}_{\rm aP} \nonumber \\
= & - \frac{1}{2} \boldsymbol{F} \wedge {}^*\boldsymbol{F} + \frac{\alpha}{2} \boldsymbol{F} \wedge \boldsymbol{F} \label{finemlag} \\
\boldsymbol{\theta}_{\text{EM}} = & \, \boldsymbol{\theta}_{\text{M}} - \delta \boldsymbol{\mathcal{L}}_\mathcal{I} + d \boldsymbol{C}_\mathcal{I} \nonumber \\
= & \, - \delta \boldsymbol{A} \wedge \left[ {}^*\boldsymbol{F} - \alpha \boldsymbol{F} \right] \label{omem} \\
\boldsymbol{\omega}_{\text{EM}} = & \boldsymbol{\omega}_{\text{M}} + d \overleftrightarrow{\boldsymbol{C}_\mathcal{I}} \nonumber \\
= & \delta_1 \boldsymbol{A} \wedge \delta_2 \left[ {}^*\boldsymbol{F} - \alpha \boldsymbol{F} \right] - \delta_2 \boldsymbol{A} \wedge \delta_1 \left[ {}^*\boldsymbol{F} - \alpha \boldsymbol{F} \right] \label{symem}
\end {align}
Note that although we did not impose any requirements on the action integral in our construction, it is readily verified that the action integral \eqref{Slag} (over a finite time interval) for the Lagrangian \eqref{finemlag} is finite, so we have a good variational principle for our final theory. Furthermore, all of the quantities \eqref{finemlag}-\eqref{symem} are covariant and Maxwell gauge invariant\footnote{In contrast, $\boldsymbol{\theta}'_{\text{M}}$ given by \eqref{modsp} is not gauge invariant on account of the presence of an unvaried $\boldsymbol{A}$ in one of the terms.}. Indeed, had we simply started with the Lagrangian \eqref{finemlag}, we would have obtained $\boldsymbol{\theta}_{\text{EM}}$ and $\boldsymbol{\omega}_{\text{EM}}$ as the natural covariant and gauge invariant symplectic potential and current associated with $\boldsymbol{\mathcal{L}}_{\rm EM}$. Thus, we may view the phase space of electromagnetism in AdS$_{1,3}$ with boundary conditions \eqref{eq:EMdualrot} as naturally arising from the Lagrangian \eqref{finemlag}.

\subsection {Phase space for Yang-Mills theory in AdS$_{1,3}$}
\label{sec:YMphase}

The construction of phase space for the Yang-Mills field in AdS$_{1,3}$ with boundary conditions \eqref{eq:YMdualrot} proceeds in exact parallel with the Maxwell case. In differential forms notation---and with the Lie algebra indices suppressed---the Yang-Mills Lagrangian is 
\begin {align}
\boldsymbol{ \mathcal{L}^{(0)}_{\text{YM}} } = - \frac{1}{2 g_{\rm YM}^2} \text{Tr} \left( \boldsymbol{F} \wedge {}^*\boldsymbol{F} \right)
\end {align}
with 
\begin {align}
\boldsymbol{F} = d\boldsymbol{A} + \boldsymbol{A} \wedge \boldsymbol{A}
\end {align}
Here, we put a ``$(0)$'' superscript on $\boldsymbol{ \mathcal{L}^{(0)}_{\text{YM}} }$ to indicate that this is the original, unmodified Yang-Mills Lagrangian.
The symplectic potential, $\boldsymbol{\theta}^{(0)}_{\text{YM}}$, and symplectic current, $\boldsymbol{\omega}^{(0)}_{\text{YM}}$, are given by exact analogs of \eqref{eq:EMtheta} and \eqref{eq:EMomega}. For the boundary condition $\tilde{B^a}_I|_{\mathcal{I}} = 0$ (i.e., the case $\alpha = \infty$), we can set $\boldsymbol{A}|_{\mathcal{I}} = 0$, and we find that properties \ref{proper1} and \ref{proper2} of subsection \ref{gencon} are satisfied. However, for finite non-zero $\alpha$, we again need to modify the symplectic potential by a corner term. Again, we could do this ``by hand,'' but it is more illuminating to follow the procedure discussed at the end of subsection \ref{gencon}.

In the Yang-Mills case, we can achieve a construction of phase space by taking $\boldsymbol{\mathcal{L}}_\mathcal{I}$ to be given by the non-abelian Chern-Simons Lagrangian
\begin {align}
\boldsymbol{\mathcal{L}}_\mathcal{I} = & - \frac{\alpha}{2 g_{\rm YM}^2} \boldsymbol{\mathcal{L}}_{\rm nCS} \nonumber \\
= & - \frac{\alpha}{2 g_{\rm YM}^2} \text{Tr} \bigg(\boldsymbol{A} \wedge d\boldsymbol{A} + \frac{2}{3} \boldsymbol{A} \wedge \boldsymbol{A} \wedge\boldsymbol{A}  \bigg)
\end {align}
and taking the corner term to be 
\begin {align}
\CT{}_{\mathcal I} = \frac{\alpha}{2 g_{\rm YM}^2} \text{Tr} \big( \boldsymbol{A} \wedge \delta \boldsymbol{A}  \big)
\label{eq:YMcorner}
\end {align}
Note that 
\begin {align}
 d \boldsymbol{\mathcal{L}}_{\rm nCS} = \text{Tr} \left(\boldsymbol{F} \wedge \boldsymbol{F} \right) = \boldsymbol{\mathcal{L}}_{\rm nP} 
\end {align}
where $\boldsymbol{\mathcal{L}}_{\rm nP}$ is the non-abelian Pontryagin Lagrangian. In exact parallel with the electromagnetic case, the final results are 
\begin {align}
\boldsymbol{\mathcal{L}}_{\rm YM} = & - \frac{1}{2g^2_{\rm YM}} \text{Tr} \left( \boldsymbol{F} \wedge {}^*\boldsymbol{F} \right) + \frac{\alpha}{2g^2_{\rm YM}} \text{Tr} \left( \boldsymbol{F} \wedge \boldsymbol{F} \right) \label{eq:YMLag}\\
\boldsymbol{\theta}_{\text{YM}} = & \,- \frac{1}{g^2_{\rm YM}}\text{Tr} \left[ \delta \boldsymbol{A} \wedge \left( {}^*\boldsymbol{F} - \alpha \boldsymbol{F} \right) \right] \\
\boldsymbol{\omega}_{\text{YM}} = & \frac{1}{g^2_{\rm YM}} \text{Tr} \big[ \delta_1 \boldsymbol{A} \wedge \delta_2 \left( {}^*\boldsymbol{F} -  \alpha \boldsymbol{F}  \right) \nonumber \\
& \quad \quad \quad \quad -  \delta_2 \boldsymbol{A} \wedge  \delta_1 \left( {}^*\boldsymbol{F} - \alpha \boldsymbol{F} \right) \big]
\end {align}
Again, the action integral \eqref{Slag} (over a finite time interval) for the Lagrangian \eqref{eq:YMLag} is finite---so we have a good variational principle---and all quantities are covariant and Yang-Mills gauge invariant. The phase space of Yang-Mills theory in AdS$_{1,3}$ with boundary conditions \eqref{eq:YMdualrot} can be viewed as naturally arising from the Lagrangian \eqref{eq:YMLag}.

\subsection {Phase space for non-linear gravity in 4-dimensional Fefferman-Graham spacetimes}
\label{sec:nonGRAphase}

The Einstein-Hilbert Lagrangian is
\begin {align}
\boldsymbol{\mathcal{L}}_{\rm EH} = \frac{1}{16 \pi G_{4}} (R [g] - 2 \Lambda) \boldsymbol{\epsilon}
\label{EHlagrang}
\end {align}
In 4 spacetime dimensions, the symplectic potential is
\begin {align}
(\theta_{\text{EH}})_{\mu \nu \rho} (g, \delta g) = \frac{1}{16 \pi G_4} \big( \nabla^\sigma {\delta g_\sigma}^\lambda - \nabla^\lambda \delta g \big) \epsilon_{\lambda \mu \nu \rho}
\label{speh}
\end {align}
where indices are raised and lowered with $g_{\mu \nu}$ and $\delta g = g^{\mu \nu} \delta g_{\mu \nu}$ on the right side of this equation. The Einstein-Hilbert symplectic current is
\begin {multline}
(\omega_{\text{EH}})_{\mu \nu \rho} = - \frac{1}{16 \pi G_4} P^{\lambda \alpha \beta \gamma  \kappa \sigma} \\
\bigg[ \big( \delta_1 g_{\alpha \beta } \nabla_\gamma \delta_2 g_{\kappa \sigma} \big) - \big( \delta_2 g_{\alpha \beta} \nabla_\gamma \delta_1 g_{\kappa \sigma} \big) \bigg]  \epsilon_{\lambda \mu \nu \rho}
\label{EHsym}
\end {multline}
where
\begin {multline}
P^{\lambda \alpha \beta \gamma \kappa \sigma}
=
g^{\lambda \kappa} g^{\sigma \alpha} g^{\beta \gamma}
-\frac{1}{2} g^{\lambda \gamma} g^{\alpha \kappa} g^{\sigma \beta}
-\frac{1}{2} g^{\lambda \alpha} g^{\beta \gamma} g^{\kappa \sigma}
\\ -\frac{1}{2} g^{\lambda \kappa} g^{\sigma \gamma} g^{\alpha \beta}
+\frac{1}{2} g^{\lambda \gamma} g^{\alpha \beta} g^{\kappa \sigma}.
\end {multline}

Our aim is to define a phase space structure on Fefferman-Graham (FG) spacetimes \eqref{fgform} subject to the boundary conditions \eqref{eq:NLGravdualrot}. We will proceed by the procedure described at the end of subsection \ref{gencon} and as carried out in the electromagnetic and Yang-Mills cases.

We begin with the case of the standard boundary conditions $\tilde{B}_{IJ}|_\mathcal I = 0$, i.e., $\alpha = \infty$. In the corresponding electromagnetic case, we could set $\boldsymbol{A} = 0$ on $\mathcal I$, in which case the Maxwell symplectic current itself satisfied $\boldsymbol{\omega}_{\rm M} |_{\mathcal I} = 0$, so no modification was needed. An analogous result holds in the gravitational case, as previously found by Papadimitriou and Skenderis \cite{PaSk2005}. When $\tilde{B}_{IJ}|_\mathcal I = 0$, the Cotton-York tensor of $g_{(0)}^{IJ}$ vanishes by \eqref{BCY}, so $g_{(0)}^{IJ}$ is conformally flat. We require the conformal class of $g_{(0)}^{IJ}$ to be fixed, i.e., we view boundary metrics of different conformal classes as defining different phase spaces. Then, in parallel with the discussion in the paragraph below \eqref{cydef} in subsection \ref{sec:NONLINGRAVBC}, by making use of the boundary diffeomorphism freedom and the freedom in defining the conformal factor $z$, we may require $g_{(0)IJ}$ itself to be fixed, so $\delta g_{(0)IJ} = 0$. It then follows that $\delta g_{(2)IJ} = 0$. It can then be seen that $\delta g_{\mu \nu}$ falls off sufficiently rapidly that $\boldsymbol{\theta}_{\rm EH} |_{\mathcal I} = 0$, so we can construct phase space with the unmodified Einstein-Hilbert symplectic potential\footnote{However, the action integral \eqref{Slag} with the unmodified Einstein-Hilbert Lagrangian will not converge, so addition of a boundary Lagrangian is still needed in order to obtain a good variational principle.}.

Next, we consider the boundary conditions $\tilde{E}_{IJ}|_\mathcal I = 0$, i.e., $\alpha = 0$. In the corresponding electromagnetic case, we had $\boldsymbol{\theta}_{\text{M}} |_{\mathcal{I}} = 0$, so we also did not need to modify the original Maxwell Lagrangian in this case. However, an analogous result does not hold in the gravitational case. In fact, since we cannot set $\delta g_{(0)IJ} = 0$, we have $\delta g_{\mu \nu} = O(1/z^2)$ and $\boldsymbol{\theta}_{\rm EH}$ itself will diverge as $O(1/z^3)$ at $\mathcal I$. Thus, we would need to modify $\boldsymbol{\mathcal{L}}_{\rm EH}$ by a boundary Lagrangian and corner terms. The Gibbons-Hawking-York boundary Lagrangian \eqref{eq:GHY} and the counterterm Lagrangian \eqref{eq:CT} introduced in \cite{BK1999} will cancel the divergent terms in $\boldsymbol{\theta}_{\rm EH}$, leaving a finite term proportional to $\tilde{E}_{IJ}$, modulo exact terms, so these can be used to provide a phase space construction. However, the resulting boundary Lagrangian and corner term do not have an obvious covariant extension into the bulk spacetime, so it is not obvious that one can obtain a covariant symplectic potential and symplectic current. Remarkably, however, as previously noted in \cite{ArArMisOlea2016, MisOlea2009}, the same effect as these boundary terms can be achieved by adding to $\boldsymbol{\mathcal{L}}_{\rm EH}$ a multiple of the Gauss-Bonnet Lagrangian
\begin {align}
\boldsymbol{\mathcal{L}}_{GB} & = \bigg(R_{\mu \nu \rho \sigma} R^{\mu \nu \rho \sigma} - 4 R_{\mu \nu} R^{\mu \nu} + R^2 \bigg) \boldsymbol{\epsilon}
\label{GBlagr}
\end {align}
Use of the Gauss-Bonnet Lagrangian in place of the Gibbons-Hawking-York and counterterm boundary Lagrangians will eliminate the need for an additional corner term and allow us to obtain a manifestly spacetime covariant symplectic potential and symplectic current.

To proceed with the construction of phase space for $\alpha = 0$, we note that the variation of the Gauss-Bonnet Lagrangian \eqref{GBlagr} yields\footnote{The pullback of $\boldsymbol{\theta}_{\text{GB}}$ to $\mathcal I$ is, up to normalization, equal to $\delta {\mathcal L}_{\rm GHY} + \delta {\mathcal L}_{\rm CT}$ modulo an exact (corner) term, where ${\mathcal L}_{\rm GHY}$ and ${\mathcal L}_{\rm CT}$ are the Gibbons-Hawking-York and counterterm Lagrangians, \eqref{eq:GHY} and \eqref{eq:CT}.}
\begin {align}
\delta \boldsymbol{\mathcal{L}}_{\rm GB} = d \boldsymbol{\theta}_{\text{GB}} 
\end {align}
Thus, the equations of motion vanish identically, corresponding to the fact that it is the integrand of a topological invariant. 
If the background spacetime about which we are varying satisfies Einstein's equation $R_{\mu \nu} = -3/L^2 g_{\mu \nu}$, the expression for $\boldsymbol{\theta}_{\text{GB}}$ simplifies to 
\begin {align}
(\theta_{\text{GB}})_{\mu \nu \rho} = & - \frac{4}{L^2} \big( \nabla^\sigma {\delta g_\sigma}^\lambda - \nabla^\lambda \delta g \big) \epsilon_{\lambda \mu \nu \rho} \nonumber \\
& + 4 \big( C^{\sigma \alpha \beta \tau} \nabla_\tau \delta g_{\alpha \beta } \big) \epsilon_{\sigma \mu \nu \rho}
\label{thetaGB}
\end {align}
Thus, if we define\footnote{We denote this Lagrangian by $\tilde{\boldsymbol{\mathcal{L}}}_{\rm G}$ rather than $\boldsymbol{\mathcal{L}}_{\rm G}$ to reserve the notation $\boldsymbol{\mathcal{L}}_{\rm G}$ for the final gravitational Lagrangian \eqref{lgfin}.}
\begin {align}
\tilde{\boldsymbol{\mathcal{L}}}_{\rm G} = \boldsymbol{\mathcal{L}}_{\rm EH} + \frac{L^2}{64 \pi G_4} \boldsymbol{\mathcal{L}}_{\rm GB}
\end {align}
then the contribution to the symplectic potential arising from the first term in \eqref{thetaGB} will cancel $\boldsymbol{\theta}_{\text{EH}}$, thereby eliminating the bad behavior of $\boldsymbol{\theta}_{\text{EH}}$ at infinity. Taking the limit to infinity of the second term in \eqref{thetaGB}, we find that the symplectic potential of $\tilde{\boldsymbol{\mathcal{L}}}_{\rm G}$ satisfies
\begin {align}
\tilde{\boldsymbol{\theta}}_{\text{G}} \big|_{\mathcal{I}} = - \frac{L}{16 \pi G_4}\tilde{E}_{IJ} \delta g_{(0)}^{IJ} \boldsymbol{\epsilon}_{(0)}
\label{tgspi}
\end {align}
where $\delta g^{IJ}_{(0)} \equiv g_{(0)}^{IK} g_{(0)}^{JL} \delta g_{(0) KL} $ and $\boldsymbol{\epsilon}_{(0)}$ is the volume element on $\mathcal{I}$ associated with the metric $g_{(0)IJ}$. Thus, for the Lagrangian $\tilde{\boldsymbol{\mathcal{L}}}_{\rm G}$ we have the desired property that $\tilde{\boldsymbol{\theta}}_{\text{G}}|_{\mathcal{I}} = 0$ for the boundary conditions $\tilde{E}_{IJ}|_\mathcal I = 0$. The corresponding symplectic current $\tilde{\boldsymbol{\omega}}_{\text{G}}$ automatically satisfies property \ref{proper2} of subsection \ref{gencon} and it also can be verified to satisfy property \ref{proper1}, so we obtain a well-defined phase space for the case $\alpha = 0$.

However, for the boundary conditions \eqref{eq:NLGravdualrot} with $\alpha \neq 0, \infty$, we have $\tilde{\boldsymbol{\theta}}_{\text{G}} |_{\mathcal{I}} \neq 0$ and a further modification of the Lagrangian is necessary. In analogy with the Maxwell and Yang-Mills cases, it might be expected that the necessary modification can be achieved with the addition of a gravitational Pontryagin term\footnote{The addition of a Pontryagin term was previously suggested in \cite{ArArMisOlea2016, MisOlea2009}.} to the Lagrangian. This turns out to be the case. The gravitational Pontryagin Lagrangian is
\begin {align}
\boldsymbol{\mathcal{L}}_P & = \text{Tr} (\boldsymbol{R} \wedge \boldsymbol{R}) =  \frac{1}{4} \epsilon^{\mu \nu \rho \sigma} R_{\mu \nu}{}^{\alpha \beta} R_{\rho \sigma \alpha \beta} \boldsymbol{\epsilon} 
\label{Plagr}
\end {align}
Its variation yields\footnote{ The pullback of $\boldsymbol{\theta}_{\text{P}}$ to $\mathcal I$ is, up to normalization, equal to the variation of the gravitational Chern-Simons Lagrangian \cite{Witten1988, JP2003} on $\mathcal I$, modulo an exact (corner) term.}
\begin {align}
\delta \boldsymbol{\mathcal{L}}_{\rm P} = d \boldsymbol{\theta}_{\text{P}}
\end {align}
where, making use of Einstein's equation for the background metric, we obtain
\begin {align} 
(\theta_{\rm P})_{\lambda \kappa \tau} = & - 6 \big(\nabla^\sigma \delta g^{\alpha}{}_{[\lambda} \big) C_{\kappa \tau] \alpha \sigma} \nonumber \\
= & 2 ({}^*C)^{\mu \nu \sigma \rho} (\nabla_\rho \delta g_{\nu \sigma}) \epsilon_{\mu \lambda \kappa \tau}
\label{sppon}
\end {align}
Taking the limit to infinity, we obtain
\begin {align}
\boldsymbol{\theta}_{\text{P}} \big|_{\mathcal{I}} = - \frac{2}{L} \tilde{B}_{IJ} \delta g_{(0)}^{IJ} \boldsymbol{\epsilon}_{(0)} 
\label{spponi}
\end {align}

Now define the new Lagrangian
\begin {align}
\boldsymbol{\mathcal{L}}_{\rm G} = & \tilde{\boldsymbol{\mathcal{L}}_{\rm G}} + \frac{\alpha L^2}{32 \pi G_4} \boldsymbol{\mathcal{L}}_P \nonumber \\
= & \boldsymbol{\mathcal{L}}_{\rm EH} + \frac{L^2}{64 \pi G_4} \boldsymbol{\mathcal{L}}_{\rm GB} + \frac{\alpha L^2}{32 \pi G_4} \boldsymbol{\mathcal{L}}_P .
\label{lgfin}
\end {align}
From \eqref{speh}, \eqref{thetaGB}, and \eqref{sppon}, we see that the symplectic potential of this Lagrangian is 
\begin {align}
(\theta_{\text{G}})_{\mu \nu \rho} = \frac{L^2}{16 \pi G_4} \bigg[ C^{\sigma \alpha \beta \tau} + \alpha ({}^*C)^{\sigma \alpha \beta \tau} \bigg] \big(\nabla_\tau \delta g_{\alpha \beta }  \big) \epsilon_{\sigma \mu \nu \rho} 
\label{thetaG}
\end {align}
From \eqref{tgspi} and \eqref{spponi}, we see that
\begin {align}
\boldsymbol{\theta}_{\text{G}} \big|_{\mathcal{I}} = -\frac{L}{16 \pi G_4} \left[\tilde{E}_{IJ} + \alpha \tilde{B}_{IJ} \right] \delta g_{(0)}^{IJ} \boldsymbol{\epsilon}_{(0)} 
\end {align}
Thus, for the boundary conditions $[\tilde{E}_{IJ} + \alpha \tilde{B}_{IJ}]|_{\mathcal{I}} = 0$, we have $\boldsymbol{\theta}_{\text{G}}|_{\mathcal{I}} = 0$. The corresponding symplectic current $\boldsymbol{\omega}_{\text{G}}$---obtained by taking a second, antisymmetrized variation of $\boldsymbol{\theta}_{\text{G}}$---automatically satisfies property \ref{proper2} of subsection \ref{gencon} and it also can be verified to satisfy property \ref{proper1}, so we obtain a well-defined phase space for all $\alpha$. The Lagrangian $\boldsymbol{\mathcal{L}}_{\rm G}$ and its symplectic potential $\boldsymbol{\theta}_{\text{G}}$ and symplectic current $\boldsymbol{\omega}_{\text{G}}$ are all spacetime covariant. Finally, it can be verified that for Fefferman-Graham metrics of the form\footnote{Here we assume the absence of the $g_{(1)IJ}$ term in the expansion \eqref{fgform} but we do not assume any other relations arising from Einstein's equation.} \eqref{fgform}, the action integral \eqref{Slag} (over a finite time interval) for the Lagrangian \eqref{lgfin} is finite. Thus, the Lagrangian \eqref{lgfin} also yields a well-defined variational principle for the boundary conditions $[\tilde{E}_{IJ} + \alpha \tilde{B}_{IJ}]|_{\mathcal{I}} = 0$.

We conclude this section by noting that there are some solutions in the literature satisfying the boundary conditions \eqref{eq:NLGravdualrot} with finite $\alpha$, although none that we are aware of that have a nonsingular boundary with topology $S^2 \times \mathbb{R}$. 
One family is the Siklos family of solutions \cite{SIKLOS1985, Podolsky1997} (also referred to as AdS pp-waves) given by the metric
\begin {align}
ds^2  = \frac{L^2}{z^2} \big[ - 2 du dv + H(u, x, z) du^2 + dx^2 + dz^2]
\end {align}
where $H$ satisfies
\begin {align}
\frac{\partial^2 H}{\partial z^2} - \frac{2}{z} \frac{\partial H}{\partial z} + \frac{\partial^2 H}{\partial x^2} = 0 .
\end {align}
Solutions in the Siklos class exist for every finite $\alpha$ but the conformal boundary has topology $\mathbb{R}^{1, 2}$. Another class of solutions is the NUT-charged exact solutions, \cite{AMR2005, CLPPPS2011} and generalizations of these solutions \cite{CLP2006, MPPPS2013, GMPPS2015}. There are planar and hyperbolic branches of these solutions with boundary topology $\mathbb{R}^{1, 2}$ and a spherical branch with boundary topology $S^2 \times \mathbb{R}$ but with a Misner string singularity present \footnote{The Schwarzschild-AdS$_{1, 3}$ black hole is a solution to the standard boundary condition $\tilde{B}_{IJ}=0$ (equivalently $\alpha=\infty$), but not a solution for any finite $\alpha$.}. Again, solutions of this type exist satisfying the boundary conditions \eqref{eq:NLGravdualrot} for all finite $\alpha$.

\section {Asymptotic symmetries and conserved charges}
\label{ASCC}

\subsection{General considerations on symmetries and charges}

In this section, we consider the asymptotic symmetries and conserved charges for Maxwell theory\footnote{The Yang-Mills case follows in exact parallel with the Maxwell case, so we will not discuss it explicitly here.} in AdS$_{1,3}$ with general boundary conditions \eqref{eq:EMdualrot} and nonlinear gravity in 4-dimensional FG spacetimes with general boundary conditions \eqref{eq:NLGravdualrot}. Since we have defined a phase space structure in these cases, we have a precise notion of asymptotic symmetries in both cases. 

\subsubsection{Electromagnetism}

In electromagnetism, we have the notion of an infinitesimal gauge transformation, 
\begin {align}
\delta_\chi \boldsymbol{A} = d \chi
\end {align}
For any gauge invariant Lagrangian, $\boldsymbol{\mathcal{L}}$, for the electromagnetic field, we have $\delta_\chi \boldsymbol{\mathcal{L}} = 0$, so these infinitesimal gauge transformations are infinitesimal local symmetries. The {\em Noether current}, $\boldsymbol{J}_\chi$, associated with this gauge transformation is defined by
\begin {align}
\boldsymbol{J}_\chi = \boldsymbol{\theta} (\boldsymbol{A}, \delta_\chi \boldsymbol{A})
\end {align}
It follows that when the equations of motion hold for $\boldsymbol{A}$, we have $d \boldsymbol{J_\chi} = 0$. Since this holds for arbitrary $\chi$, it follows that \cite{Wald1990} $\boldsymbol{J}_\chi = d \boldsymbol{Q}_\chi$ for some $\boldsymbol{Q}_\chi$, called the {\em Noether charge form}, which is locally constructed from $\boldsymbol{A}$ and $\chi$. Assuming that $\boldsymbol{\theta}$ has been constructed in a gauge invariant way \cite{Prabhu2015}, it then follows that the symplectic current associated with $\delta_\chi \boldsymbol{A}$ and an arbitrary perturbation $\delta \boldsymbol{A}$ satisfying the linearized equations is 
\begin {align}
\boldsymbol{\omega} (\delta \boldsymbol{A}, \delta_\chi \boldsymbol{A}) = & \delta \boldsymbol{\theta}(\boldsymbol{A}, \delta_\chi \boldsymbol{A}) - \delta_\chi \boldsymbol{\theta}(\boldsymbol{A}, \delta \boldsymbol{A}) \nonumber \\
= & \delta \boldsymbol{J}_\chi = d \delta \boldsymbol{Q}_\chi
\label{funidem}
\end {align}
where the second equality follows from the definition of $\boldsymbol{J}_\chi$ and the gauge invariance of $\boldsymbol{\theta}$.

If $\chi$ is such that the symplectic product
\begin {align}
\Omega(\delta \boldsymbol{A}, d \chi) = \int_\Sigma \boldsymbol{\omega} (\delta \boldsymbol{A}, d \chi)
\end {align}
vanishes for all $\delta \boldsymbol{A}$, then the transformation generated by $\chi$ is said to be {\em gauge}. However, if $\Omega(\delta \boldsymbol{A}, d \chi)$ is nonvanishing for some $\delta \boldsymbol{A}$, then it is said to be a {\em symmetry}. It is clear from \eqref{funidem} that any $\chi$ that is of compact support or vanishes sufficiently rapidly at infinity is gauge, so only  $\chi$ that are suitably nonvanishing at infinity can be a symmetry. 

The charge, $\mathscr{Q}_\chi$ associated with $\chi$ is given by
\begin {align}
\mathscr{Q}_\chi = \int_C \boldsymbol{Q}_\chi
\end {align}
where $C = \Sigma \cap \mathcal{I}$. We assume that $C$ is compact, so $\mathscr{Q}_\chi$ is well defined. Since the pullback of $\boldsymbol{\omega}$ to $\mathcal I$ vanishes by our general requirement of property \ref{proper2} of subsection \ref{gencon}, it is clear from \eqref{funidem} that the pullback of $d \delta \boldsymbol{Q}_\chi$ to infinity vanishes, from which it follows that $\mathscr{Q}_\chi$ is conserved, i.e., independent of cross-section $C$ of $\mathcal{I}$. Finally, we have
\begin {align}
\delta \mathscr{Q}_\chi = \int_C \delta \boldsymbol{Q}_\chi = \int_\Sigma d \delta \boldsymbol{Q}_\chi = \Omega(\delta \boldsymbol{A}, d \chi)
\label{nchvar}
\end {align}
from which it is clear that $\mathscr{Q}_\chi$ is nontrivial (i.e., it is not constant on phase space) if and only if $\chi$ is a symmetry.

\subsubsection{Gravity}

The corresponding results in the gravitational case were given in \cite{IyerWald1994} and many subsequent references. For a diffeomorphism covariant Lagrangian, the infinitesimal local symmetries are generated by vector fields $\xi^\mu$. The Noether current is 
\begin {align}
\boldsymbol{J}_\xi = \boldsymbol{\theta} (g,\mathcal{L}_\xi g ) - \xi \cdot \boldsymbol{\mathcal L}
\end {align}
and the Noether charge is again given by $\boldsymbol{J}_\xi = d \boldsymbol{Q}_\xi$. Assuming that $\boldsymbol{\theta}$ has been covariantly defined, the relation corresponding to \eqref{funidem} is 
\begin {align}
\boldsymbol{\omega} (g; \delta g, \mathcal{L}_\xi g) = d \left[ \delta \boldsymbol{Q}_\xi - \xi \cdot \boldsymbol{\theta}  \right]
\label{funidgrav}
\end {align}
where it is assumed that $\xi^\mu$ remains fixed under all metric variations.
Again, if $\Omega (g; \delta g, \mathcal{L}_\xi g)$ vanishes for all $\delta g$, then  the transformation generated by $\xi^\mu$ is said to be {\em gauge}, but if $\Omega (g; \delta g, \mathcal{L}_\xi g)$ is nonvanishing for some $\delta g$, then it is said to be a {\em symmetry}. 

A function $H_\xi$ on phase space is said to be the {\em Hamiltonian charge} associated to $\xi^\mu$ if it satisfies
\begin {align}
\delta H_\xi = \Omega (g; \delta g, \mathcal{L}_\xi g) = \int_\Sigma \boldsymbol{\omega} (\delta g, \mathcal{L}_\xi g)
\end {align}
whenever $g$ satisfies the equations of motion. Clearly, $H_\xi$ is nontrivial if and only if $\xi^\mu$ is a symmetry. By \eqref{funidgrav}, we have 
\begin {align}
\delta H_\xi = \int_C \left[ \delta \boldsymbol{Q}_\xi - \xi \cdot \boldsymbol{\theta}  \right]
\label{varhxi}
\end {align}
For the particular gravitational phase spaces we have constructed for the boundary condition \eqref{eq:NLGravdualrot} for each value of $\alpha$ (including $\alpha = \infty$), we have $\boldsymbol{\theta} |_{\mathcal I} = 0$ in the FG metric setting\footnote{In more general circumstances, such as asymptotically flat spacetimes, $\boldsymbol{\theta}$ does not vanish at infinity and the term involving $\boldsymbol{\theta}$ would contribute to $H_\xi$. See \cite{WaldZou1999} for further discussion.}, so we have 
\begin {align}
\delta H_\xi = \int_C \delta \boldsymbol{Q}_\xi
\label{dhxiads}
\end {align}
However, we cannot automatically ``remove the $\delta$'s'' from this equation because the expression for $\boldsymbol{Q}_\xi$ itself may diverge at $\mathcal I$. In fact, we will find this to be the case for the boundary conditions with $\alpha = \infty$, where $\boldsymbol{Q}_\xi$ is given by the Einstein-Hilbert expression. For $\alpha = \infty$, we can proceed by defining $H_\xi = 0$ for exact AdS spacetime and using \eqref{dhxiads} to define $H_\xi$ on all other spacetimes in the $\alpha = \infty$ phase space. Thus, for $\alpha = \infty$, we have
\begin {align}
H_\xi = \int_C \left[ \boldsymbol{Q}_\xi - \boldsymbol{Q}^0_\xi \right]
\label{hxiads0}
\end {align}
where $\boldsymbol{Q}^0_\xi$ is the Noether charge expression in exact AdS. On the other hand, for $\alpha \neq \infty$, we will find that $\boldsymbol{Q}_\xi$ is given by a modified expression that is well defined at $\mathcal I$ and we have for $\alpha \neq \infty$
\begin {align}
H_\xi = \int_C \boldsymbol{Q}_\xi 
\label{hxiads}
\end {align}
As in the electromagnetic case, since the pullback of $\boldsymbol{\omega}$ to ${\mathcal{I}}$ vanishes, it follows that $H_\xi$ is conserved, i.e., independent of choice of cross-section $C$ of $\mathcal{I}$.

\subsection {Symmetries and charges for Maxwell theory in AdS$_{1,3}$}

We now apply the general results of the previous section to determine the symmetries and charges for Maxwell fields in AdS$_{1,3}$ for all values of $\alpha$ in the boundary conditions \eqref{eq:EMdualrot}.

Consider, first, the case of the standard boundary conditions, $\alpha = \infty$, corresponding to $\tilde{B}_I |_{\mathcal{I}} = 0$. In this case, the Lagrangian is the standard Maxwell Lagrangian, and the symplectic potential is given by \eqref{eq:EMtheta}. Thus, the Noether current is
\begin {align}
\boldsymbol{J}_\chi = - (d \chi) \wedge {}^*\boldsymbol{F} .
\label{nocurem}
\end {align}
and the Noether charge 2-form is
\begin {align}
\boldsymbol{Q}_\chi = - \chi \wedge {}^*\boldsymbol{F} .
\label{noQem}
\end {align}
Therefore, in order to have a non-vanishing charge, we must have $\chi |_{\mathcal I} \neq 0$. However,
as discussed in subsection \ref{sec:EMphase}, in order to have a well-defined phase space for $\alpha = \infty$, we must impose the gauge condition $\boldsymbol{A}|_{\mathcal{I}} = 0$. Therefore, the only allowed gauge transformations that are nonvanishing at infinity 
must have $\chi|_{\mathcal I} = {\rm const}$. This is the only asymptotic symmetry, and the corresponding conserved charge $\mathscr Q$ is just the ordinary total electric charge $Q_e$. For a globally regular, source-free solution to Maxwell's equations, we have $Q_e = 0$.

Now, consider the general boundary conditions
\begin {align}
\big[ \tilde{E}_I + \alpha \tilde{B}_I \big]_{\mathcal{I}} = 0 
\label{embc}
\end {align}
for any $\alpha \neq \infty$. It is worth noting first that integration of the time component of this boundary condition over a cross-section $C$ at infinity yields
\begin {align}
Q_e + \alpha Q_m = 0
\label{eq:QeQm}
\end {align}
where $Q_m$ is the total magnetic charge. For a solution to Maxwell's equations with a globally regular $\boldsymbol{A}$, we have $Q_m = 0$, and for a source-free solution we have $Q_e = 0$, but if Dirac strings and charged sources were allowed, \eqref{eq:QeQm} would place a restriction on $Q_e$ and $Q_m$. 

For the boundary conditions \eqref{embc}, the symplectic potential is 
\begin {align}
\boldsymbol{\theta} = - \delta \boldsymbol{A} \wedge \left[ {}^*\boldsymbol{F} - \alpha \boldsymbol{F} \right]  
\end {align}
(see \eqref{omem}). The Noether current and Noether charge are thus
\begin {align}
\boldsymbol{J}_\chi = - (d \chi) \wedge \left[ {}^*\boldsymbol{F} - \alpha \boldsymbol{F} \right]
\label{nocuremalp}
\end {align}
and 
\begin {align}
\boldsymbol{Q}_\chi = - \chi \left[ {}^*\boldsymbol{F} - \alpha \boldsymbol{F} \right] .
\label{noQmalp}
\end {align}
For the phase space with boundary conditions \eqref{embc}, no gauge choice is needed at ${\mathcal I}$, so gauge transformations with $\chi$ nonvanishing at ${\mathcal I}$ are allowed. Nevertheless, it follows immediately from \eqref{noQmalp} that under the boundary conditions \eqref{embc} we have $\boldsymbol{Q}_\chi|_{\mathcal I} = 0$ and, hence, $\mathscr{Q}_\chi = 0$ for all $\chi$. After all, we see that for the boundary conditions \eqref{embc}, there are no asymptotic gauge symmetries, i.e., all transformations generated by $\chi$ are pure gauge, including the transformations where $\chi$ is nonvanishing at infinity.

In fact the vanishing of $\mathscr{Q}_\chi$ for all $\chi$ can be shown from a general argument not requiring us to obtain an explicit formula for $\boldsymbol{Q}_\chi$. From \eqref{funidem} together with the phase space requirement that the pullback of $\boldsymbol{\omega}$ to ${\mathcal I}$ vanish, we see that the pullback of $d \delta \boldsymbol{Q}_\chi$ to ${\mathcal I}$ vanishes. Thus, $\delta \boldsymbol{Q}_\chi$ is a closed form on ${\mathcal I}$. But, the fact that it is closed for all $\chi$ implies \cite{Wald1990} that it is exact, $\delta \boldsymbol{Q}_\chi = d \boldsymbol{\sigma}$, where $\boldsymbol{\sigma}$ is a one-form locally constructed out of $\chi$ and $\boldsymbol{A}$. However, \eqref{nchvar} then immediately implies that $\delta \mathscr{Q}_\chi = 0$, which, in turn, implies that $\mathscr{Q}_\chi $ is a constant for all $\chi$.

\subsection {Symmetries and charges for nonlinear gravity in 4-dimensional FG spacetimes}

In parallel with the electromagnetic case, we first consider the case $\alpha = \infty$, corresponding to $\tilde{B}_{IJ} |_{\mathcal{I}} = 0$. We have the unmodified Einstein-Hilbert Lagrangian in this case, for which the Noether charge is 
\begin {align}
({Q}_\xi)_{\mu \nu} = - \frac{1}{16 \pi G_4} \epsilon_{\mu \nu \lambda \rho} \nabla^\lambda \xi^\rho .
\label{nceh}
\end {align}
As shown in \cite{HIM2005}, \eqref{hxiads0} in the FG setting evaluates to 
\begin {align}
H_\xi = - \frac{L}{8 \pi G_4} \int_{C} \tilde{E}_{IJ} \xi^I \tilde{\eta}^J d\tilde{S}
\label{eq:ADM1}
\end {align}
where $\tilde{\eta}^J$ is the unit normal to $C$ and $d \tilde{S}$ is the induced volume element on $C$ with respect to the unphysical boundary metric $g_{(0)}^{IJ}$.
Thus $\xi^\mu$ must be nonvanishing on $\mathcal I$ in order to have a nonvanishing Hamiltonian charge. However, in order to define the phase space for $\alpha = \infty$, we had to require $g_{(0)}^{IJ}$ to be fixed. This is compatible with a nonvanishing $\xi^\mu$ on $\mathcal I$ if and only if $\xi^\mu$ is a conformal Killing field of $g_{(0)}^{IJ}$. Thus, the asymptotic symmetries in this case are precisely the conformal Killing fields of the boundary metric $g_{(0)}^{IJ}$.

For $\alpha \neq \infty$, the Lagrangian is given by \eqref{lgfin}. For theories in $D$ dimensions with a Lagrangian of the form of a function of curvature but not derivatives of curvature, i.e.,
\begin{align}
\boldsymbol{\mathcal L}  ={}& f(g_{\mu \nu}, R_{\mu \nu \lambda \rho}) \boldsymbol{\epsilon}
\end{align} 
the Noether charge $(D-2)$-form is \cite{AzCoOgTaTe2009}
\begin{multline}
    Q_{\mu_1 \cdots \mu_{D-2}} ={}  \epsilon_{\nu \sigma \mu_1 \cdots \mu_{D-2}}  \Big[-\frac{\partial f}{\partial R_{\nu \sigma \lambda \rho}} \nabla_{[\lambda} \xi_{\rho]} \\
    - 2 \nabla_\rho \Big(\frac{\partial f}{\partial R_{\nu \sigma \lambda \rho}}\Big) \xi_\lambda \Big].
\end{multline}
modulo the inherent ambiguity of adding an exact form to $\boldsymbol{Q}$.
For the Lagrangian \eqref{lgfin} with $g_{\mu \nu}$ satisfying Einstein's equation, $R_{\mu \nu} = -3/L^2 g_{\mu \nu}$, this evaluates to
\begin{align}
(Q_{\rm G})_{\mu \nu} = - \frac{L^2}{32 \pi G_4} \epsilon_{\mu \nu \lambda \rho} \bigg[C^{\lambda \rho \alpha \beta} + \alpha ({}^*C)^{\lambda \rho \alpha \beta} \bigg] \nabla_\alpha \xi_\beta
\label{ncalpha}
\end{align} 
where we have used the freedom in defining $\boldsymbol{Q}_{\rm G}$ to discard a term proportional to $d \boldsymbol{\xi}$. Taking the limit to $\mathcal I$, we obtain 
\begin{align}
(Q_{\rm G})_{IJ} = -\frac{L}{8 \pi G_4} (\epsilon_{(0)})_{IJ}{}^K \big(\tilde{E}_{K L} + \alpha \tilde{B}_{KL} \big) \xi^L
\end{align}
Thus, we see that $\boldsymbol{Q}_{\rm G}|_{\mathcal I} = 0$, which implies
\begin{align}
H_\xi = \int_C \boldsymbol{Q}_{\rm G} = 0
\end{align}
Therefore, for the boundary conditions \eqref{eq:NLGravdualrot} with $\alpha \neq \infty$, there are no asymptotic symmetries, i.e., all boundary-condition-preserving diffeomorphisms are gauge, including those that are nonvanishing at $\mathcal I$.

In fact, in close parallel with the electromagnetic case, the vanishing of $H_\xi$ follows directly from the vanishing of the pullbacks of $\boldsymbol{\theta}$ and $\boldsymbol{\omega}$ to ${\mathcal I}$ together with the fact that---unlike the case where $\alpha = \infty$---the vector field $\xi^\mu$ is unrestricted. By \eqref{funidgrav} and the vanishing of the pullbacks of $\boldsymbol{\theta}$ and $\boldsymbol{\omega}$, we see that $ \delta \boldsymbol{Q}_\xi$ is a closed form on $\mathcal I$. Since it is closed for all $\xi^\mu$, it follows \cite{Wald1990} that it is exact, in which case it follows immediately that $\delta H_\xi$ vanishes.  

\subsection{Linearization instability and black hole entropy}

The considerations of the previous subsection have important implications for linearization instability and black hole entropy in FG spacetimes with boundary conditions \eqref{eq:NLGravdualrot} with $\alpha \neq \infty$.

\subsubsection{Linearization instability}

Given a boundary condition in \eqref{eq:NLGravdualrot} with $\alpha \neq \infty$, the fact that all boundary-condition-preserving diffeomorphisms are gauge and all Hamiltonian charges vanish is reminiscent of what occurs in a closed universe, i.e., for a spacetime with a compact Cauchy surface. For a closed universe with a Killing field, it is well known from the work of Fischer and Marsden \cite{FM1979} and Moncrief \cite{Mon1975} that there is a ``linearization instability'' in the sense that there exist ``spurious" linearized perturbations such that they do not correspond to the linearization of a one-parameter family of exact solutions to the Einstein equation. 

The reason for this phenomenon---and its applicability to the present case of FG spacetimes with boundary conditions \eqref{eq:NLGravdualrot} with $\alpha \neq \infty$---can be seen as follows. The second order variation of Hamiltonian charge, $\delta^2 H_\xi$, about a background solution, $g$, can be obtained by taking a second variation of \eqref{varhxi}. For the case of a general vector field $\xi^\mu$, this does not produce an interesting relation. However, if $\xi^\mu$ is a Killing field of $g$, then since $\mathcal{L}_\xi g = 0$, we obtain
\begin {align}
\delta^2 H_\xi = \int_\Sigma \boldsymbol{\omega} (g; \delta g, \mathcal{L}_\xi \delta g)
\end {align}
In the case of an asymptotically flat spacetime---where the Killing field $\xi^\mu$ would be an asymptotic symmetry---this formula yields a useful expression for the second order variation of the Hamiltonian charge $H_\xi$ in terms of the first order perturbation $\delta g$. However, if $\xi^\mu$ is gauge---as occurs for a closed universe or for the phase space of the FG spacetimes with non-standard boundary conditions---then $H_\xi$ vanishes for all solutions, and we obtain
\begin {align}
\int_\Sigma \boldsymbol{\omega} (g; \delta g, \mathcal{L}_\xi \delta g) = 0 .
\label{linin}
\end {align}
This yields a constraint on the first order perturbations that is not automatically satisfied by solutions to the linearized equations. Solutions to the linearized equations that do not satisfy \eqref{linin} cannot correspond to exact solutions. This is the linearization instability phenomenon.

If the background metric has $k$ independent Killing fields, then \eqref{linin} yields a $k$-parameter family of constraints that are quadratic in the perturbation $\delta g$. Generically, the solutions to these constraints would be expected to comprise a conical space of co-dimension $k$ in the space of all linearized solutions. However, the situation for AdS$_{1,3}$ for the boundary conditions \eqref{eq:NLGravdualrot} for all $\alpha \neq \infty$ is much more severe, since the constraint \eqref{linin} for the timelike Killing field $\xi^\mu = t^\mu = (\partial/\partial t)^\mu$ by itself precludes {\em all} nonsingular, non-gauge linearized solutions. This is a consequence of the fact that the canonical energy (i.e., the left side of \eqref{linin}) for linearized perturbations of AdS$_{1,3}$ is essentially the IW energy, which is positive definite. More precisely, for a linearized perturbation with angular dependence given by $l,m$ spherical harmonics, we have
\begin {align}
{\mathcal E}_t \equiv & \int_\Sigma \boldsymbol{\omega} (g; \delta g, \mathcal{L}_t \delta g) \nonumber \\
= & \frac{(l-1)l(l+1)(l+2)}{16 \pi G_4}  \left[ \frac{1}{4} E_{IW} (\Phi_{S{\textbf{k}}}^{\rm G}) + E_{IW} (\Phi_{V{\textbf{k}}}^{\rm G}) \right]
\end {align}
where $\Phi_{V{\textbf{k}}}^{\rm G}$ and $\Phi_{S{\textbf{k}}}^{\rm G}$ are the potentials given by \eqref{phivg} and \eqref{phisg} (suitably duality rotated to correspond to the linear combinations satisfying independent boundary conditions) and the IW energy $E_{IW}$ is given by \eqref{IWen} with $A$ given by \eqref{eq:Awaveeq}. Since $E_{IW}$ is positive definite, the vanishing of ${\mathcal E}_t$ implies the vanishing of $\Phi_{V{\textbf{k}}}^{\rm G}$ and $\Phi_{S{\textbf{k}}}^{\rm G}$ for all $l \geq 2$, which, in turn, implies that the perturbation is pure gauge for $l \geq 2$. But the only non-pure-gauge $l = 0$ perturbation satisfying the $\alpha \neq \infty$ boundary conditions is the linearized Taub-NUT-AdS solution, which is singular at the center and contains a Misner string. There are no nonsingular, non-gauge $l=1$ linearized perturbations that satisfy the $\alpha \neq \infty$ boundary conditions. Thus, all nontrivial, nonsingular linearized perturbations of AdS$_{1,3}$ for the boundary conditions \eqref{eq:NLGravdualrot} fail to satisfy \eqref{linin}. Consequently, for $\alpha \neq \infty$, all such perturbations are spurious, i.e., they do not correspond to the linearization of a one-parameter family of exact solutions. Thus, AdS$_{1,3}$ exhibits an extreme form of linearization instability.

\subsubsection{Black hole entropy}

Although the Lagrangian \eqref{lgfin} yields the same equations of motion as the Einstein-Hilbert Lagrangian, it gives rise to a different symplectic structure and, as we have seen in the previous subsection, it gives rise to the formula \eqref{ncalpha} for Noether charge that is quite different from the formula \eqref{nceh} arising from the unmodified Einstein-Hilbert Lagrangian. Black hole entropy is directly related to Noether charge, so the entropy of a stationary black hole in an FG spacetime with boundary conditions \eqref{eq:NLGravdualrot} with $\alpha \neq \infty$ will differ from the usual Bekenstein-Hawking formula. From the formula \eqref{ncalpha} for Noether charge, we see that for $\alpha \neq \infty$, the entropy of a stationary black hole with nondegenerate Killing horizon is given by \cite{IyerWald1994}
\begin{align}
S = - \frac{L^2}{16 G_4} \int_\mathcal{B} \big[C_{\lambda \rho \alpha \beta} + \alpha ({}^*C)_{\lambda \rho \alpha \beta} \big] n^{\lambda \rho} n^{\alpha \beta} dA
\label{sbhalpha}
\end{align} 
where $\mathcal B$ is the bifurcation surface of the black hole (which we assume to be smooth and compact) and $n^{\mu \nu}$ is the binormal to $\mathcal{B}$, normalized so that $n^{\mu \nu} n_{\mu \nu} = -2$. However, using Einstein's equation and the fact that both extrinsic curvatures of $\mathcal B$ vanish, it follows from the Gauss relation that
\begin {align}
C^{\mu \nu \rho \sigma} n_{\mu \nu} n_{\rho \sigma} = - 2 R^{(2)} - \frac{4}{L^2} .
\end {align}
Furthermore, when the extrinsic curvatures of $\mathcal B$ vanish, we have
\begin{align}
\int_{\mathcal B} {}^*C_{\lambda \rho \alpha \beta} n^{\lambda \rho} n^{\alpha \beta} dA = 0
\label{ponflux}
\end{align} 
Thus, the entropy formula \eqref{sbhalpha} becomes simply
\begin {align}
S = \frac{A_{\mathcal B}}{4 G_4} + \frac{\pi L^2}{2 G_4} \chi(\mathcal{B})
\label{sbhalpha2}
\end {align}
where $A_{\mathcal B}$ is the area of $\mathcal B$ and $\chi(\mathcal{B})$ is its Euler characteristic. It is worth pointing out that, although the initial formula \eqref{sbhalpha} for black hole entropy had a very different appearance from the Bekenstein-Hawking formula, our final formula \eqref{sbhalpha2} differs from the Bekenstein-Hawking formula by only a constant.

All of the above discussion applies to the case where the spacetime metric is smooth. Moreover, even if the spacetime has Misner strings, our formula for the Noether charge $\boldsymbol{Q}_{\rm G}$ continues to make sense and still yields $\boldsymbol{Q}_{\rm G}|_{\mathcal I} = 0$, so we still have $H_\xi = 0$. However, the integral \eqref{ponflux} now no longer vanishes and is equal to the ``normal-bundle flux'' at the horizon. For example, for the case of the 4-dimensional Taub-NUT-AdS solution with the metric form given in eq.(1)-(2) of \cite{HeKuMa2019}, we obtain 
\begin{align}
\int_{\mathcal B} {}^*C_{\lambda \rho \alpha \beta} n^{\lambda \rho} n^{\alpha \beta} dA = 32 \pi \kappa n
\label{ponfluxkn}
\end{align} 
where $\kappa$ is the surface gravity of the black hole associated with $\partial_t$ and $n$ is the NUT parameter. Thus, for the Taub-NUT-AdS black hole, our entropy formula becomes
\begin {align}
S_{\rm NUT}^\alpha = \frac{A_{\mathcal B}}{4 G_4} + \frac{\pi L^2}{G_4} - \frac{2 \pi \alpha L^2 n \kappa}{G_4}
\end {align}
where we have set $\chi(\mathcal{B}) = 2$, since the horizon is a sphere.

In the presence of a stationary black hole with bifurcate Killing horizon $\mathcal B$, integration of the right side of \eqref{funidgrav} over a hypersurface $\Sigma$ extending from $\mathcal B$ to infinity now picks up a boundary term from $\mathcal B$ as well as from $C$. If we choose $\xi^\mu$ to be the horizon Killing field, then the left side of \eqref{funidgrav} vanishes. Since all Hamiltonian charges at infinity vanish, the boundary contribution from $C$ also vanishes. Thus, the analog of the first law of black hole mechanics with boundary conditions \eqref{eq:NLGravdualrot} with $\alpha \neq \infty$ becomes simply
\begin {align}
\delta S = 0
\label{firstlaw}
\end {align}
for all perturbations, i.e., stationary black holes are extrema of entropy in phase space. Of course, the phase space consists of solutions satisfying \eqref{eq:NLGravdualrot} at a given $\alpha$, so only variations that keep $\alpha$ fixed are permitted. 

If a Misner string is present in the spacetime, then one would need to excise a small tube around the string before integrating \eqref{funidgrav} over $\Sigma$. The integration of the right side of \eqref{funidgrav} over $\Sigma$ would then pick up an additional boundary term from the tube. Even if the metric perturbation, $\delta g$, is smooth, the quantity $\nabla_\alpha \xi_\beta$ appearing in the expression for $\boldsymbol{Q}_{\rm G}$ will be singular on the string, so the boundary term from the tube may make a nontrivial contribution in the limit as the tube shrinks down to the string. Thus, even for smooth metric perturbations, \eqref{firstlaw} may be modified by a boundary term from the tube for black holes with NUT charge.

Finally, we note that in the presence of a black hole, the linearization instability argument given above no longer applies. In particular, for $\xi^\mu$ taken to be the horizon Killing field, in place of \eqref{linin} we now obtain
\begin {align}
\int_\Sigma \boldsymbol{\omega} (g; \delta g, \mathcal{L}_\xi \delta g) = - \frac{\kappa}{2 \pi} \delta^2 S
\end {align}
which yields a formula for the second order variation of $S$ rather than a quadratic constraint on the linearized perturbation $\delta g$.

\begin{acknowledgments}
We wish to thank Pau Figueras, Donald Marolf, Savdeep Sethi, Simone Speziale, Claude Warnick, and, especially, Edward Witten for helpful discussions and correspondence. X. Wang would like to thank the organizers and participants of the workshop \textit{Timelike Boundaries in Classical and Quantum Gravity} at the Simons Center for Geometry and Physics in December 2025. This research was supported in part by NSF grant PHY 24-03584 to the University of Chicago.
\end{acknowledgments}

\appendix

\section {Linearized Weyl tensor in AdS$_{1, 3}$}
\label{sec:appfall}

In this appendix, we consider linearized metric perturbations $h_{\mu \nu}$ off of AdS$_{1, 3}$ background. We give formulas for the linearized Weyl tensor $C_{\mu \nu \rho \lambda}^{(1)} [h]$ in terms of the IW potentials $\Phi_V^{\rm G}$ and $\Phi_S^{\rm G}$.  

In AdS$_{1, d}$ the linearized Einstein equation takes the form
\begin {multline}
 - \frac{d}{L^2} h_{\mu \rho} =  + \frac{1}{2L^2} \big[g_{\mu \rho} h -  (1+d) h_{\rho \mu} \big] \\
 - g^{\nu \delta} \big( \nabla_\mu \nabla_{[\rho} h_{\delta] \nu}  - \nabla_\nu  \nabla_{[\rho} h_{\delta] \mu} \big) 
\end {multline}
where $g_{\mu \nu}$ here is the background AdS metric and $\nabla_{\mu}$ is its compatible covariant derivative, and $h = g^{\mu \nu} h_{\mu \nu}$. For solutions to the linearized Einstein equation, the linearized Weyl tensor takes the form
\begin {align}
        C_{\mu \nu \rho \lambda}^{(1)} &  = + \frac{1}{L^2} \big(g_{\mu [\rho} h_{\lambda] \nu} - g_{\nu[\rho} h_{\lambda] \mu} \big) \nonumber \\
        & \hspace{50 pt} - \big( \nabla_\mu  \nabla_{[\rho} h_{\lambda] \nu}  - \nabla_\nu  \nabla_{[\rho} h_{\lambda] \mu} \big) 
\end {align}

\newcommand{\AdS}{\mathrm{AdS}}
\newcommand{\hsn}{\widehat\nabla}
\newcommand{\hD}{\widehat D}
\newcommand{\hboxx}{\widehat\Box}
\newcommand{\eps}{\varepsilon}
\newcommand{\phS}{\phi^{\mathrm G}_{\mathrm S}}
\newcommand{\phV}{\phi^{\mathrm G}_{\mathrm V}}
\newcommand{\cS}{C^{(1,\mathrm S)}}
\newcommand{\cV}{C^{(1,\mathrm V)}}
\newcommand{\St}{\mathcal S}
\newcommand{\Zs}{\mathcal Z}
\newcommand{\Vt}{\mathcal V}

We now restrict our consideration to AdS$_{1, 3}$ and express $C_{\mu \nu \rho \lambda}^{(1)}$ in terms of the potentials $\Phi_V^{\rm G}$ and $\Phi_S^{\rm G}$. In order to do so, it is useful to introduce the following quantities
\begin{equation}
 \begin{aligned}
 \St_{ab}& \equiv\left(\hsn_a\hsn_b-\frac12g_{ab}\hboxx\right)(r \Phi^{\mathrm G}_{\mathrm S}),\\
 \Zs& \equiv \left(\hboxx-\frac2{L^2}\right)(r \Phi^{\mathrm G}_{\mathrm S}),\\
 \Vt_{ai}& \equiv \widehat\eps_i{}^{j}\hD_j
 \left(\eps_{ab}\hsn^b (r \Phi^{\mathrm G}_{\mathrm V})\right)
 \end{aligned}
 \label{eq:basic}
\end{equation}
where $\hboxx = g^{ab}\hsn_a\hsn_b$ and our index labeling conventions were stated at the end of section \ref{sec:intro}. Then, we have the IW decomposition into the scalar part and the vector part
\begin{equation}
 C^{(1)}_{\mu\nu\rho\sigma}
 =C^{(1,\mathrm S)}_{\mu\nu\rho\sigma}
 +C^{(1,\mathrm V)}_{\mu\nu\rho\sigma}
 \label{eq:split}
\end{equation}
The scalar part $C^{(1,\mathrm S)}_{\mu\nu\rho\sigma}$ is given by
\begin{align}
 \cS_{abcd} &=\frac1{L^2}\bigl(g_{a[c}\St_{d]b}-g_{b[c}\St_{d]a}\bigr) -\hsn_a\hsn_{[c}\St_{d]b}+\hsn_b\hsn_{[c}\St_{d]a} \label{eq:S_abcd} \\
 \cS_{abci}
 &=\hD_i\!\left[(\hsn_{[a}-u_{[a})\St_{b]c}\right] \label{eq:S_abci} \\
 \cS_{abij} & =0 \\
 \cS_{aibj} &=-\frac12\hD_i\hD_j\St_{ab}
 +\frac{r^2}{2}\gamma_{ij}\Bigl[
 \frac2{L^2}\St_{ab}-\frac12\hsn_a\hsn_b\Zs \nonumber \\
 & \hspace{15 pt} -\frac12u_a\hsn_b\Zs  -\frac12u_b\hsn_a\Zs 
 -u^c\bigl(\hsn_c\St_{ab} \nonumber \\
 & \hspace{15 pt} -\hsn_a\St_{bc}-\hsn_b\St_{ac}\bigr)
 \Bigr] \label{eq:S_aibj}\\
 \cS_{aijk}
 &=r^2\gamma_{i[j}\hD_{k]}\!\left(
 \frac12\hsn_a\Zs-u^b\St_{ab}\right)
 \label{eq:S_aijk}\\
 \cS_{ijkl}
 &=r^4\Bigl[
 \frac{\Zs}{2L^2}-\frac12u^a\hsn_a\Zs
 +u^au^b\St_{ab}-\frac12u^au_a\Zs
 \Bigr] \nonumber \\
 & \hspace{15 pt} (\gamma_{ik}\gamma_{jl}-\gamma_{il}\gamma_{jk})  -\frac{r^2}{4}\Bigl(
 \gamma_{jl}\hD_i\hD_k\Zs-\gamma_{jk}\hD_i\hD_l\Zs \nonumber \\
 & \hspace{15 pt} -\gamma_{il}\hD_j\hD_k\Zs +\gamma_{ik}\hD_j\hD_l\Zs
 \Bigr)
 \label{eq:S_ijkl}
\end{align}
where we have defined $ u_a \equiv \hsn_a\log r$ for convenience. On the other hand, the nonvanishing components of the vector part $C^{(1,\mathrm V)}_{\mu\nu\rho\sigma}$ are given by
\begin{align}
 \cV_{abci}
 &=\frac1{L^2}g_{c[a}\Vt_{b]i}
 -(\hsn_{[a}-u_{[a})
 \left(\hsn_c\Vt_{b]i}+u_{b]}\Vt_{ci}\right),
 \label{eq:V_abci}\\
 \cV_{abij}
 &=-\frac12\Bigl\{
 \hD_i\bigl(\hsn_a\Vt_{bj}-\hsn_b\Vt_{aj}-2u_a\Vt_{bj}+2u_b\Vt_{aj}\bigr)
 \notag\\[-2pt]
 &\hspace{15 pt}
 -\hD_j\bigl(\hsn_a\Vt_{bi}-\hsn_b\Vt_{ai}-2u_a\Vt_{bi}+2u_b\Vt_{ai}\bigr)
 \Bigr\},
 \label{eq:V_abij}\\
 \cV_{aibj}
 &=\frac12\hD_j\bigl(\hsn_a\Vt_{bi}-u_a\Vt_{bi}+u_b\Vt_{ai}\bigr)
 \nonumber \\
 & \hspace{15 pt} +\frac12\hD_i\bigl(\hsn_b\Vt_{aj}-u_b\Vt_{aj}+u_a\Vt_{bj}\bigr),
 \label{eq:V_aibj}\\
 \cV_{aijk}
 &=\hD_i\hD_{[j}\Vt_{a k]}
 +r^2\gamma_{i[k}\Bigl[
 \frac2{L^2}\Vt_{a j]}
 -u^b\bigl(\hsn_b\Vt_{a j]} \nonumber \\
 & \hspace{15 pt} -\hsn_a\Vt_{b j]}\bigr)
 -2u_au^b\Vt_{b j]}
 \Bigr],
 \label{eq:V_aijk}\\
 \cV_{ijkl}
 &=-r^2\Bigl[
 \gamma_{jk}\hD_{(i}(u^a\Vt_{a l)})
 -\gamma_{jl}\hD_{(i}(u^a\Vt_{a k)})
 \nonumber \\
 & \hspace{15 pt} +\gamma_{il}\hD_{(j}(u^a\Vt_{a k)})
 -\gamma_{ik}\hD_{(j}(u^a\Vt_{a l)})
 \Bigr].
 \label{eq:V_ijkl}
\end{align}
(All omitted components in the above formulas follow from the Weyl tensor symmetry properties.)

\section {Boundary conditions in AdS$_{1, 4}$ and AdS$_{1, 5}$}
\label{sec:apphigher}

\subsection {Maxwell theory in AdS$_{1, 4}$ and AdS$_{1, 5}$}
\label{sec:apphigher1}

As discussed at the beginning of subsection \ref{sec:EMBC}, the values of $\nu$ for the IW electromagnetic potentials $\Phi_S^{\rm EM}$ and $\Phi_V^{\rm EM}$  in different spacetime dimensions are given by $\nu_S^2 = (d-4)^2/4$ and $\nu_V^2= (d-2)^2/4$ (see \eqref{eq:EMnu}). Thus, taking account of the fact that the vector part is present only for $d \geq 3$, we see that $\nu \geq 1$---and, consequently, the IW boundary conditions are unique---in all cases except the following: (i) AdS$_{1, 3}$, where $\nu_S = \nu_V = 1/2$, (ii) AdS$_{1, 4}$, where 
$\nu_S = 0$, and (iii) AdS$_{1, 5}$ where $\nu_S = 1/2$. The AdS$_{1, 3}$ case was analyzed extensively in the main text. In this appendix, we consider the AdS$_{1, 4}$ and AdS$_{1, 5}$ cases and determine which of the IW boundary conditions are local in the fields and which are AdS-invariant.

Consider, first, AdS$_{1, 5}$. We have $\nu_V = 3/2$, hence there is a unique IW boundary condition for the vector part of the Maxwell field (namely, $a^V_{3/2} = 0$ in \eqref{eq:nonintnu}). On the other hand, $\nu_S = 1/2$, and therefore we have exactly the same one-parameter family of IW boundary conditions \eqref{eq:EMphiqS} for $\Phi_S^{\rm EM}$ as in AdS$_{1, 3}$, namely
\begin {align}
\big[ n^\mu \partial_\mu \Phi_S^{\rm EM} + q^S_{\frac{1}{2}} \Phi_S^{\rm EM} \big]_{\mathcal{I}} = 0 
\label{bcem15}
\end {align}
This expression can be reformulated as a local condition on the fields by using the fact that in AdS$_{1, 5}$, we have
\begin {align}
     F_{tr} = & - \frac{L}{r^3} \hat{D}^2 \Phi_S^{\rm EM}  \\ 
     F_{ai} = & \frac{1}{r^{2}} \epsilon_{ab} \hat{\nabla}^b \hat{D}_i \left( r \Phi_S^{\rm EM} \right) + \partial_a \bigg[\mathcal{O} \bigg(\frac{1}{r^3}\bigg)_V \bigg] \\
     F_{ij} = & \mathcal{O} \bigg(\frac{1}{r^3}\bigg)_V
\end {align}
where the terms labeled with a subscript $V$ arise from the vector part of the Maxwell field.
It can then be seen that the boundary condition \eqref{bcem15} is equivalent to 
\begin {align}
\big[ n^\mu \partial_\mu \big(\tilde{E}_t \big) + q_{\frac{1}{2}}^S \big( \tilde{E}_t \big) \big] \big|_{\mathcal{I}} = 0 
\end{align}
where $\tilde{E}_I = r F_{I \mu} n^\mu$. 

From an analysis of the properties of these boundary conditions (together with the unique IW condition $a^V_{3/2} = 0$ for the vector part) under boost transformations analogous to \eqref{eq:r}, we find that the unique AdS-invariant condition is the Dirichlet condition $q^S_{1/2} = \infty$, i.e., $a^S_{1/2} = 0$. Note that this contrasts with the situation in AdS$_{1, 3}$, where the unique AdS-invariant boundary condition with  $a^V_{1/2} = 0$ is the Neumann condition $q^S_{1/2} = 0$. Nevertheless, it can be seen from the asymptotic behavior of $F_{ai}$ and $F_{ij}$ that the Dirichlet condition $a^S_{1/2} = 0$ is equivalent to 
\begin {align}
 F_{IJ} \big|_{\mathcal{I}} = 0 ,
\label{d5adsem}
\end {align}
which is a direct analog of the usual boundary condition $\tilde{B}_I|_{\mathcal{I}} = 0$ in AdS$_{1, 3}$. 

Finally, for the AdS-invariant boundary conditions \eqref{d5adsem}, we find that properties \ref{proper1} and \ref{proper2} of subsection \ref{gencon} are satisfied using the unmodified Maxwell symplectic current, so there is no difficulty in providing a well-defined phase space structure to solutions. 

Next, we consider AdS$_{1, 4}$. The vector part has $\nu_V = 1$, so the IW boundary condition is unique. However, the scalar part has $\nu_S = 0$, so there is a one-parameter family of IW boundary conditions for the scalar part. In analogy with \eqref{eq:bfsBC}, these boundary conditions are given by
\begin {align}
\bigg[ r \big[\text{log} (r/L) - q^S_0/2 \big] \partial_r \big(r^{\frac{1}{2}}\Phi_S^{\rm EM} \big) - r^{\frac{1}{2}} \Phi_S^{\rm EM} \bigg]_{\mathcal{I}} = 0
\end {align}
This can be reformulated as the local condition
\begin {align}
\bigg[r \big[\text{log} (r/L) - q^S_0/2 \big] \partial_r  \tilde{E}_t  - \tilde{E}_t  \bigg]_{\mathcal{I}} = 0
\end {align}
where, again, $\tilde{E}_I = r F_{I\mu} n^\mu$. The only AdS-invariant boundary condition is the Dirichlet condition $q^S_0 = \infty$, which can be shown to be equivalent to 
\begin {align}
 F_{IJ} \big|_{\mathcal{I}} = 0.
\end {align}
Again, we find that the unmodified Maxwell symplectic current with these boundary conditions satisfies properties \ref{proper1} and \ref{proper2} of subsection \ref{gencon}, hence we obtain a well-defined phase space structure.

\subsection {Linearized gravity in AdS$_{1, 4}$ and AdS$_{1, 5}$}
\label{sec:apphigher2}

The formulas for $\nu_S$, $\nu_V$, and $\nu_T$ in linearized gravity were given in \eqref{eq:rescaledGT}. We have $\nu_T = d/2$, and since the tensor part exists only for $d \geq 4$, we have $\nu_T \geq 1$ in all cases. The formulas for $\nu_S$ and $\nu_V$ in the linearized gravitational case are identical to the electromagnetic case, so---in addition to AdS$_{1, 3}$---we need to specify boundary conditions only for the scalar parts of the linearized gravitational field in AdS$_{1, 4}$ and AdS$_{1, 5}$.

For AdS$_{1, 5}$, the one-parameter family of IW boundary conditions is again given by
\begin {align}
\big[ n^\mu \partial_\mu \Phi_S^{\rm G} + q^S_{\frac{1}{2}} \Phi_S^{\rm G} \big]_{\mathcal{I}} = 0 
\label{bcg15}
\end {align}
Defining
\begin {align}
E_{IJ}^{(1)} =  C_{I \mu J \nu}^{(1)} n^\mu n^\nu 
\end {align}
we find that to leading order
\begin {align}
E_{tt}^{(1)} = - \frac{3}{8} (\hat{D}^2+4) \hat{D}^2 \Phi_S^{\rm G} .
\end {align}
The boundary conditions \eqref{bcg15} can then be reformulated as the local conditions
\begin {align}
\bigg[ n^\mu \partial_\mu E_{tt}^{(1)} + q^S_{\frac{1}{2}} E_{tt}^{(1)} \bigg]_{\mathcal{I}}  = 0
\end {align}
In parallel with the electromagnetic case, we find that the only AdS-invariant boundary condition is $q^S_{1/2} = \infty$. From the asymptotic behavior of the linearized Weyl tensor in AdS$_{1, 5}$, it can be seen that this boundary condition is equivalent to 
\begin {align}
 r C_{rIJK}^{(1)} \bigg|_{\mathcal{I}} = 0
\end {align}
which is a direct analog of the usual boundary condition $\tilde{B}^{(1)}_{IJ}|_{\mathcal{I}} = 0$ in AdS$_{1, 3}$. A well-defined phase space structure with this boundary condition can then be obtained from the unmodified Einstein-Hilbert symplectic current.

For AdS$_{1, 4}$, the scalar part has $\nu_S = 0$ and the one-parameter family of IW boundary conditions is 
\begin {align}
\bigg[ r \big[\text{log} (r/L) - q^S_0/2 \big] \partial_r \big(r^{\frac{1}{2}}\Phi_S^{\rm G} \big) - r^{\frac{1}{2}} \Phi_S^{\rm G} \bigg]_{\mathcal{I}} = 0
\end {align}
This can be reformulated as the local condition
\begin {align}
\bigg[r \big[\text{log} (r/L) - q_0^S/2 \big] \partial_r E_{tt}^{(1)} - E_{tt}^{(1)} \bigg]_{\mathcal{I}} = 0
\end {align}
where, again, $E_{IJ}^{(1)} =  C_{I \mu J \nu}^{(1)} n^\mu n^\nu $.
The only AdS-invariant boundary condition is the Dirichlet condition $q^S_0 = \infty$, which can be shown to be equivalent to 
\begin {align}
 r C_{rIJK}^{(1)}  \bigg|_{\mathcal{I}} = 0
\end {align}
Again, we find that the unmodified Einstein-Hilbert symplectic current with these boundary conditions satisfies properties \ref{proper1} and \ref{proper2} of subsection \ref{gencon}, hence we obtain a well-defined phase space structure.

\section {Conservative case of Friedrich's Boundary Conditions}
\label{sec:appInh}

In a landmark paper, Friedrich \cite{Friedrich1995} proved well-posed evolution for Einstein's equation with negative cosmological constant in four-dimensional spacetimes of FG type. Analogous boundary conditions for well-posed evolution were subsequently given for Maxwell fields in AdS$_{1, 3}$ by Friedrich and Nagy \cite{FriNagy1999}. The general boundary conditions considered in these works were ``maximally dissipative'' and do not have a conserved energy. However, a special case of these boundary conditions---at the limit of the allowed range of parameters in their boundary conditions---is ``conservative''. In this appendix, we show that this limit of the Friedrich boundary conditions is essentially the same as the AdS-invariant IW boundary conditions considered in subsections \ref{sec:EMBC} and \ref{sec:NONLINGRAVBC}.

\subsection{Electromagnetic boundary conditions}

Friedrich and Nagy studied boundary conditions for Maxwell fields at a timelike boundary \cite{FriNagy1999}, and in four dimensions these conditions can be applied to the electromagnetic field at AdS conformal infinity
\begin {align}
 \left[\Phi_{2} - a \Phi_{0} - c \bar{\Phi}_0 \right] \big|_{\mathcal I} =  d
 \label{eq:FN1999}
\end {align}
where 
\begin {equation}
\Phi_2 = F_{\mu \nu} \bar{m}^\mu n_{\rm NP}^\nu, \quad \Phi_0  =  F_{\mu \nu} l^\mu m^\nu
\end {equation}
where $\{ l^\mu, n_{\rm NP}^\mu, m^\mu, \bar{m}^\mu \}$ is a suitable Newman-Penrose null tetrad with respect to the unphysical metric. In \eqref{eq:FN1999} $a$, $c$, and $d$ are complex functions on $\mathcal I$ satisfying $|a| + |c| \leq 1$. We are interested here only in the homogeneous conditions $d=0$. The ``covariant'' boundary conditions are $a= 0$ and $c$ constant. The ``conservative'' boundary conditions are the limiting case $|c| = 1$. Thus, the special case of the Friedrich-Nagy boundary conditions that we shall consider is 
\begin {align}
\left[ \Phi_{2} - c \bar{\Phi}_0 \right] \big|_{\mathcal I} = 0
\label{frconsem}
\end {align}
where $c$ is a complex constant with $|c| = 1$.

In terms of our rescaled fields $\tilde{E}_I$ and $\tilde{B}_I$, \eqref{frconsem} becomes
\begin {align}
- \frac{1}{\sin \theta} \tilde{B}_\phi - & \tilde{E}_\theta - i \tilde{B}_\theta + i \frac{1}{\sin \theta} \tilde{E}_\phi   \nonumber \\
= & \, c \left[\frac{1}{\sin \theta} \tilde{B}_\phi - \tilde{E}_\theta + i \tilde{B}_\theta + i \frac{1}{\sin \theta} \tilde{E}_\phi \right]
\label{frbcem1}
\end {align}
where evaluation on $\mathcal I$ is understood. Writing
\begin {align}
\alpha = i \frac{1 +c}{1-c}
\label{alphac}
\end {align}
it can be seen that $\alpha$ takes real values when $|c|=1$ and \eqref{frbcem1} becomes
\begin {align}
\left[ \tilde{E}_i + \alpha \tilde{B}_i \right] \big|_{\mathcal I} = 0
\label{frbcem2}
\end {align}
where $i$ denotes angular components. This differs from our boundary conditions \eqref{eq:EMdualrot} only in that the time component of \eqref{eq:EMdualrot} is not included. However, it follows from Maxwell's equations that when \eqref{frbcem2} holds, we have
\begin {align}
\partial_t \left[ \tilde{E}_t + \alpha \tilde{B}_t \right] \big|_{\mathcal I} = 0 .
\end {align}
It follows that when \eqref{frbcem2} holds, if $\left[ \tilde{E}_t + \alpha \tilde{B}_t \right] |_{\mathcal I} = 0$ at $t=0$ then this sum will vanish everywhere on $\mathcal I$. For $\alpha = \infty$, the same argument applies to $\tilde{B}_{I}$ alone.
Thus, the Friedrich-Nagy conservative boundary condition \eqref{frconsem} together with the restriction that the initial data at $t=0$ satisfies $\left[ \tilde{E}_t + \alpha \tilde{B}_t \right] \big|_{\mathcal I} = 0$ is precisely equivalent to our boundary condition \eqref{eq:EMdualrot}.

\subsection{Gravitational boundary conditions}

Friedrich's general boundary conditions for nonlinear gravity are \cite{Friedrich1995}
\begin {align}
\Psi_4 - a \Psi_0 - c \bar{\Psi}_0 = d
\label{eq:F1995}
\end {align}
where $a, c, d$ have the same restrictions as in the electromagnetic case and 
\begin {align}
\Psi_4 & =  \Omega^{-1} \tilde{C}_{\mu \nu \rho \lambda } (n_{\rm NP}^\mu) (\bar{m}^\nu) (n_{\rm NP}^\rho) (\bar{m}^\lambda) \\
\Psi_0 & =  \Omega^{-1} \tilde{C}_{\mu \nu \rho \lambda} (l^\mu) (m^\nu) (l^\rho) (m^\lambda)
\end {align}
where $\tilde{C}_{\mu \nu \rho \lambda }$ is the unphysical Weyl tensor. The homogeneous, covariant, conservative boundary conditions are 
\begin {align}
\left[ \Psi_{4} - c \bar{\Psi}_0 \right] \big|_{\mathcal I} = 0
\label{frconsgrav}
\end {align}
where $c$ is a complex constant with $|c| = 1$.

Evaluating $\Psi_4$ and $\Psi_0$ in terms of $\tilde{E}_{IJ}$ and $\tilde{B}_{IJ}$ in parallel with the electromagnetic case and, again, defining $\alpha$ by \eqref{alphac}, we find that the boundary conditions \eqref{frconsgrav} become
\begin {align}
\left[ \hat{E}_{ij} + \alpha \hat{B}_{ij} \right] \big|_{\mathcal I} = 0
\label{frbcgrav2}
\end {align}
where $\hat{E}_{ij}$ and $\hat{B}_{ij}$ are the trace free parts of $\tilde{E}_{ij}$ and $\tilde{B}_{ij}$, i.e.,
\begin {align}
\hat{E}_{ij} = \tilde{E}_{ij} - \frac{1}{2} \gamma_{ij} \gamma^{kl}\tilde{E}_{kl} , \quad
\hat{B}_{ij} = \tilde{B}_{ij} - \frac{1}{2} \gamma_{ij} \gamma^{kl}\tilde{B}_{kl}
\end {align}
where $\gamma_{ij}$ is the unphysical metric on the cross-sections of $\mathcal I$ at constant $t$.
Note that we have 
\begin {align}
\hat{E}_{ij} =0 \quad \Leftrightarrow \quad J_{ij} = 0
\end {align}
where $J_{ij}$ is the quantity introduced by Comp\`ere, Fiorucci, and Ruzziconi \cite{CFR2019}, who proposed the ``Neumann" boundary condition $J_{ij} = 0$. Thus, we see that the boundary condition $J_{ij} = 0$ of \cite{CFR2019} is precisely the same as Friedrich's boundary condition for the case $\alpha = 0$.

Friedrich's boundary condition \eqref{frbcgrav2} differs from our conditions \eqref{eq:NLGravdualrot} only in that the $tI$ components are not included in \eqref{frbcgrav2}.  (We have not listed the trace of $\tilde{E}_{ij} + \alpha \tilde{B}_{ij}$ among the components not included, since---by virtue of the tracelessness of $\tilde{E}_{IJ}$ and $\tilde{B}_{IJ}$---it will vanish once the $tt$ component of \eqref{eq:NLGravdualrot} is satisfied.) However, when Einstein's equation holds, it follows from the Bianchi identity that
\begin {align}
\nabla^{(0)I} \big[\tilde{E}_{IJ} + \alpha \tilde{B}_{IJ} \big] \big|_{\mathcal{I}} =0
\end {align}
where $\nabla_I^{(0)}$ is the derivative operator associated with the boundary metric $g_{(0)IJ}$.
Using the tracelessness of the Weyl tensor, it then follows that if \eqref{frbcgrav2} holds and if $\left[ \tilde{E}_{tI} + \alpha \tilde{B}_{tI} \right] |_{\mathcal I} = 0$ at $t=0$, then this $tI$-component sum will vanish everywhere on $\mathcal I$.
At $\alpha = \infty$, the same argument applies to $\tilde{B}_{IJ}$ alone. Thus, the Friedrich conservative boundary condition \eqref{frconsgrav} together with the restriction that the initial data at $t=0$ satisfies $\left[ \tilde{E}_{tI} + \alpha \tilde{B}_{tI} \right] |_{\mathcal I} = 0$ is precisely equivalent to our boundary condition \eqref{eq:NLGravdualrot}.

Friedrich's proof of well-posedness applies to the special case of the boundary conditions \eqref{frconsgrav}. Imposition of the additional restriction $\left[ \tilde{E}_{tI} + \alpha \tilde{B}_{tI} \right] |_{\mathcal I} = 0$ on initial data at $t=0$ does not affect the well-posedness. Therefore, it follows from Friedrich's results \cite{Friedrich1995} that the initial value formulation with our boundary conditions \eqref{eq:NLGravdualrot} is well-posed. As already has been mentioned several times above, this result also has been proven independently by Sou\^{e}tre \cite{Souetre2025}.


\bibliography{final}

\end{document}